\documentclass[11pt,letterpaper]{article}
\usepackage[margin=1in]{geometry}
\usepackage[T1]{fontenc}
\usepackage{lmodern}
\usepackage{amsmath,amssymb,amsthm,mathtools,booktabs,array,enumitem}
\usepackage{microtype,xurl}
\usepackage[hypertexnames=false,colorlinks=true,linkcolor=blue,citecolor=blue,urlcolor=blue]{hyperref}
\usepackage[numbers,sort&compress]{natbib}
\newtheorem{theorem}{Theorem}[section]
\newtheorem{lemma}[theorem]{Lemma}
\newtheorem{proposition}[theorem]{Proposition}
\newtheorem{corollary}[theorem]{Corollary}
\theoremstyle{definition}

\newcommand{\R}{\mathbb R}
\newcommand{\E}[1]{\mathbb E\!\left[#1\right]}
\newcommand{\Prb}[1]{\mathbb P\!\left[#1\right]}
\newcommand{\eps}{\varepsilon}

\newcommand{\ind}{\mathbf 1}

\DeclareMathOperator*{\argmax}{arg\,max}

\hypersetup{pdftitle={The Query Complexity of Envy-Free Division: Cake, Chores, and Mixed Resources},pdfauthor={Alireza Farhadi and Mohammadtaghi Hajiaghayi}}
\title{The Query Complexity of Envy-Free Division: Cake, Chores, and Mixed Resources}
\author{Alireza Farhadi\\Profound\and Mohammadtaghi Hajiaghayi\\University of Maryland, College Park}
\date{}
\begin{document}
\maketitle
\begin{abstract}
We study the query complexity of envy-free division for cake, chores,
and resources with both positive and negative values. For arbitrary
nonnegative nonatomic chore costs, we give a deterministic exact protocol
using $2^{n\log_2 n+O(n)}$ queries and cuts. For both cake and chores,
a Las Vegas protocol uses $O(n^{12}\log(2n))$ expected queries and
$O(n^{11}\log(2n))$ expected cuts. Conversely, every exact protocol with
fixed positive success probability requires $\Omega(n^2)$ expected queries
in general, even with arbitrarily many output cuts at no query cost.
For mixed cake and chores, no uniform finite
expected-query bound exists even when every Cut returns the leftmost root
or every Cut returns the rightmost root. This holds for any fixed positive
multiplicative approximation factor and success
probability, already when all net totals are one. An additive impossibility
holds on the same normalized domain.
\end{abstract}
\clearpage
\section{Introduction}\label{sec:introduction}
Envy-free division seeks an allocation in which no agent prefers another
agent's share to its own. For cake, agents seek greater value; for chores,
they seek smaller cost. A resource containing both desirable and
undesirable parts requires signed valuations. When preferences are
accessible through queries, the algorithmic question is how much
information is needed to construct an envy-free allocation. We study
complete allocations into finite unions of intervals and query bounds
that depend only on the number of agents.

In the Robertson--Webb model, a query asks one agent either to evaluate an
interval or to cut off a specified value. Aziz and Mackenzie~\cite{AM16}
gave a bounded protocol for general cake cutting. Dehghani, Farhadi,
Hajiaghayi, and Yami~\cite{DFHY18} established boundedness for chores.
Recent preprints give $n^{O(1)}2^n$ deterministic cake queries~\cite{YB26}
and $\widetilde O(n^{30})$ expected queries for exact randomized cake
cutting~\cite{MS26}. The remaining questions include how far these bounds
can be reduced, what randomization can achieve for chores, and whether
bounded-query guarantees extend to signed valuations.

For chores, we give a deterministic exact protocol using
$2^{n\log_2 n+O(n)}$ queries. For both cake and chores, we give a Las Vegas
protocol using $O(n^{12}\log(2n))$ expected queries and
$O(n^{11}\log(2n))$ expected cuts. These upper bounds apply to arbitrary
finite nonnegative nonatomic measures, without density, common-support,
or strict-positivity assumptions. The randomized bound improves the known
cake bound and gives a polynomial expected-query guarantee for chores.

We also prove two lower bounds. For either cake or chores, every exact
protocol with a fixed positive success probability requires
$\Omega(n^2)$ expected queries in general, even with arbitrarily many
output cuts at no query cost. The quantitative bound is linear in the
required success probability. For mixed cake and chores, no uniform finite
expected-query bound exists in the signed bracketed model, even with
an oracle that always returns the leftmost root, or always returns the
rightmost root. This
impossibility holds for every fixed positive multiplicative approximation
factor and success probability, and has an additive counterpart. It
already applies when every agent values the whole resource at one;
positive and negative values may have arbitrarily large total variation.

The deterministic chore bound comes from separating compensation from
reassignment. Partial allocations can then be combined without requiring
their entire permission graphs to agree; a factorial potential bounds
the number of original allocations needed. For the randomized bound, we
select compensation pieces by a rule stable under deletion of an
unprotected sample, reduce rounding discrepancy to a linear dependence
on the number of agents, and control exceptional endpoints across repeated
correction attempts. Together these refinements reduce the sample budget
needed to realize numerical compensation with disjoint pieces.
The lower bounds preserve query transcripts while changing envy comparisons.
For nonnegative valuations we use localized perturbations. For mixed
valuations, a moving frontier separates the recorded query coordinates
from a final interval that every acceptable allocation must split.
A bound on the cumulative values preserves every first-root answer,
including replies at the endpoint. Each refinement makes a positive
fraction of the remaining executions fail permanently. Finitely many
refinements give one fixed hard input, independently of the protocol's
realized random tape.

\clearpage
\begin{table}[t]
\centering\small
\renewcommand{\arraystretch}{1.14}
\begin{tabular}{@{}>{\raggedright\arraybackslash}p{.19\linewidth}>{\raggedright\arraybackslash}p{.35\linewidth}>{\raggedright\arraybackslash}p{.40\linewidth}@{}}
\toprule
Setting & Prior benchmark & This paper\\
\midrule
Deterministic\newline exact chores & A bounded function of $n$~\cite{DFHY18} (SODA 2018) & $Q,C\le2^{n\log_2n+O(n)}$; Theorem~\ref{thm:deterministic-chores}.\\
\midrule[.25pt]
Randomized exact cake & $\E{Q}=\widetilde O(n^{30})$~\cite{MS26} (preprint) & $\E{Q}=O(n^{12}\log(2n))$, $\E{C}=O(n^{11}\log(2n))$; Theorem~\ref{thm:general-upper}.\\
\midrule[.25pt]
Randomized exact chores & A bounded expected-query guarantee follows from~\cite{DFHY18} & The same polynomial bounds as cake; Theorem~\ref{thm:general-upper}.\\
\midrule[.25pt]
Exact cake and\newline chore lower\newline bounds & $\Omega(n^2)$ deterministic cake queries~\cite{Procaccia09} (IJCAI 2009) & $\E{Q}\ge\max\{n(np-1)/2,n^2p/4-1\}$, even with arbitrarily many output cuts at no query cost; Theorem~\ref{thm:nonnegative-lower}.\\
\midrule[.25pt]
Mixed cake and chores & Connected EF existence for signed valuations with positive net totals~\cite{BV26} (SODA 2026) & No uniform finite $\E{Q}$, even for any fixed positive multiplicative factor and success probability; also for fixed additive $\eps<1$. Theorem~\ref{thm:mixed-mc}.\\
\bottomrule
\end{tabular}
\caption{Four general query-complexity results. $Q$ counts all queries and $C$
counts Cut calls. Randomized upper bounds are Las Vegas and use ideal-real
sampling. The mixed lower bound holds for both always-leftmost and always-rightmost
Cut rules, net totals one, and no common total-variation ceiling. Theorems
state the precise oracle and randomness assumptions.}
\label{tab:comparison}
\end{table}

\section{Model and main results}\label{sec:model}\label{sec:results}
The resource is $I=[0,1]$. An allocation is a complete partition
$(A_1,\ldots,A_n)$ into finite unions of intervals. Cake values $v_i$
and chore costs $c_i$ are finite nonnegative nonatomic Borel measures.
Envy-freeness means $v_i(A_i)\ge v_i(A_j)$ for cake and
$c_i(A_i)\le c_i(A_j)$ for chores, for every $i,j$.
For mixed cake and chores, $v_i$ is a finite signed nonatomic measure
and larger values are preferred. Individual parts can be good for one
agent and bad for another.

An \emph{Eval} query asks for one agent's value of a specified interval.
For a nonnegative measure, a \emph{Cut} query specifies $[a,b]$ and
$t\in[0,v_i([a,b])]$ and returns any $x\in[a,b]$ with
$v_i([a,x])=t$. For signed measures the legal target interval is
$[\min\{0,v_i([a,b])\},\max\{0,v_i([a,b])\}]$.
The signed oracle may return any legal root; it is not required to return
an extremal root. Each lower-bound theorem specifies the allowed root
policy. We count all oracle calls as $Q$ and Cut calls as $C$.
These counts are distinct from the number of boundaries in the final
allocation. Exact real arithmetic and internal computation are uncharged.
Our upper bounds are query bounds, without a polynomial bit-complexity
claim.

A randomized success is a legal finite return satisfying the required
fairness condition. Incorrect returns, illegal calls and nonreturn count
as failures. Expected costs include unsuccessful trials. A Las Vegas
protocol returns correctly almost surely. The positive algorithms use
independent uniform-real samples and exact real computation; the lower
bounds allow arbitrary measurable ideal-real randomness.

A measurable randomized protocol has an input-independent tape on a
probability space $(\Omega,\mathcal F,\mathbb P)$. Finite transcripts,
query actions, and finite interval-output encodings have their standard
Borel structures, using countable disjoint unions over lengths and types.
The next-action rule is jointly measurable in the tape and the finite
transcript. Its actions include requests, finite returns, and query-free
nonreturn; this convention does not require nonreturn to be decidable.

\clearpage
\subsection{Deterministic chores}
\begin{theorem}\label{thm:deterministic-chores}
For $n\ge3$ finite nonnegative nonatomic chore costs, a deterministic
Eval/Cut protocol returns a complete exactly envy-free allocation using
\[
 Q,C\le 2^{n\log_2 n+O(n)}.
\]
The guarantee holds for every legal Cut response. The one- and two-agent
cases have constant-size protocols.
\end{theorem}
Appendix~\ref{r8c:appendix} gives the construction and the explicit bound
$n!\,2^{25n+18}$. The construction defers reassignment until compensation
has been paid and bounds the number of snapshots represented by each live
allocation using a factorial potential.
A $2^{O(n)}$ bound, and in particular the $n^{O(1)}2^n$ bound known for cake,
remain open for chores.

\subsection{Randomized cake and chores}
\begin{theorem}\label{thm:general-upper}
For every profile of $n$ finite nonnegative nonatomic measures, interpreted
as cake values or as chore costs, a Las Vegas Eval/Cut protocol returns a
complete exactly envy-free allocation with
\[
 \E{Q}=O(n^{12}\log(2n)),\qquad
 \E{C}=O(n^{11}\log(2n)).
\]
The protocol uses exact real arithmetic and independent uniform-real
samples. The bounds hold uniformly over profiles and legal measurable
nonanticipating Cut policies, including all rejected work.
\end{theorem}
An oracle policy is a measurable function of the answered query transcript
and the current query. The full proof is in
Appendix~\ref{r6u:appendix}. For cake, the result improves a known polynomial
expected-query bound. The same construction supplies a polynomial
expected-query bound for chores.

\subsection{A randomized lower bound}
\begin{theorem}\label{thm:nonnegative-lower}
Fix $n\ge2$ and $p\in(0,1]$. Suppose a measurable randomized interval-query
protocol returns a complete exactly envy-free allocation with probability
at least $p$ on every normalized finite-step cake profile, or on every
such chore profile, whose densities lie in $[1/2,3/2]$. Its unconditional
expected query count on the uniform profile satisfies
\[
 \E{Q_{\mathrm{unif}}}\ge
 \max\left\{0,\frac{n(np-1)}2,\frac{n^2p}4-1\right\}.
\]
The bound already holds for the expected number of answered oracle calls.
The protocol may use ideal-real randomness and return allocations with
arbitrarily many intervals, choosing their endpoints without oracle calls.
\end{theorem}
In particular, the lower bound is $\Omega_p(n^2)$ for every fixed $p>0$,
with a coefficient linear in $p$. The protocol must work on the entire
stated class, although the expected cost is charged on the uniform input.
The result covers unsuccessful and nonterminating executions. It extends
the quadratic lower bound to randomized failure and chores without
increasing its order. Appendix~\ref{r6lb:appendix} gives the proof.

\subsection{Mixed cake and chores}
For signed valuations, define \emph{$\alpha$-multiplicative envy-freeness},
where $0<\alpha\le1$, by the relative-envy condition
\begin{equation}\label{eq:signed-relative-ef}
 v_i(A_j)-v_i(A_i)
 \le (1-\alpha)\max\{|v_i(A_i)|,|v_i(A_j)|\}
 \qquad(i,j\in[n]).
\end{equation}
This definition handles either sign and zero values without dividing by a
net total. For nonnegative cake values it is the usual guarantee
$v_i(A_i)\ge\alpha v_i(A_j)$~\cite{BK23}; for nonnegative chore costs it
is $c_i(A_i)\le c_i(A_j)/\alpha$. At $\alpha=1$ it is exact EF.
Our hard instances have total value one for every agent. Any allocation
satisfying the relative guarantee then gives each agent positive own value,
so the condition is equivalent to the usual cake inequality.
Lemma~\ref{r6m:relative-equivalence} gives the proof. Hardness on this
subclass suffices for the general signed domain.

We state additive approximation separately. On profiles normalized by
$v_i(I)=1$, additive $\eps$-envy-freeness means
$v_i(A_i)\ge v_i(A_j)-\eps$. This tolerance is in net-value units.
\begin{theorem}\label{thm:mixed-mc}
Fix $n\ge3$, a success probability $p\in(0,1]$, and a multiplicative factor
$\alpha\in(0,1]$. For finite signed nonatomic measures,
no measurable randomized protocol can guarantee both
$\alpha$-multiplicative envy-freeness with probability at least $p$ on
every input and a uniform finite expected-query bound. This already holds
when $v_i(I)=1$ for every agent. On that normalized subclass, the same
conclusion holds for additive $\eps$-envy-freeness for every fixed
$0\le\eps<1$.

The oracle uses signed bracketed Eval/Cut queries. The impossibility
holds when every Cut returns the leftmost root, and also when every Cut
returns the rightmost root. No common total-variation ceiling is imposed.
Ideal-real randomness is allowed, and output cut locations may be chosen
without queries, with no bound on their number.
\end{theorem}
The theorem concerns the information needed to find an allocation.
It does not rule out existence, bound the minimum number of output cuts,
or require a single input to have infinite expected cost. It already
holds for finite piecewise-constant densities. Appendix~\ref{r6m:appendix}
proves both approximation statements together through the inequality
$\alpha v_i(A_j)-v_i(A_i)\le\eps$, for $0\le\eps<\alpha\le1$.
The absence of a common variation ceiling allows increasingly large
positive and negative masses to cancel while every net total stays one.
The root rule is prescribed before the input and random tape are chosen.
The conclusion also rules out uniformly bounded protocols that must work
for every legal root selector. It does not cover a class with a fixed
common bound on total variation.

\section{Proof overview}\label{sec:proof-ideas}
\subsection{Deterministic chores: deferred compensation}
A partial chore allocation leaves a residue that must eventually be
assigned. Reassigning existing pieces can reduce an agent's cost and
create enough slack to absorb this residue. The difficulty is preserving
every other agent's envy comparison during the reassignment. We record
these obligations relative to the original pieces and pay them before
changing the allocation.

Each live partial allocation has a directed graph of permitted recipient
changes. A package contains one such live allocation and the allocations
already frozen to pay its obligations. Its weight is the number of
original allocations it contains. A compensation step combines packages
that agree on just one source component. Their other components may
differ. An Offset SubCore call chooses the donors and the surviving live
allocation; only the donors change their physical allocations. The
survivor acquires a paid permission. A cycle of paid permissions can then
be contracted without moving any material.

The key bound uses a capacity for each component. An unpaid component of
size $d$ has capacity $u(d)=8^{d-1}d!$; paying it multiplies this capacity
by $2(d+1)$. A package's capacity is the product over its components.
Grouping package weights between consecutive powers of two ensures that
combining $d+1$ packages costs at most this multiplier, regardless of
which package survives. The inequality
\[
 [2(a+1)u(a)]\,[2(b+1)u(b)]\le u(a+b)
\]
shows that contracting paid components preserves the capacity bound.
Every package consequently has weight at most $u(n)$. There are only
$\operatorname{poly}(n)4^n$ possible classes of next operations, so
$n!2^{O(n)}$ original allocations suffice to keep the process moving
until a terminal class permits recursive completion.

The accounting uses actual cost reductions from the original baselines,
rather than the whole costs of removed suffixes. Frozen allocations never
change again, and every removed piece remains in the unallocated domain.
The terminal margins therefore cover the entire remaining resource.
Preprocessing, compensation, and recursive endpoint recording preserve
the bound $n!2^{O(n)}=2^{n\log_2n+O(n)}$.

\subsection{Randomized upper bounds: exact repair}
The randomized protocol alternates between repairing approximate
envy-freeness and creating a comparison that remains safe after the
remaining resource is allocated. Repair takes whole existing bundles and
an unallocated buffer. An individual-agent quantile grid and independent
mixture samples produce a finite value table. Small witness pieces perturb
an assignment problem; a maximizing assignment for cake, or minimizing
assignment for chores, determines credits or taxes. Disjoint physical
payments then make that assignment exactly envy-free while leaving at least
three quarters of the buffer. We use the compensation framework of
Mackenzie--Suzuki~\cite{MS26}, with the following improvements.

First, we retain a large compensation domain deterministically. For each
row pair, maximize the absolute determinant of two cell-value vectors,
then their total mass, allowing repeated cells. Remove their coarse
parents from later selection.
The two cells control every linear combination of the rows, including the
normalizations determined after selection. The total removed value is
small. Moreover, if an inserted point is unprotected in the augmented run
and its coarse parent survives in the original run, deleting that point
leaves the selections unchanged. An
exchangeability argument bounds the measure of the exceptional coordinates
by the number of protected endpoints divided by the sample count. Outside
this set, partial-cell discrepancy is controlled by the largest whole-cell
discrepancy. This avoids conditioning a fixed-measure sampling estimate on
a domain selected from the same samples.

Second, whole-cell rounding incurs only linear discrepancy. Keep each
recipient's own-value equation exact, and drop a pairwise-discrepancy
equation only when it has at most $4k$ fractional coordinates, where $k$
is the active population. A rank and incidence count guarantees such an
equation whenever too many fractional coordinates remain. Supported sorted-slot
rounding and trimming preserve fixed zero and one coordinates for real
recipients, so each dropped equation
stays within $4k$ times the largest cell disagreement. Fresh circular cuts
fill the remaining own-value deficits. A maximum-cell envelope gives
constant success probability per correction attempt; a stopped retry
argument controls the first exceptional endpoint without a factor equal
to the number of retries. All inherited boundaries of earlier accepted
corrections enter the error estimate.

These changes allow separation parameter $\lambda=\Theta(n)$,
$L=\Theta(n^4\log(2n))$ witnesses per pair and batch, and
$m=\Theta(n^9\log(2n))$ samples per trial. No density, common-support,
or hunger assumption is needed. An accepted Snapshot trial either completes
the allocation or supplies an envy-free partial allocation with a certified
positive margin. Prescribed shrinking and
whole-bundle transfers turn that margin into a new domination comparison,
following the global framework of Psomas~\cite{Psomas26}. A potential over
all original observers increases at committed allocations and is at most
$n(n-1)$. Every iteration uses an expected constant number of trials, each
costing $O(nm)$ queries and $O(m)$ Cuts, including rejected work. This gives
$\E{Q}=O(n^{12}\log(2n))$ and $\E{C}=O(n^{11}\log(2n))$ for both cake
and chores. Explicit zero-value and small-population cases give almost sure
termination with a complete allocation.

\subsection{Randomized lower bound: indistinguishable perturbations}
On the uniform profile, every exact envy-free allocation gives each agent
length $1/n$. We perturb a uniformly chosen agent's density by a narrow
bump inside a wider circular arc, subtracting the bump's mass uniformly
over that arc. The density remains normalized and between $1/2$ and
$3/2$. The wider arc has length just above $1/n$. Unless a recorded query
endpoint meets it, all answers remain unchanged. A query detects the
perturbation with probability at most approximately $2/n^2$: it must
query the perturbed agent and meet the arc with one of its endpoints.

If the narrow bump avoids output boundaries, its owner is well defined.
Every other perturbed agent strictly envies that owner; reversing the
perturbation gives the chore version. Perturbing two distinct agents
removes the exception that the perturbed agent owns the bump. Each agent
is then selected with probability $2/n$, and the detection probability
is at most approximately $4/n^2$ per query.

One sufficiently small bump width preserves strictly unfair baseline
returns and avoids finite output endpoint lists with arbitrarily high
probability, even when their expected size is unbounded. Exact replay
preserves the other failures. Averaging the success guarantee over the
fixed perturbed inputs and then taking the wider arc's length down to
$1/n$ gives the two terms of the lower bound. Retaining failed executions
is essential: it makes the bound linear in the required success
probability, rather than applying only to protocols that always succeed.

\subsection{Mixed resources: signed cancellation}
For signed valuations, a finite query history can conceal a final interval
that every fair allocation must split. We construct such histories while
requiring every Cut to return the leftmost root.

Maintain a frontier $\ell<1$. The densities are constant on finitely many
intervals before $\ell$, and all agents have positive constant density
after it. Each resolved interval is either bad for everyone or good for
only one agent and sufficiently bad for all the others. This condition
ensures that, if one agent receives the entire final interval, the average
envy of the other agents toward it exceeds the permitted tolerance.
The earlier intervals may be divided arbitrarily; the obstruction does
not depend on their remaining whole.

To answer a query to agent $j$, leave that agent's entire valuation
unchanged and move the frontier beyond the query coordinates smaller than
one. Give every other agent negative density on the newly resolved
interval and restore its net total using a positive density on the
remaining tail. All cumulative functions stay strictly below one before
the endpoint. Earlier roots inside the resolved prefix are unchanged;
an earlier root at one is still first because its absolute target level
is one. Thus arbitrary legal brackets and negative targets are allowed.

For a randomized protocol, we perform the same construction on measurable
sets of tapes. At each query, a row occurring on at least a $1/n$ fraction
of the current tapes is selected, and a fixed frontier captures at least
half their coordinates. After a fixed number of queries, another
coordinate quantile captures the endpoints of their finite outputs.
Those outputs leave the remaining interval unsplit and hence fail.
Their transcripts and bundle values remain fixed in all later refinements.

Each batch therefore makes a positive fraction of the remaining tapes
fail permanently. Finitely many batches reduce the probability of a
correct return within any prescribed query horizon below $p/2$ on one
fixed input. A protocol with success probability at least $p$ must then
exceed that horizon with probability at least $p/2$. This rules out a
uniform finite expected-query bound. Reflection transfers the result to
an always-rightmost oracle. The same argument supplies the multiplicative
and additive comparisons by changing the density ratios in the resolved
intervals.

\section{Related work}\label{sec:comparison}
\paragraph{Bounded protocols.}
The general cake protocol of Aziz and Mackenzie~\cite{AM16} and the chore
protocol of Dehghani, Farhadi, Hajiaghayi, and Yami~\cite{DFHY18} establish
boundedness with disconnected shares. Our deterministic construction uses
the chore compensation framework and the cake SubCore primitive, with the
order and tie conventions of Ye and Bai~\cite[Appendix F]{YB26}.
The latter preprint gives $n^{O(1)}2^n$ deterministic cake queries.
Improving the chore bound to this scale remains open. The snapshot
budgets in~\cite{DFHY18} count intermediate components of the protocol;
Table~\ref{tab:comparison} records its general boundedness guarantee.

\paragraph{Randomized upper bounds.}
Mackenzie and Suzuki~\cite[Theorem 1 and Remark 1.1]{MS26} give a polynomial
expected-query protocol for exact cake cutting. We retain its numerical
compensation and physical realization framework, and handle both
preference directions. Our smaller sample budget follows from stable
compensation selection, linear-discrepancy rounding, and endpoint accounting
across repeated correction attempts. The assignment sensitivity argument
uses~\cite{MMWY21}; the rounding argument builds on the sorted-slot
construction of Shmoys and Tardos~\cite{ST93}. Global completion uses the domination and
bundle-transport framework of Psomas~\cite{Psomas26}. The three recent cake papers
\cite{MS26,YB26,Psomas26} are cited as versioned preprints.

\paragraph{Lower bounds and approximation.}
Procaccia~\cite{Procaccia09} proves a quadratic deterministic cake-query
lower bound. Farhadi and Hajiaghayi~\cite{FH18} give a proportional chore
lower bound in their model. Our exact EF lower bound permits randomized
failure and arbitrary finite output fragmentation. A cubic lower bound
is announced in~\cite[Footnote 2]{AM17} and the thesis's repository
abstract~\cite{Mac17};
the cited public thesis version omits the announced proof. We distinguish
that announcement from the proved benchmarks in Table~\ref{tab:comparison}.
For connected approximate cake division, Br\^anzei and Nisan~\cite{BN22}
give query bounds for additive approximation, and Barman and
Kulkarni~\cite{BK23} give polynomial-time algorithms with additive and
multiplicative guarantees. Our positive upper bounds require exact EF and
permit disconnected shares.

\paragraph{Signed valuations.}
Barman and Verma~\cite{BV26} establish connected envy-free existence for the
relevant signed valuations with positive net totals. Our mixed theorem
separates existence from uniformly bounded query complexity. The lower bound holds even with an always-leftmost or always-rightmost
Cut oracle; its additive tolerance is measured in net-value units.
It does not imply hardness for total-variation-normalized approximation.
Protocols based on an initial decomposition into regions of constant
sign~\cite{OM2020} assume access to information that bounded interval queries
need not reveal.

\label{main:lastpage}
\clearpage
\phantomsection
\addcontentsline{toc}{section}{References}
\bibliographystyle{alphaurl-safe}
\bibliography{references}
\clearpage
\appendix
\section{Deterministic chore division by deferred compensation}
\label{r8c:appendix}

The protocol maintains a separate graph of paid permissions for each live
snapshot. Paying a permission does not immediately change that snapshot's
allocation. This separates compensation from component contraction: roots
can be combined when they agree on one source component, even if all their
other components differ. A factorial capacity and grouping by powers of two
bound the number of original snapshots that any live package can contain.

\begin{theorem}\label{r8c:theorem}
For $n$ finite nonnegative nonatomic Borel costs on an interval, a
deterministic ordinary Eval/Cut protocol produces a complete exactly
envy-free allocation into finite unions of intervals. For $n\ge3$, its
query and cut counts satisfy
\[
 Q,C\le n!\,2^{25n+18}\le 2^{n\log_2 n+25n+18}.
\]
The protocol permits every legal Cut reply and includes all preprocessing,
unsuccessful primitive trials, and recursive calls. Exact comparisons and
finite computation are free; no time or bit-complexity bound is asserted.
For one or two agents, the standard direct protocols suffice.
\end{theorem}

The Core and near-equal splitter are proved in
Appendix~\ref{r6core:appendix}. The source SubCore proof is
Lemma~\ref{dc:subcore}; the parent-dependent offset implementation is
Theorem~\ref{r7offset:theorem}. The whole-piece augmentation and two-group
completion used below are proved in Appendix~\ref{r6c:completion}.

\subsection{Ordinary queries, preprocessing, and a protected label}
\label{r8c:preprocessing}

All domains are recorded ordered finite unions, with a fixed convention
assigning each shared endpoint to one part.  Every new endpoint is
recorded for the current observers.  Cached component costs locate a
strict prefix or suffix target in one physical interval.  One ordinary
Cut supplies the cutting observer's value, and at most $n-1$ Evals record
the other observers.  Zero, full, and known-endpoint targets reuse
records.  Null components remain present.  Thus no disconnected domain
is treated as a unit-cost interval query.

At a completed envy-free boundary, an agent with zero cost for the whole
remaining domain can receive that domain: her additional own cost is zero,
and every other agent's rival cost only increases.  An empty domain is
finished immediately.  One agent receives everything, and two agents use
an equal-own-cost cut followed by cheaper choice.  A continuing call thus
has $n\ge3$ and positive reference costs.

The Core of Theorem~\ref{r6core:theorem} partitions its input into $n$
equal-cost parents for a designated cutter and returns an envy-free
allocation of one prefix of a distinct parent to each agent, including
at least one whole parent.  If the cutter's input, own allocation, and
residue have costs $t,b,u$, respectively, then
\begin{equation}\label{r8c:core-progress}
 u\le(1-1/n)t,\qquad u\le(n-1)(t/n-b).
\end{equation}
Indeed, one allocated parent has cost $t/n$, and every other allocated
prefix costs at least $b$ to the cutter.  If $u>0$, the whole-parent
recipient is a noncutter and her piece has cutter gap at least $u/(n-1)$.

Fix the integers $J,H,L$ in Section~\ref{r8c:ledger} before querying.
Generate $J$ Core snapshots on successive residues with one cutter $c$,
unless a complete or zero-cost exit occurs.  In snapshot $s$, let
$P_{s,p}$ be the actual prefix originally returned to agent $p$, and put
$b_{s,i}=C_i(P_{s,i})$.  The larger initial cutter parents are not
$P_{s,p}$.  Record every equalizer
\[
 T_{s,p,i}\subseteq P_{s,p},\qquad C_i(T_{s,p,i})=b_{s,i},
\]
using $P_{s,p}$ itself for its owner.  Original envy-freeness makes each
target feasible.  Keep all marks, including coincident and zero marks
and marks of agents later declared ineligible.  The differences between
consecutive nested prefixes form at most $n$ original atom sets per
piece.  An atom set can span several recorded intervals.

Track every original gap
$g_{s,i,p}=C_i(P_{s,p})-b_{s,i}$ and every observer's value of every
original atom.  There are at most
\[
 V=J(n^3+n^2)
\]
observer--value pairs.  Choose $L$ so that
$H^2((n-1)/n)^L\le1$; the explicit choice in
Section~\ref{r8c:ledger} has this property.  While a tracked value $v$ for observer $i$ lies
strictly between $C_i(R)/H$ and $HC_i(R)$ for the current outer residue,
make $L$ Core calls with cutter $i$.  By
\eqref{r8c:core-progress}, $v$ becomes large and remains large after
every subsequent residue reduction.  Each tracked pair triggers at most
once, giving at most $VL$ additional Core calls.  These additional
allocations remain fixed and do not enlarge the tracked universe.

Let $R_0$ be the resulting outer residue and set $r_i=C_i(R_0)>0$.
Every tracked value is either small, at most $r_i/H$, or large, at least
$Hr_i$.  Keep these reference costs fixed during the package stage.  Set
\begin{equation}\label{r8c:eligible}
 E_{s,p}=\{i:g_{s,i,p}\le r_i/H\}.
\end{equation}
It contains $p$.  Order its agents by decreasing equalizer-prefix
containment, breaking geometric ties deterministically and putting $p$
first at its whole prefix.  We retain both each physical mark and its
index in this local order.

\begin{lemma}[A common omitted pair]\label{r8c:pivot}
If $H>n-1$ and $J=(S-1)(n-1)+1$, there are $S$ original snapshots
and a label $p_*\ne c$ such that $c\notin E_{s,p_*}$ in every selected
snapshot.
\end{lemma}
\begin{proof}
Every original snapshot has a noncutter label whose cutter gap is at least
its post-snapshot residual cost divided by $n-1$.  This is at least
$r_c/(n-1)>r_c/H$, so separation makes the gap large.  Color the snapshot
by the least such noncutter label and apply the pigeonhole principle.
Unselected original snapshots stay allocated unchanged.
\end{proof}

Every later live root is one of these selected originals.  The label
$p_*$ therefore always omits $c$, without synchronizing any other gap bit.

\subsection{Paid permissions and physical packages}
\label{r8c:packages}

A package consists of one live original snapshot and zero or more frozen
originals. Different packages contain disjoint originals. Every live root
keeps its original identity allocation $P_{s,p}$; it changes allocation only
when it freezes. Frozen roots never change again. Unselected originals and
all clearance allocations remain fixed outside the live pool.

Each live root has a directed permission graph containing the self edges.
The maintained partition consists of strongly connected blocks, each
with a representative. A block has at most one paid outgoing edge, from
its representative to an agent outside that block. Normalize the partition
by contracting directed cycles of these outgoing edges until its quotient
is acyclic. The resulting blocks are the strongly connected components
of the permission graph. Normalization changes only graph records: all
permissions and all physical allocations remain unchanged. The component
containing $p_*$ always keeps $p_*$ as representative. Other choices are
fixed deterministically.

Write $\mathcal C_s$ for the resulting partition in root $s$. These
partitions need not agree across roots. For each label $p$, maintain an
allowed set $R_{s,p}$, a forbidden set $F_{s,p}$, and an eligible frontier
prefix $Z_{s,p}$. The allowed set is exactly the outgoing-neighbor set in
the permission graph. If $C$ is the component of $p$, then
\begin{itemize}[leftmargin=*]
\item $R_{s,p}\subseteq E_{s,p}$ and $F_{s,p}\subseteq C$ are disjoint;
\item every allowed equalizer contains $Z_{s,p}$, and every atom between
$Z_{s,p}$ and $P_{s,p}$ is small to every observer;
\item every eligible mark strictly before the frontier in the decreasing
prefix order belongs to $C$;
\item allowed recipients lie in $C$, except possibly the single paid
outgoing target, which is the frontier's eligible agent.
\end{itemize}
Initially $R_{s,p}=\{p\}$, $F_{s,p}=\varnothing$, and $Z_{s,p}=P_{s,p}$.
Contracting components preserves these properties. In particular, for a
component with no paid outgoing edge its representative has only internal
allowed recipients and an internal frontier.

A frozen root has a permutation $\pi_s$ of the recipients and an allocated
prefix $Y_{s,p}\subseteq P_{s,p}$ at each label. All but possibly one label
use the matched original equalizer. At the exceptional donor label, the
prefix is the complement of an actual suffix returned by SubCore; its
recipient's own cost is at most her original baseline. For a live root,
$\pi_s$ is the identity and $Y_{s,p}=P_{s,p}$. Put
$U_{s,p}=P_{s,p}\setminus Y_{s,p}$. The entire unallocated domain is
\begin{equation}\label{r8c:material-state}
 \mathcal U=R_0\sqcup\bigsqcup_{s,p}U_{s,p},
\end{equation}
where $\sqcup$ denotes disjoint union. Thus live roots have empty physical
complements, although their paid permissions can carry nonzero reserves.

For distinct observer and recipient $i,j$, define
\begin{equation}\label{r8c:reserve}
 h_s(i,j,p)=
 \begin{cases}
 [b_{s,i}-C_i(T_{s,p,j})]_+,&j\in R_{s,p},\\
 0,&j\in F_{s,p},\\
 [b_{s,i}-C_i(Z_{s,p})]_+,&\text{otherwise}.
 \end{cases}
\end{equation}
For each package, let $D_{i,j,p}$ be the sum of the frozen roots' actual
negative deviations at label $p$ allocated to $j$, and let $L_{i,p}$ be
the sum of their actual own-cost reductions at that label:
\[
 D_{i,j,p}=\sum_{s\text{ frozen}}\ind\{\pi_s(p)=j\}
                 [b_{s,i}-C_i(Y_{s,p})]_+,\qquad
 L_{i,p}=\sum_{s\text{ frozen}}\ind\{\pi_s(p)=i\}
                 (b_{s,i}-C_i(Y_{s,p})).
\]
The invariant is
\begin{equation}\label{r8c:invariant}
 L_{i,p}\ge D_{i,j,p}+h_s(i,j,p)\qquad(i\ne j),
\end{equation}
where $s$ is the live root. For a closed package the reserve is zero.
All quantities vanish initially.

\begin{lemma}[Physical margins]\label{r8c:surplus}
For a valid package, let $A_i$ be its actual bundle for $i$ and
$B_i=\sum_s b_{s,i}$. Then
\begin{equation}\label{r8c:own}
 C_i(A_i)=B_i-\sum_p L_{i,p},
\end{equation}
and, for every $i\ne j$,
\begin{equation}\label{r8c:positive-surplus}
 C_i(A_j)-C_i(A_i)
 \ge\sum_s[C_i(A_{s,j})-b_{s,i}]_+.
\end{equation}
Closing a live root at its identity allocation preserves these claims.
They also hold after summing disjoint packages.
\end{lemma}
\begin{proof}
Each equalizer gives its recipient exactly her original baseline; only
actual donor complements reduce own cost. This proves
\eqref{r8c:own}. The total negative deviation of a rival's bundle is at
most $\sum_p(D_{i,j,p}+h_s(i,j,p))$, and hence at most
$\sum_pL_{i,p}$. Subtracting \eqref{r8c:own} leaves all its positive
deviations. A live identity allocation has no negative deviation and no
own reduction. Closing it therefore only removes its nonnegative reserve.
\end{proof}

\subsection{Paying one source component}
\label{r8c:advance-section}

In a normalized root choose a component with no paid outgoing edge; such a
component exists because the quotient is acyclic. Let $p$ be its
representative and $C$ its vertex set. Scan the recorded original atoms
through the first eligible agent $q\notin C$, or through the last eligible
mark if there is no such agent. The scan has three outcomes: an all-small
outside suffix, a first large atom, or exhaustion. The frontier properties
ensure that any previously passed atoms are small. All tests use cached
values.

Suppose $d+1$ live roots agree on $(p,C)$, where $d=|C|$, and all have the
first outcome. Their outside agents $q_s$ and all other components may
differ. Offer the original suffixes
\[
 W_s=P_{s,p}\setminus T_{s,p,q_s}.
\]
Each $W_s$ consists of at most $n$ atoms small to every observer. For any
$i\notin C$,
\begin{equation}\label{r8c:protected}
 C_i(W_s)\le g_{s,i,p}.
\end{equation}
Indeed, an eligible outside agent has its equalizer at or below the first
outside mark. An ineligible agent has gap at least $Hr_i$, whereas the
suffix costs at most $nr_i/H$; use $H^2>n$.

Apply clipped Offset SubCore with participants $C$ to these $d+1$ parents,
using utilities $[C_i(D)-g_{s,i,p}]_+$ on suffixes of parent $W_s$.
It assigns distinct nominal donor roots $s_j$ to $j\in C$ and leaves one
root $w$ unused. For the actual retained suffix $D_j\subseteq W_{s_j}$,
put $a_j=[C_j(D_j)-g_{s_j,j,p}]_+$. Its guarantees are
\begin{align}
 a_i&\ge[C_i(D_j)-g_{s_j,i,p}]_+ &&(i,j\in C),\label{r8c:payments}\\
 a_i&\ge[C_i(W_w)-g_{w,i,p}]_+ &&(i\in C).\label{r8c:threshold}
\end{align}
The offset primitive operates on recorded finite unions; locating a target
uses cached component totals and a genuine original Cut in one interval.
Every new endpoint is recorded for the active observers. We use the
conservative complete recorded-call bound $5n^2 2^{n+1}$ below, including
retained-piece records and unsuccessful primitive trials.

\begin{lemma}[Deferred payment]\label{r8c:advance}
The $d+1$ input packages can be replaced by one package with live root
$w$. The survivor keeps its original allocation, while its source
component acquires the paid edge $p\to q_w$. The physical margin and
frontier invariants are preserved.
\end{lemma}
\begin{proof}
If $a_j>0$, use an allowed simple path $j\to p$ in donor root $s_j$.
Together with the new edge $p\to j$ it gives a permutation cycle; use
self edges elsewhere. Give $j$ the actual complement $P_{s_j,p}\setminus
D_j$, and give every other changed recipient its matched equalizer.
For $j=p$ the permutation is the identity. The actual own reduction is
\[
 b_{s_j,j}-C_j(P_{s_j,p}\setminus D_j)
   =C_j(D_j)-g_{s_j,j,p}=a_j.
\]
If $a_j=0$, make no modification and freeze that nominal donor at its
original identity allocation. Thus a negative affine value is never
executed or counted as a credit. Every nominal donor freezes, including
those with zero credit, and all its frozen ancestors enter the output
package once.

On the unused root change only the permission records:
\[
 R'_{w,p}=R_{w,p}\cup\{q_w\},\qquad
 F'_{w,p}=F_{w,p}\cup(C\setminus R_{w,p}),\qquad
 Z'_{w,p}=T_{w,p,q_w}.
\]
The old frontier and every allowed equalizer contain the new frontier;
all newly passed atoms are small. The new target is outside $C$, while
all earlier eligible marks belong to $C$. This proves the updated
frontier properties without changing the survivor's allocation.

For completeness, fix $i\ne j$ and let $e_{i,j,p}$ be the additional
deficit of the modified donor assigned to $j$, or zero if that assignment
is absent or stripped. Extend $a_i$ by zero outside $C$. Let
$\widehat D,\widehat L$ be the sums of the input frozen ledgers. Every
other new frozen contribution uses an old allowed edge, so
\begin{equation}\label{r8c:group-update}
 D'_{i,j,p}\le\widehat D_{i,j,p}
       +\sum_{s\ne w}h_s(i,j,p)+e_{i,j,p},\qquad
 L'_{i,p}=\widehat L_{i,p}+a_i.
\end{equation}
If $i,j\in C$, \eqref{r8c:payments} pays $e_{i,j,p}$, and the new
reserve at an internal recipient is at most the old reserve. If $i\in C$
and $j\notin C$, no donor is assigned to $j$; the new reserve is the
threshold in \eqref{r8c:threshold}. If $i\notin C$,
\eqref{r8c:protected} makes every new deficit and the new frontier
reserve zero. At every other label there is no new own credit or extra
deficit; any equalizer deficit is covered by that donor's old reserve.
Consequently
\[
 e_{i,j,p}+h'_w(i,j,p)\le a_i+h_w(i,j,p).
\]
Adding the input inequalities \eqref{r8c:invariant} to
\eqref{r8c:group-update} proves the output invariant. In particular, no
credit is spent simultaneously on a donor deficit and an increased
survivor reserve.

Every changed root is partitioned into its actual retained prefixes and
complements within the same original pieces. All released ordinary
material lies in passed small atoms. This proves the physical identity
\eqref{r8c:material-state}, including zero-credit assignments.
\end{proof}

A paid edge may complete a directed cycle of components. Contract that
cycle, remove its now-internal pending edges from the component quotient,
and retain them in the permission graph. The merged component is unpaid.
This changes neither the physical allocation nor any reserve. No roots
outside the current group are discarded or synchronized.

\subsection{A factorial capacity and a sufficient pool}
\label{r8c:schedule}

Define unpaid and paid component capacities by
\[
 u(k)=8^{k-1}k!,\qquad v(k)=2(k+1)u(k)\quad(k\ge1).
\]
For a live package $K$, let $w(K)$ be its number of original snapshots,
including frozen ancestors. Its capacity $\Phi(K)$ is the product of
$u(|C|)$ over unpaid components and $v(|C|)$ over paid components in its
live root. We maintain $w(K)\le\Phi(K)$.

\begin{lemma}[Capacity under payment and contraction]\label{r8c:capacity}
For positive $a,b$, $v(a)v(b)\le u(a+b)$. Contracting a cycle of paid
components does not decrease capacity. Every live root has
$\Phi(K)\le u(n)$. Paying an unpaid component of size $d$ multiplies its
capacity by $2(d+1)$.
\end{lemma}
\begin{proof}
The inequality $(a+1)!(b+1)!\le2(a+b)!$ is equality at $b=1$; increasing
$b$ multiplies its left-hand side by $b+2$ and its right-hand side by
$a+b+1\ge b+2$. It implies $v(a)v(b)\le u(a+b)$ directly. Repeatedly
combine components, using $u(k)\le v(k)$, to obtain
$\prod_t v(k_t)\le u(\sum_t k_t)$ whenever there are at least two
components. This proves both the cycle-contraction claim and the global
bound for a partition with at least two components. A single full
component has no external target, so it is unpaid and has capacity $u(n)$.
The payment multiplier follows from the definition of $v$.
\end{proof}

Group weights into buckets $2^b\le w(K)<2^{b+1}$. If $d+1$ packages
lie in one bucket and $K_w$ is the unused root selected by SubCore, then
\begin{equation}\label{r8c:bucket}
 \sum_Kw(K)<2(d+1)w(K_w)\le2(d+1)\Phi(K_w).
\end{equation}
Thus the output of Lemma~\ref{r8c:advance} satisfies its capacity bound
when the survivor's source component is marked paid. Subsequent
contractions preserve that bound. Initially each package has weight and
capacity one.

Since $u(n)\le2^{4n^2}$, only $4n^2+1$ buckets are needed. Normalize
each root, choose a deterministic unpaid component, and perform its scan.
Classify the root as follows:
\begin{itemize}[leftmargin=*]
\item an ordinary class records $(p,C,b)$, where $p$ is the source
representative, $C$ its component, and $b$ the weight bucket;
\item an atomic class records $(p,A,B)$, where $A$ is the set of eligible
marks preceding the first large atom and $B$ is its nonempty large-value
mask;
\item an exhausted class records $(p,A)$, where $A=E_{s,p}$.
\end{itemize}
The terminal operations for the last two classes are proved next. There
are at most
\begin{equation}\label{r8c:class-count}
 \Lambda=n2^n(4n^2+1)+n4^n+n2^n
\end{equation}
classes in total. Choose
\begin{equation}\label{r8c:pool}
 S=n\Lambda u(n)+1.
\end{equation}

\begin{lemma}[Existence of a next operation]\label{r8c:counts}
Starting with $S$ selected snapshots, every nonterminal pool contains a
class with at least $n+1$ live roots. An ordinary operation reduces the
number of live roots. Hence at most $S$ ordinary operations precede a
terminal class.
\end{lemma}
\begin{proof}
Ordinary operations combine their input packages into one output package;
they do not discard any package outside the group. The total weight of
the live pool therefore remains $S$. Each package has weight at most
$u(n)$ by Lemma~\ref{r8c:capacity}. If every class had at most $n$ live
roots, the total weight would be at most $n\Lambda u(n)<S$, a
contradiction. Choose any $d+1$ members of a populated ordinary class,
where $d=|C|$, and apply Lemma~\ref{r8c:advance}. Inequality
\eqref{r8c:bucket} preserves the weight invariant. This freezes $d\ge1$
live roots and leaves one survivor. Normalization is finite because every
cycle contraction reduces its root's component count. The stated bound
on ordinary operations follows. A populated terminal class supplies all
its required roots, since its participant set has size at most $n$.
\end{proof}

This counting argument requires neither equal package sizes nor agreement
on whole component partitions. In particular, it has no loss from retaining
a largest common target class and no repeated synchronization factor.

\subsection{Terminal classes and the entire remaining domain}
\label{r8c:closed-batches}

For a terminal class with source label $p$ and participant set $A$, select
$|A|+1$ roots. In each root, $A$ lies inside that root's chosen source
component; those source components need not be equal. Use affine Offset
SubCore, with utility $C_i(D)-g_{s,i,p}$, on the full suffixes described
below. Every participant's whole offered parent has nonnegative utility.
The primitive supplies distinct donor roots, envy-free affine utilities,
and at least one whole assigned parent. Install each donor's $p\to j$
cycle using its own internal allowed return path. Execute every affine
assignment, including zero-own and empty ones. Close the unused root and
all live packages outside the selected group at their unchanged identity
allocations, retaining every frozen ancestor. Thus every package is closed.

The ledger proof is the same donor calculation as
\eqref{r8c:group-update}, but there is no surviving reserve. Affine
envy-freeness pays participants' rival deficits. Eligible nonparticipants
have their equalizers below the offered suffix boundary; ineligible
nonparticipants are protected by their large original gap and the
small-atom bound, apart from the explicitly extracted large atom below.
The scan boundary ensures that they too have nonpositive offered affine
utility. Thus all actual deficits are paid, and
\eqref{r8c:positive-surplus} holds for the full closed allocation.

\begin{lemma}[Atomic extraction]\label{r8c:atomic}
For a populated atomic class $(p,A,B)$, a terminal call produces an actual
unallocated domain $X$ with
\begin{equation}\label{r8c:X}
 C_i(X)\ge Hr_i\quad(i\in B),\qquad
 C_i(X)\le n^2r_i/H\quad(i\notin B).
\end{equation}
Both $A$ and $B$ are nonempty proper subsets of the agents. The remaining
unallocated domain $R=\mathcal U\setminus X$ is disjoint from $X$ and
satisfies
\begin{equation}\label{r8c:residue-bound}
 r_i\le C_i(R)\le(1+Jn^2/H)r_i.
\end{equation}
\end{lemma}
\begin{proof}
Let $F_s$ be the first large atom in each selected root. Offer the entire
original suffix through its left boundary. The eligible agents whose
marks are at or to the right of its right boundary are exactly $A$.
They include $p$ and lie in the chosen component. Their whole affine
utilities are nonnegative. Every other eligible mark lies at or to the
left of the left boundary. For an ineligible observer whose mark were
strictly to its right, the gap would consist only of the already passed
small atoms, contradicting $H^2>n$. These facts justify the terminal
ledger calculation for every observer.

Let $X$ be the union of the actual assigned suffixes. The whole assigned
parent contains an entire $F_s$, giving the lower bound for every $i\in B$.
For $i\notin B$, every offered parent consists of at most $n$ atoms
small to $i$, and at most $n$ parents are assigned. This gives the upper
bound. The affine whole-parent guarantee is essential: no assignment is
stripped merely because its own affine value is zero.

The scan ends at an eligible mark at or below the large atom's left
boundary. This agent is outside $A$, and its small original gap contains
the atom, so it is outside $B$ as well. Thus both sets are proper, while
$p\in A$ and the definition of a large atom gives $B\ne\varnothing$.

Each live root was physically unchanged before this terminal call. Its
actual retained prefix and actual removed suffix partition each original
piece. At the donor label all removed material belongs to $X$; all other
newly released material uses passed small atoms. Previously frozen
complements also use small atoms. Charge each original atom once to this
physical union. There are at most $Jn^2$ such atoms, while $R_0$ remains
untouched and lies in $R$. This proves \eqref{r8c:residue-bound} and the
exact disjoint decomposition of all remaining material.
\end{proof}

\begin{lemma}[Exhausted termination]\label{r8c:terminal}
For a populated exhausted class $(p,A)$, $A$ is nonempty and proper.
After its terminal call, every observer outside $A$ has a margin against
every recipient in $A$ at least $(H-n/H)r_i$. The entire unallocated
domain satisfies \eqref{r8c:residue-bound} and can be recursively divided
among the strictly smaller population $N\setminus A$.
\end{lemma}
\begin{proof}
Offer the complete suffix through the last eligible mark. Here
$p\in A\subseteq C_s$ in every selected root, and every offered suffix
is all-small. If one selected source component is proper, then $A$ is
proper. Otherwise the representative is $p_*$, and the omitted cutter
$c$ from Lemma~\ref{r8c:pivot} does not belong to $A$.

For $i\notin A$ and each assigned $j\in A$, the distinct donor root
$s_j$ has $g_{s_j,i,p}\ge Hr_i$. Its removed suffix costs at most
$nr_i/H$. Its actual retained piece assigned to $j$ therefore has
positive deviation at least $(H-n/H)r_i$ from observer $i$'s baseline.
Lemma~\ref{r8c:surplus} preserves this deviation as an aggregate margin,
including all frozen ancestors. All released material is small, so the
same atom count as above proves \eqref{r8c:residue-bound}.

Our parameters give $H\ge Jn^2$ and $H-n/H\ge2$. Each margin thus
covers the entire remaining domain. Recursively divide that exact domain
among $N\setminus A$. Their internal comparisons follow by adding envy-free
allocations; comparisons to $A$ are paid by these whole-domain margins.
Agents in $A$ receive nothing further, so their comparisons cannot worsen.
\end{proof}

\subsection{Completion and full query accounting}
\label{r8c:completion}
\label{r8c:ledger}

Set $S$ as in \eqref{r8c:pool} and choose
\[
 J=(S-1)(n-1)+1,\qquad H=2^{n+5}Jn^3,\qquad
 L=n(2n^2+28n+16).
\]
These choices give $H>n-1$, $H^2>n$, $H\ge Jn^2$, and $H\ge n+2$.
The estimates below also give $H^2\le2^{2n^2+28n+16}$.
Since $((n-1)/n)^n\le1/2$ for $n\ge2$, this $L$ satisfies the
clearance requirement $H^2((n-1)/n)^L\le1$.

In the atomic branch put $\alpha=2^{n+2}$. The scale satisfies
$H\ge2\alpha$ and $H\ge n^2\alpha$. Hence
\eqref{r8c:X} and \eqref{r8c:residue-bound} imply
\[
 C_i(X)\ge\alpha C_i(R)\ (i\in B),\qquad
 C_i(R)\ge\alpha C_i(X)\ (i\notin B).
\]
Apply the two-group completion of Lemma~\ref{r6c:discrepancy} to these
actual disjoint recorded domains. Its preparation uses at most
$19\cdot2^n$ cuts and $19n2^n$ recorded queries; its two children have
positive complementary populations strictly smaller than $n$. The
exhausted branch has one strictly smaller child. Attaching the fixed
clearance allocations then covers the full input domain without envy.

We give conservative counts for the physical implementations, including
all endpoint records. The Core bound
$n^2+n^3[\binom{2n-2}{n-1}-1]\le2n^3 4^n$, together with its
initial partition and equalizer records, bounds a complete snapshot by
$5n$ times that Core allowance, hence by $10n^4 4^n$ queries. The
recorded offset primitive, including its masks and final retained-piece
tables, costs at most $5n^2 2^{n+1}$.
Preprocessing makes at most
$N_{\mathrm{snap}}=J+J(n^2+n^3)L$ snapshot calls. The physical prefix and
terminal batch use at most $S+1$ offset calls. From
\eqref{r8c:class-count}--\eqref{r8c:pool},
\[
 S\le8n^4 32^n n!,\qquad J\le8n^5 32^n n!,\qquad
 H\le256n^8 64^n n!\le n!2^{14n+8}.
\]
Using $n!\le n^n\le2^{n^2}$ gives the stated bound on $H^2$ and
$L\le46n^3$. Consequently, for $n\ge3$,
\begin{align*}
 N_{\mathrm{snap}}&\le93Jn^6,\\
 N_{\mathrm{snap}}Q_{\mathrm{snap}}&\le7440n^{15}128^n n!
                         \le n!2^{22n+13},\\
 (S+1)Q_{\mathrm{off}}&\le160n^7 64^n n!
                         \le n!2^{13n+8},\\
 19n2^n&\le n!2^{2n+5}.
\end{align*}
Their sum is at most $E(n)=n!2^{22n+15}$. The same local allowance
bounds cuts, since every cut is a counted query. All masks, graph tests,
contractions, weight buckets, and cached suffix values require no further
queries.

It remains to include the actual recursive calls. If a child of size $k$
is executed inside a population of size $m$, charge its native query
count plus $m$ records for each of its new Cut endpoints. Its endpoint
provenance ensures that inherited endpoints require no new query. Thus a
child with native bounds $(Q_k,C_k)$ costs at most $Q_k+mC_k$ queries and
$C_k$ cuts to its parent. This charges all ambient observer recording.

The recursion has at most $2n-1$ nodes: two-child steps partition the
active population, and one-child steps discard at least one agent. Along
an ancestor chain the population strictly decreases, so the total
recording charge per native cut is at most $n^2$. Since $E$ is monotone,
the recursive query count is therefore at most
\[
 (2n-1)(E(n)+n^2E(n))\le n!2^{25n+17}.
\]
The direct one- and two-agent cases fit the same allowance. Finally,
recording the initial $n$ total costs adds $n$ Eval queries, absorbed in
$n!2^{25n+18}$. Every output coordinate is an input endpoint or an actual
Cut reply. Assigning the left endpoint to one fixed agent adds no query
and changes no cost by nonatomicity. This proves
Theorem~\ref{r8c:theorem} on the closed interval.

\section{Canonical prefix states and deterministic chore primitives}
\label{r6core:appendix}

We give the prefix Core used in the deterministic chore construction, its
cached compensation interface, and a whole-cell near-equal partition primitive.
Costs throughout this section are finite, nonnegative, nonatomic measures.
All queries are individual ordinary interval queries.  A Cut may return any
legal root, including any point of a plateau.  An ordered finite union is
represented by its component intervals and their recorded costs.  Its prefix
consists of initial components and, possibly, a prefix of one further component.
Endpoint costs are recorded for every observer in the enclosing call.  No
query to a difference of measures, an infinitesimally perturbed measure, or a
moving window is used.

\begin{theorem}[Chore Core]\label{r6core:theorem}
Let $n\ge2$, let $R$ be a recorded ordered finite union, and fix a cutter $c$.
There is a deterministic protocol with the following alternatives.
If some agent has zero cost for $R$, it gives $R$ to that agent and empty
bundles to the others, completing an envy-free allocation.  Otherwise it cuts
$R$ into $n$ parents of equal cutter cost and returns an envy-free allocation
of one prefix from each of the $n$ distinct parents, with at least one parent
allocated whole.  The residue is the exact complementary union of suffixes.
Writing
\[
 T_n=n^2\left(\binom{2n-2}{n-1}-1\right),
\]
the protocol uses at most $n-1+T_n$ Cut queries and $n^2+nT_n$ total queries,
including all exploratory queries and endpoint recording.  In the prefix
alternative, if $R'$ is the residue, then
\[
 C_c(R')\le\frac{n-1}{n}C_c(R).
\]
If $j$ receives a whole parent and $A_i$ denotes agent $i$'s returned bundle,
then
\[
 C_c(A_j)-C_c(A_c)\ge\frac{C_c(R')}{n-1}.
\]
The bounds assume that inherited component costs are cached; uncached input
components must be evaluated separately.
\end{theorem}

The proof uses continuous existence only to certify the success of a finite
dynamic program.  The protocol does not compute a fixed point.  Its
Cut bound is $O(n^{3/2}4^n)$ and its total query bound is
$O(n^{5/2}4^n)$; these are bounds for this primitive, not for a complete
single-exponential fair-division protocol.

\subsection{Hungry existence and an anchored order}

A cost is \emph{hungry} on a parent if it is positive on every nondegenerate
subinterval in its component order.  For this existence argument, parameterize
each parent by $[0,1]$.  Its prefix-cost functions are then continuous and
strictly increasing.  The parents need not have equal costs.

\begin{lemma}[Local matching]\label{r6core:local-matching}
Let a bipartite graph on $k$ agents and $k$ parents contain a perfect matching.
For every permitted pair $(i,j)$ let
$f_{ij}:[0,\delta]\to\R_{\ge0}$ be continuous and strictly increasing, with
$f_{ij}(0)=0$, where $\delta>0$.
There are nonnegative $y_j$ with $\sum_jy_j=\delta$ and a perfect matching
$\pi$ of permitted pairs such that $\pi(i)$ minimizes $f_{ij}(y_j)$ among
agent $i$'s permitted parents.
\end{lemma}
\begin{proof}
Relabel an existing perfect matching as $(i,i)$.  Let
\[
 Y=\{y\ge0:\textstyle\sum_jy_j=\delta\},\qquad
 D=\{d\ge0:\textstyle\sum_jd_j=k\}.
\]
For $y\in Y$, allow each agent to choose any probability distribution
supported on her permitted minimizers of $f_{ij}(y_j)$, and let $D(y)$ be the
set of resulting column sums.  This correspondence is nonempty, compact,
convex-valued, and upper hemicontinuous.  Indeed, finite continuous
minimization has a closed graph, row distributions lie in compact simplices,
and projecting the compact graph of their product preserves the closed-graph
property.  The same properties hold for
\[
 Y(d)=\argmax_{z\in Y}\sum_jz_jd_j.
\]
Kakutani's theorem~\citep{Kakutani41}, applied on $Y\times D$ to
$(y,d)\mapsto Y(d)\times D(y)$, gives $y\in Y(d)$ and $d\in D(y)$.

Put $Z=\{j:y_j=0\}$ and $P=\{j:y_j>0\}$.  Each of the $|Z|$ agents whose
original matched parent is in $Z$ can attain incremental cost zero, whereas
every permitted parent in $P$ has positive incremental cost.  Those agents
place all demand in $Z$, so $\sum_{j\in Z}d_j\ge|Z|$.  Since $y$ maximizes
the linear form, every $j\in P$ has $d_j=M=\max_\ell d_\ell$.  Hence
\[
 k=|P|M+\sum_{j\in Z}d_j\ge |P|M+|Z|.
\]
Because $P$ is nonempty, $M\le1$.  The average column sum is one, so every
column sum equals one.  The row distributions therefore form a doubly
stochastic matrix supported on permitted minimizers.  Its support satisfies
Hall's condition: for any set of rows, its unit total per row must be carried
by neighboring columns of capacity one.  A perfect matching in the support
proves the claim.
\end{proof}

An envy-free prefix state on equally sized agent and parent sets consists of
one prefix from each parent and a bijection of these prefixes to the agents.
Every assigned prefix must minimize its owner's cost among all prefixes in
the state.  Call such a state \emph{anchored} if its agents admit an order in
which each nonwhole assigned prefix has the same cost, to its owner, as an
earlier assigned prefix.  Whole prefixes need no earlier anchor.

\begin{lemma}[Anchored existence]\label{r6core:existence}
For any balanced finite sets of hungry agents and parents, an anchored
envy-free prefix state exists.
\end{lemma}
\begin{proof}
For each bijection, the envy-free constraints define a closed subset of the
cube of prefix endpoints.  The finite union over bijections is compact and
nonempty, since all-empty prefixes are feasible.  Choose a state maximizing
the sum of its endpoints.

Suppose a nonempty group $S$ of owners has only nonwhole prefixes and every
member strictly prefers her own prefix to every prefix outside $S$.
Write $b_i$ for each member's current own cost.  Inside $S$, permit $(i,j)$
exactly when parent $j$'s current prefix costs $b_i$ to agent $i$; the current
ownership is a perfect matching.  If the old endpoint of $j$ is $x_j$, use
\[
 f_{ij}(y)=C_i(P_j(x_j+y))-b_i
\]
on permitted pairs.  Choose $\delta>0$ below all available endpoint slack and
small enough that moving any endpoint by at most $\delta$ changes any relevant
cost by less than half of every positive old comparison gap from a member of
$S$ to a nontied internal or outside prefix.  There are finitely many gaps,
and continuity provides this choice; when there are none, only endpoint caps
need be respected.

Lemma~\ref{r6core:local-matching} now enlarges these prefixes by a total of
$\delta$ and reassigns them along old tie edges.  Its minimum condition
preserves comparisons on the old ties.  The choice of $\delta$ preserves
strict comparisons.  Outside owners retain their own prefixes and see every
changed prefix enlarge, so their envy-free inequalities persist by
nonnegativity.  This contradicts maximality.

Consequently at least one prefix is whole: otherwise $S$ could be all owners.
List all whole-prefix owners first.  If some owners remain, their group
cannot satisfy the preceding strict-separation condition.  Since the state
is envy-free, some remaining owner must tie an already listed prefix.
Append that owner and repeat.  This produces the required anchored order.
\end{proof}

\subsection{Canonical states on balanced subsets}

The following finite arguments also apply to lexicographically ordered formal
costs.  They require only that every observer's cost increase strictly along
each parent's ordered prefix states.  For a prefix vector $x$, write
$b_i(x)=\min_p C_i(x_p)$, which is independent of the chosen envy-free matching.

\begin{lemma}[Sign classes and lattice]\label{r6core:lattice}
Let $x,y$ be envy-free states on the same balanced sets, with matchings
$\pi_x,\pi_y$.  Put
\[
 J_+=\{p:x_p<y_p\},\qquad A=\{i:b_i(x)<b_i(y)\}.
\]
Then $\pi_x(A)=\pi_y(A)=J_+$.  Coordinatewise minimum and maximum of the
two prefix vectors are envy-free; their own costs are respectively the
minimum and maximum of the two old own costs.
\end{lemma}
\begin{proof}
For $i\in A$, envy-freeness implies
\[
 C_i(x_{\pi_x(i)})=b_i(x)<b_i(y)\le C_i(y_{\pi_x(i)}),
\]
so $\pi_x(A)\subseteq J_+$.  Conversely, for $p\in J_+$ and
$i=\pi_y^{-1}(p)$, strict within-parent increase gives
\[
 b_i(x)\le C_i(x_p)<C_i(y_p)=b_i(y).
\]
Thus $\pi_y^{-1}(J_+)\subseteq A$.  Bijection and finite cardinality force
equality in both inclusions.  Swapping $x,y$ gives the same identity for
$B=\{i:b_i(y)<b_i(x)\}$ and $J_-=\{p:y_p<x_p\}$.  The remaining agents
are matched to equal-coordinate parents by both matchings.

For the coordinatewise maximum, use $\pi_y$ on $A$ and $\pi_x$ on the other
agents.  Its images partition the parents into the three sign classes, and
each assigned cost is $\max\{b_i(x),b_i(y)\}$.  Every other maximum prefix
costs at least both old minima.  For the coordinatewise minimum, use $\pi_x$
on $A$, $\pi_y$ on $B$, and $\pi_x$ on the remaining agents.  Its assigned
cost is the smaller old minimum; every other minimum prefix costs at least
that value.  These are the required envy-free matchings.
\end{proof}

\begin{lemma}[Anchored maximality]\label{r6core:canonical}
An anchored envy-free prefix vector is coordinatewise at least every
envy-free prefix vector on the same balanced subsets.  In particular, all
anchored states on those subsets have the same prefix vector.
\end{lemma}
\begin{proof}
Let $x$ be anchored and $y$ be any envy-free state.  If some parent grows from
$x$ to $y$, the sign-class identity says that its original $x$-owner $i$
satisfies $b_i(y)>b_i(x)$.  Its old prefix cannot have been whole.  Let $p'$
be an earlier anchor, so $C_i(x_{p'})=b_i(x)$.  Then
\[
 C_i(y_{p'})\ge b_i(y)>b_i(x)=C_i(x_{p'}),
\]
and $p'$ also grows.  Repeating with the original owner of this anchor gives
a strictly decreasing sequence of positions in the anchoring order.  Its
first position is a whole prefix, which cannot grow.  This contradiction
proves $y\le x$.  Applying the conclusion in both directions proves
uniqueness for two anchored vectors.
\end{proof}

Strictness is essential here.  We do not claim that coordinatewise maximum
preserves envy-freeness for arbitrary unperturbed plateau states.  Nor do we
discard arbitrary smaller partial states in favor of larger ones.  The
states retained below are anchored and hence canonical.

\subsection{The dynamic program and its hungry correctness}

On $n$ labeled parents and $n$ agents, store one state for every nonempty pair
$(I,P)$ of equally sized subsets.  Process the pairs in increasing size.
A singleton state assigns its sole parent whole.  For a state of size
$k\ge2$, try each $(i,p)\in I\times P$ in a fixed order:
\begin{enumerate}[leftmargin=*,label=\arabic*.]
\item Read the stored state on $(I\setminus\{i\},P\setminus\{p\})$.
Let $t$ be agent $i$'s minimum cost of its prefixes.
\item If $C_i(P_p)\le t$, use $P_p$ whole.  Otherwise obtain a prefix of
$P_p$ costing exactly $t$ to $i$.
\item Retain the smaller state's matching, append the new pair, and test
all envy-free inequalities.  Store the first successful candidate.
\end{enumerate}
Every stored state is anchored: retain the smaller state's order, append
$i$, and, when its prefix is not whole, use an earlier minimizer attaining
$t$ as its anchor.  Every comparison uses recorded costs.

\begin{lemma}[Hungry dynamic program]\label{r6core:dp}
For hungry costs, every state succeeds and stores the canonical anchored
prefix vector for its two subsets.
\end{lemma}
\begin{proof}
Induct on the subset size.  By Lemma~\ref{r6core:existence}, choose an
anchored witness for the present subsets and let $(i,p)$ be its last pair.
Deleting that pair leaves an anchored envy-free state on the smaller
subsets.  Lemma~\ref{r6core:canonical} identifies its prefix vector with
the stored smaller vector.

The stored smaller matching can differ from the witness restriction.  This
does not change any old agent's own cost: it is the minimum of the identical
collection of prefixes.  Thus all comparisons of old agents to the new
prefix have the witness values, independently of that matching.

If the last witness prefix is whole, its cost to $i$ is at most every earlier
prefix, and the whole-parent branch reproduces it.  Otherwise its anchor and
envy-freeness imply that $t$ equals its witness own cost.  Hunger makes the
target root unique, so the trimming branch reproduces the same prefix.
The candidate for $(i,p)$ therefore succeeds.  Every successful candidate is
anchored, and Lemma~\ref{r6core:canonical} completes the induction.
\end{proof}

\subsection{Persistent records for arbitrary legal roots}

For each parent maintain one ordered list of prefix records, shared by
\emph{all} states and failed candidates.  A record has a physical endpoint,
all actual prefix costs, and one rational secondary label $\theta_i$ for each
observer.  Compare costs lexicographically as
\[
 \bigl(C_i(\text{physical prefix}),\theta_i(\text{record})\bigr).
\]
Physical endpoints are nondecreasing along the list.  Formal costs increase
strictly for every observer.  Initialize the empty record with $(0,0)$ and
the whole record with $(C_i(P),1)$.  Distinct records may have the same
physical endpoint, including on physically empty parents.  The whole and
empty records remain distinct formal records even in that case.

Run the dynamic program using formal costs.  Its trim targets are formal
costs of already stored records.  Such a target is never below the empty
record.  When a whole-parent comparison fails, the target lies below the
whole record.  If the parent already has the target formal cost for the
trimming observer, reuse that record.  Otherwise find adjacent records
strictly bracketing it.  Let their physical endpoints be $l\le r$ and let
$v$ be the target's real cost.

If a neighbor has real cost $v$, reuse that physical endpoint and insert a
new formal record between the two neighbors.  Otherwise their real costs
strictly bracket $v$.  Cached component costs locate a component containing
the target; one ordinary Cut obtains a physical root, with existing
component-boundary roots reused when available.  Nonnegativity and strict
bracketing imply that every legal returned root lies between $l$ and $r$.
Record its cost for every observer.  Give the trimming observer exactly the
target secondary label.  For each other observer, choose a rational label
that makes its new formal value strictly intermediate between its two
neighbors: a midpoint when all three real values tie, a label above or below
the tied neighbor when only one ties, and any rational label when the real
value lies strictly between.  Such a choice always exists.

The secondary labels are arbitrary rationals, not restricted to $[0,1]$.
Existing labels never change.  An insertion at a previously recorded
physical coordinate needs no additional evaluation.  This procedure
preserves the record invariants, including for zero-cost observers and
duplicate physical coordinates.  A formally envy-free allocation is
physically envy-free because a lexicographic inequality implies the
corresponding weak inequality of real costs.

\begin{lemma}[Simultaneous finite realization]\label{r6core:realization}
The formal-record dynamic program succeeds for arbitrary nonnegative
nonatomic costs and every legal sequence of ordinary Cut replies.
\end{lemma}
\begin{proof}
Suppose the first failed subset state has been fully tried.  At that point
the global table contains finitely many records.  Choose one $\epsilon>0$
such that all numbers
\[
 C_i(\text{physical prefix})+\epsilon\theta_i(\text{record})
 \tag{\ref{r6core:realization}$^*$}
\]
preserve every formal comparison and equality between records, including
records on different parents.  This is possible: pairs with unequal real
parts have finitely many positive separations, while their secondary parts
form a finite set; equal real parts are ordered exactly by their secondary
parts.

On a new abstract interval for each parent, place all its records at distinct
ordered positions.  Interpolate the displayed prefix values linearly
between consecutive records, separately for every observer.  The empty
value is zero, and formal strictness makes every increment positive.  This
defines a single hungry profile realizing the entire finite table.  The
construction is permitted even when several old records shared a physical
endpoint, since the new intervals are used only in this contradiction.

Replay the ordinary hungry dynamic program on this profile.  Every earlier
state, candidate order, minimum, whole-parent branch, and envy-free decision
is reproduced.  Every target root is also reproduced: the stored trimming
record copied both parts of its target exactly, and hungry roots are unique.
Later insertions never changed an earlier record.  Consequently this is one
simultaneous replay of every completed state and attempted candidate, not
separate realizations chosen for separate trials.

It follows that the hungry program first fails at the same state,
contradicting Lemma~\ref{r6core:dp}.  The protocol itself never constructs
or queries the hungry profile and does not need to know $\epsilon$.
\end{proof}

\subsection{Query ledger and cutter progress}

There are $\binom nk^2$ balanced subset pairs of size $k$, with at most $k^2$
last-pair trials each.  Each trial requires at most one new Cut.  Therefore
the number of trim requests, including failed candidates, is at most
\begin{align*}
 \sum_{k=2}^n k^2\binom nk^2
 &=n^2\sum_{j=1}^{n-1}\binom{n-1}{j}^2\\
 &=n^2\left(\binom{2n-2}{n-1}-1\right)=T_n.
\end{align*}
Here $k\binom nk=n\binom{n-1}{k-1}$ and the final equality is
Vandermonde's identity.  The number of nonempty subset states is
$\binom{2n}{n}-1$.

A new physical trim requires one Cut and at most $n-1$ recording Evals:
the trimmer's cost is already fixed by the Cut target.  Reused coordinates
need no new query.  All subsequent comparisons, matching data, and anchors
use these records.  For disconnected input parents, known component costs
locate the component of a trim arithmetically; a union-value query is not
charged as one interval query.

For the full Core, a zero-total row gives the complete exit in the theorem.
Otherwise cut $R$ into $n$ equal-cost parents for $c$, recording the
$n-1$ new endpoints.  These initial cuts and at most $n$ initial total
evaluations cost at most $n(n-1)+n=n^2$ queries.  All parent costs then follow
from the recorded components and endpoints.  Apply the prefix program.
It returns a formally envy-free anchored state, so some allocated parent is
whole.  Prefixes on distinct parents are disjoint, and their suffix
complements exactly constitute $R'$.

Put $a=C_c(R)/n$ and $b=C_c(A_c)$.  Every allocated prefix costs at least
$b$ to $c$, and one costs $a$.  Thus
\[
 C_c(R')\le na-\bigl(a+(n-1)b\bigr)=(n-1)(a-b),
\]
which proves the claimed gap.  Allocating that whole parent also gives
$C_c(R')\le(n-1)a$.  Together with the query ledger, this proves
Theorem~\ref{r6core:theorem}.

\subsection{Cached many-parent compensation}

\begin{corollary}[Cached SubCore contract]\label{r6core:subcore}
Let $K\ge1$ participants have nonnegative nonatomic costs, now treated as
cake values, on more than $K$ recorded ordered parents.  Start with zero
benchmarks and distinct empty inherited suffixes, using coherent empty and
whole formal labels $0$ and $1$.  With $N$ enclosing observers, the
many-parent SubCore returns distinct suffix parents, cake envy-freeness,
comparison with every unused whole parent, and at least one whole assigned
parent.  Its residue is the exact complementary collection of prefixes and
unused parents.  With cached input component costs it uses at most
\[
 S_K=2^{K+1}-2K-2\quad\text{Cuts},\qquad NS_K\quad\text{total queries}.
\]
\end{corollary}
\begin{proof}
Apply the many-parent routine proved in Lemma~\ref{dc:subcore}.  The stated
fresh-zero initialization is coherent for its persistent suffix records and
does not assume an arbitrary positive external benchmark can be initialized
for free.  Its recurrence, from the distinct recursive selected-set sizes,
is
\[
 S_0=S_1=0,\qquad
 S_K=S_{K-1}+K(K-1)+\sum_{j=1}^{K-2}S_j
     =2S_{K-1}+2(K-1).
\]
It solves to the displayed expression.  The $K=1$ call chooses a greatest
whole parent using cached costs and makes no cut.  Every new endpoint in a
larger call requires at most $N-1$ additional observer evaluations beyond
its Cut response.  The full chain argument, unused-parent comparison, and
ordinary-root implementation are included in Lemma~\ref{dc:subcore}; no
optimized SubCore statement beyond that in-paper proof is needed here.
\end{proof}

\subsection{Whole-cell multicolor rounding}

The discrepancy finisher needs a positive lower share for each observer and
color, rather than an exactly preserved designated row.  The following
rounding argument is the multicolor discrepancy principle of
Gla\ss er--Reitwie\ss ner--Witek~\citep[Theorem~2; full-version
Theorem~3.1]{GRW11}, following Doerr--Srivastav~\citep{DS03}.  We include its
finite-kernel proof for real recorded weights and its ordinary-query
implementation.  We make no bit-complexity claim.

\begin{lemma}[Whole-cell rounding]\label{r6core:rounding}
Let finitely many cells have nonnegative weights $w_i(Q)$ for $r\ge1$
observers, with $\sum_Qw_i(Q)=1$ and $w_i(Q)\le1/K$.  For any integer
$m\ge2$, exact finite linear algebra assigns each cell wholly to one of
$m$ colors so that
\[
 \left|\sum_{Q:\,\operatorname{color}(Q)=j}w_i(Q)-\frac1m\right|
 \le\frac{2r}{K}
 \quad\text{for every }i,j.
\]
\end{lemma}
\begin{proof}
Start with variables $x_{Qj}=1/m$.  Always preserve the cell equations
$\sum_jx_{Qj}=1$ and nonnegativity.  Initially also maintain all $rm$
observer/color equations $\sum_Qw_i(Q)x_{Qj}=1/m$.  A zero variable never
returns; a cell with one positive variable is integral and fixed forever.

Let $C$ be the number of fractional cells, $V$ their positive variables, and
$D$ the currently maintained observer/color equations.  Then $V\ge2C$.
If a nonzero direction on these variables preserves all cell sums and
maintained equations, move a positive distance along it until some positive
variable reaches zero.  A nonzero direction with zero sums in its affected
cells has both signs.  Thus this move exists, eliminates a variable, and
keeps all variables in $[0,1]$.

Otherwise the constraint matrix on the $V$ variables has independent columns,
so $V\le C+D$ and hence $D\ge V/2$.  Each variable appears in at most $r$
observer/color equations, one per observer for its own color.  Total
incidence is at most $rV$, so some maintained equation contains at most $2r$
positive fractional variables.  Drop that equation.

At the time of dropping, the equation is exact.  All its integral-cell
contributions remain fixed, its zero variables never return, and at most
$2r$ remaining variables can change.  Each changes by at most one and has
weight at most $1/K$, so its final discrepancy is at most $2r/K$.
Equations never dropped stay exact.  Every step reduces either positive
support or the number of maintained equations, and one of the two steps is
available whenever a fractional cell remains.  The procedure therefore
terminates with an integral color assignment and the stated bounds.
\end{proof}

\begin{lemma}[Near-equal physical partition]\label{r6core:near-equal}
Let $U$ be a recorded ordered finite union, let $m\ge2$, and let designated
observers have nonnegative nonatomic costs.  Let $r$ be the number of those
observers with positive total cost for $U$, and let $N$ be the number of
enclosing observers whose endpoint costs are recorded.  A deterministic
ordinary-query procedure partitions all of $U$ into $m$ finite unions
$P_1,\ldots,P_m$ satisfying, for every designated observer $i$,
\[
 \frac{C_i(U)}{2m}\le C_i(P_j)\le\frac{3C_i(U)}{2m}.
\]
If $r=0$, it makes no cuts.  Otherwise it uses at most $r(4rm-1)$ Cuts and
$Nr(4rm-1)$ total queries beyond the cached input costs.
\end{lemma}
\begin{proof}
A zero-total observer has zero cost on every subset and needs no constraint.
If all designated totals are zero, assign all of $U$ to one color and leave
the others empty.  Otherwise normalize each positive row by its known total
and set $K=4rm$.  For each of the $r$ rows make its $K-1$ positive quantile
cuts and merge all their boundaries.  Nonnegativity makes the row's quantile
endpoints monotone for every choice of plateau roots.  Each cell in the
common refinement has normalized cost at most $1/K$ to every row.  Component
joins can be retained as cell boundaries; their costs are already recorded.

Record every new endpoint for all $N$ observers and apply
Lemma~\ref{r6core:rounding} to the normalized cell weights.  Its error is
$2r/K=1/(2m)$.  Assign every cell whole to its chosen color.  This yields
the two displayed bounds, is a complete physical partition, and introduces
no further cuts.  A color may be a disconnected union of recorded cells.
Each quantile requires at most one Cut and $N-1$ additional Evals; locating
its component uses the cached component costs.  This proves the ledger.
\end{proof}

The splitter does not preserve any designated row exactly.  The finisher
uses only its displayed lower bounds; any other caller needing an exact
designated share requires a different interface.

\section{The many-parent cake SubCore primitive}\label{r6base:subcore}
\label{dc:subcore-section}

This appendix includes the inherited primitive proof underlying our compensation routine. It follows the Aziz--Mackenzie SubCore framework~\cite{AM16}, with the tie and equal-value-chain conventions of Ye--Bai~\cite[Appendix F]{YB26}. All parent component values are cached, and a strict prefix or suffix target is located from those component totals and implemented by one ordinary interval Cut; zero, full, and previously recorded targets reuse endpoints. Every fresh endpoint is recorded for the enclosing observers. Null material remains part of the domain.

We temporarily interpret the same nonnegative measures as values to
maximize. The following subroutine supplies the removed pieces in the
chore compensation argument.

\begin{lemma}[Many-parent SubCore]\label{dc:subcore}
Given $k\ge1$ participants and $m>k$ disjoint recorded ordered parents,
there is a deterministic procedure assigning each participant a suffix
of a distinct parent such that: the assignments are cake envy-free; every
participant weakly prefers her assignment to every wholly unused parent;
and at least one assigned parent is whole. The other $m-k$ parents remain
whole and unused. For $k=1$ no cut is required. For $k\ge2$ at most
$2^{k+1}-2k-2$ cuts are required, independent of $m$. With $N\ge k$ enclosing
observers and cached parent component values, at most $N(2^{k+1}-2k-2)$ queries
record the new endpoints. The result permits zero-valued or empty parents
and every legal Cut reply.
\end{lemma}

We prove the more general recursive interface used by the procedure.
The input additionally supplies benchmarks $b_i$ and a cake envy-free
allocation $H$ giving every agent a suffix of a distinct input parent of
value exactly $b_i$. The output meets the benchmarks and all the conclusions
of the lemma. Initially $b_i=0$ and $H$ consists of distinct empty suffixes.

First suppose all suffix comparisons are strictly monotone along each
parent. Process participants in a fixed order $1,\ldots,k$, maintaining a
tentative allocation of those already processed. If the new agent's
favorite whole input parent is unallocated, give it to her. Otherwise at
stage $t$ let $X$ be the tentative allocation of the first $t-1$ agents.
Its $t-1$ parents are \emph{contested}; all other input parents are
\emph{uncontested}. For each $i\le t$ put
\[
\beta_i=\max\{b_i,\max_{P\text{ uncontested}}V_i(P)\}.
\]
Ask each processed agent to trim every contested parent of value at least
$\beta_i$ to a suffix of value $\beta_i$. On each contested parent select
a rightmost trimmer, favoring its owner in the inherited allocation $H$
when she ties. Let $T$ be the set of selected agents. Give each member of
$T$ one suffix from a parent on which she was selected. Such choices use
distinct parents, since each parent selected only one agent.

While $|T|<t-1$, ignore the original trims of $T$ and form $\mathcal C'$
by retaining on each contested parent the suffix at its rightmost remaining
trim, or the whole parent if no trim remains. Recursively run SubCore on
$T$ and $\mathcal C'$, with their current own values as benchmarks and
their current suffixes as inherited allocation. Let its output be $Y$.
If an unused member of $\mathcal C'$ has a remaining trim, give it to one
of its rightmost remaining trimmers. Otherwise perform the chain operation
proved next to free such a parent, and give that parent to its rightmost
remaining trimmer. Add this agent to $T$ and update its members' benchmarks
to their own current values. When $|T|=t-1$, keep these assignments and give
the sole remaining agent among $1,\ldots,t$ her favorite uncontested parent.
There is no additional recursive call after $|T|=t-1$.

\begin{lemma}[Equal-value chain]\label{dc:subcore-chain}
During a recursive stage above, if an unused parent $P\in\mathcal C'$
has no remaining trim, agents in $T$ can be reassigned without changing
their own values so that $P$ becomes allocated and a parent with a remaining
trim becomes unused. Cake envy-freeness and comparisons with unused parents
are preserved.
\end{lemma}
\begin{proof}
Every owner in $X$ trimmed her own parent, because its value is at least
her initial $\beta_i$. Let $i_0$ own $P$ in $X$, and let $i_{r+1}$ be
the owner in $X$ of the parent assigned to $i_r$ in $Y$. Stop at the first
$i_h\notin T$. Initially $i_0\in T$, since $P$ has no remaining trim.
The chain has no repeated agent: $X$ and $Y$ are injective assignments and
$P$ is unallocated in $Y$. It therefore terminates outside $T$.

Walking backwards, we claim $Y_{i_{r-1}}\subseteq X_{i_r}$ for
$1\le r\le h$. At $r=h$ the outside agent's remaining trim lies within
$X_{i_h}$, and the entire parent in $\mathcal C'$ begins at or after that
trim. At an earlier index, induction and envy-freeness of $X$ give
\[
V_{i_r}(Y_{i_r})\le V_{i_r}(X_{i_{r+1}})
                 \le V_{i_r}(X_{i_r}).
\]
If $X_{i_r}\subsetneq Y_{i_{r-1}}$, strict monotonicity along the same
parent would make $Y_{i_{r-1}}$ more valuable than
$Y_{i_r}$, contradicting envy-freeness of $Y$. This proves the claim.

Now walk forwards. The unused-parent guarantee for $Y$ gives
$V_{i_0}(Y_{i_0})\ge V_{i_0}(P)\ge V_{i_0}(X_{i_0})$.
Together with the preceding inequalities this forces equality, hence
$X_{i_0}=P$ and $Y_{i_0}=X_{i_1}$. If
$Y_{i_{r-1}}=X_{i_r}$, envy-freeness of $Y$ and the same upper bounds
give equality again: agent $i_r$ is indifferent between $X_{i_r}$ and
$Y_{i_r}$, and $Y_{i_r}=X_{i_{r+1}}$. Finally
$Y_{i_{h-1}}=X_{i_h}$ equals its parent in $\mathcal C'$, since that
parent lies between it and the outside agent's trim. It has a remaining
trim. Give each $i_r$, $r<h$, her suffix $X_{i_r}$. Their own values are
unchanged. The multiset of allocated suffixes adds only $P$ and removes
$X_{i_h}$; every agent weakly preferred her old assignment to both these
whole parents. All required comparisons persist.
\end{proof}

\begin{proof}[Correctness and complexity of Lemma~\ref{dc:subcore}]
Induct on the participant count and, within a call, on $t$. The easy branch
gives the new agent a favorite whole parent, meeting her benchmark; previous
agents did not envy an unused parent. In a contested stage, every old owner
has $V_i(X_i)\ge\beta_i$ and so trims her own parent; every contested
parent has at least one trim. A nontrimmer values that parent below her
$\beta_i$. Consequently every selected rightmost suffix is worth at most
$\beta_i$ to every agent $i\le t$, and its selected owner values it
exactly $\beta_i$. The initial allocation of $T$ is envy-free.

An agent $i$ whose benchmark exceeds every uncontested value must lie in
$T$. Her inherited suffix $H_i$ belongs to a contested parent; she trims
that parent exactly to $H_i$. Every other agent $j$ values $H_i$ at most
$b_j\le\beta_j$ by envy-freeness of $H$, so no trim is to its right.
The specified tie rule selects $i$. Such agents never leave $T$.

At a recursive stage, ignoring more original trims only enlarges parents,
so the current allocation is a valid inherited allocation. Every agent
outside $T$ retains her original $\beta_i$ and values each parent in
$\mathcal C'$ at most $\beta_i$. The recursive result meets the current
benchmarks and compares favorably to unused parents. The chain lemma,
when necessary, frees a parent at a remaining trim. Its new outside owner
values it exactly $\beta_i$, and the enlarged allocation is envy-free in
both directions. Updating the benchmarks can only increase them.

At the end all contested parents are assigned to $T$. The sole outside
agent can meet her benchmark on an uncontested parent by the preceding
necessity argument. She receives value $\beta_i$ and values every contested
suffix at most $\beta_i$. Every member of $T$ receives at least her initial
$\beta_i$, so does not envy any uncontested parent. This proves all
comparisons, the benchmarks, and the whole-parent guarantee. Every produced
suffix is whole or starts at a recorded trim.

For nonnegative measures use persistent suffix records with formal values
$(V_i(P),\theta_i)$. On each original parent initialize the whole record
with secondary value $1$ and the empty suffix with $0$, even when the
parent is physically empty. Formal values decrease strictly in the ordered
list from whole to empty. A formal trim reuses its target record or inserts
one between adjacent records by the bracketed endpoint procedure for persistent
prefix records in Appendix~\ref{r6core:appendix}. All old records persist in descendants; new records copy
the trimming agent's target secondary value and choose intermediate rational
values for other observers. The proof just given uses only ordering,
strict monotonicity within a parent, and exact trim targets. It therefore
applies to these formal records verbatim. In particular equality along the
chain identifies the same suffix record. Formal envy-freeness implies
real envy-freeness, and a whole input record is physically whole. No
positive-benchmark initialization is required at the top level: distinct
empty records provide the inherited allocation at formal benchmark $(0,0)$.

Let $S(k)$ bound all cuts in a call with $k$ agents, uniformly over its
number of parents. The first $k-1$ stages cost at most $S(k-1)$.
At stage $k$, there are at most $k(k-1)$ initial trims. During its
while loop the selected set strictly grows, so each recursive
cardinality $j\in\{1,\ldots,k-2\}$ occurs at most once; the last
same-size call is omitted and the chain operation uses no query.
Thus the original generic source recurrence~\cite[Appendix F.4]{YB26} is
\[
 S(0)=S(1)=0,\qquad
 S(k)\le S(k-1)+k(k-1)+\sum_{j=1}^{k-2}S(j).
\]
The equality envelope obeys $S(k)=2S(k-1)+2(k-1)$ and therefore
$S(k)=2^{k+1}-2k-2$. Induction bounds every actual call by this envelope.
Favorite scans and the chain use recorded values. Each genuine new trim
requires one Cut and at most $N-1$ recording evaluations. Reused endpoints
require no query. This proves the stated bound, including exploratory cuts.
\end{proof}

\begin{corollary}[Threshold guarantee]\label{dc:subcore-threshold}
If at least $k$ offered parents have value at least $z_i$ for participant
$i$, then her SubCore assignment has value at least $z_i$.
\end{corollary}
\begin{proof}
If one such parent is unused, compare with it. Otherwise exactly $k$
parents are assigned, all of them have value at least $z_i$, and one is
assigned whole. Cake envy-freeness supplies the comparison with that share.
The high-parent sets may differ between participants.
\end{proof}

\section{SubCore with parent-dependent cost offsets}
\label{r7offset:appendix}

This appendix supplies the compensation primitive used by the component
batches.  The transformation is an implementation of the many-parent
SubCore of Aziz--Mackenzie and Ye--Bai, rather than a new valuation oracle.
The full source algorithm, equal-value-chain argument, and fresh-zero
interface are given in Lemma~\ref{dc:subcore}; see also
Ye and Bai~\cite[Algorithm~6 and Appendix~F]{YB26}.

Let $W_1,\ldots,W_m$ be disjoint recorded ordered suffix parents and let
$k<m$ agents participate.  There are $N\ge k$ enclosing observers whose
costs at new endpoints are recorded.  Their original costs $C_i$ are finite,
nonnegative, and nonatomic.  For each observer and parent, a known constant
$g_{si}\ge0$ is given.  On suffixes $D\subseteq W_s$ define
\[
 u_i(s,D)=C_i(D)-g_{si},\qquad
 v_i(s,D)=\max\{u_i(s,D),0\}.
\]
The transformed utilities are compared only on individual parent suffixes.
They are not asserted to be additive on unions of different parents.
Original costs account for every physical allocation and complement.

\begin{theorem}[Offset SubCore]\label{r7offset:theorem}
The following two deterministic routines permit every legal ordinary Cut
reply, including arbitrary plateau roots.
\begin{enumerate}[leftmargin=*,label=\textup{(\roman*)}]
\item\label{r7offset:clipped}
For the utilities $v$, the routine returns one suffix $D_i$ to each
participant, from distinct parents, with
\[
 v_i(D_i)\ge v_i(D_j)\quad\text{for all participants }i,j,
\]
and
\[
 v_i(D_i)\ge v_i(W_s)\quad\text{for every wholly unused parent }s.
\]
At least one returned suffix is its entire offered parent.
\item\label{r7offset:affine}
Suppose $u_i(W_s)\ge0$ for every participant $i$ and every offered whole
parent $s$.  The same guarantees hold for $u$, and every returned own
utility $u_i(D_i)$ is nonnegative.  A whole output with affine own utility
zero is a physical output and remains assigned.
\end{enumerate}
Each routine uses at most
\[
 S_k=2^{k+1}-2k-2\quad\text{Cuts},\qquad NS_k\quad\text{total queries}
\]
beyond the cached input component costs and offsets.  In both cases the
unallocated material is exactly the complementary prefixes and the unused
whole parents.  No cut to a difference measure or perturbed measure is used.
\end{theorem}

\subsection{The monotone-parent source interface}

The recursive SubCore interface supplies an inherited cake envy-free suffix
assignment $H$ on distinct parents, with benchmarks $b_i$ equal to the
inherited own utilities.  At a contested stage it trims a parent for agent
$i$ to the target
\[
 \beta_i=\max\{b_i,\ \max_{W\text{ uncontested}}\text{utility}_i(W)\}.
\]
The algorithm and its proof use comparisons of individual suffix utilities,
monotonicity along one parent, exact target equalities, and suffix
containment.  In the equal-value chain, the backwards step deduces
containment by comparing two suffixes of the same parent.  The forwards step
uses envy-freeness and comparison with unused parents to prove equality of
own utilities.  Neither step adds utilities from different parents or
identifies a difference of utilities with the utility of a set difference.
Thus the proof of Lemma~\ref{dc:subcore} applies to strictly increasing
continuous parent utility functions with a coherent inherited assignment.

For weak monotonicity, retain the source's permanent formal suffix records.
A record consists of a physical endpoint, its actual utility to every
observer, and rational secondary labels.  Compare utilities
lexicographically.  For every observer they decrease strictly as the
suffix's left endpoint moves right through the record list.  Distinct
formal records may share a physical endpoint.  All records, including
exploratory records, persist; old values and labels never change.

A formal trim target is the formal utility of an existing record.  Reuse a
matching record if available.  Otherwise find adjacent records bracketing
it.  If one endpoint has the target primary utility, reuse that physical
endpoint and insert a new record.  Give the trimming observer the target
secondary label.  For another observer, choose a rational intermediate
label: a midpoint when both neighboring primary values tie, a value above
or below the single tied neighbor when only one ties, and any rational
value when the new primary value is strictly intermediate.  These choices
are always possible.  Secondary labels need not lie in $[0,1]$.

This strict-order implementation supports the same finite source proof.
Alternatively, after any finite completed table, a constant row shift and
one sufficiently small common $\epsilon>0$ realize all formal values as
hungry additive suffix values on abstract parents.  Each record's value is
its shifted primary value plus $\epsilon$ times its secondary label.
There are finitely many comparisons, so one $\epsilon$ preserves them all;
linear interpolation between distinct abstract record positions gives
positive densities.  Existing records are immutable, so this is one
realization of the whole table.  It is used only in the proof, not by the
physical protocol.  The two implementations below specify the actual
original-measure query for every formal trim.

\subsection{Clipped utilities and physical zero targets}

For $v$, physical empty has value zero for every observer.  Initialize its
secondary label to zero and the whole-parent label to one, also for zero
utility or physically empty parents.  This supplies the fresh-zero inherited
assignment on distinct parents.

Suppose a formal trim has primary target $t$.  If it cannot reuse a
bracketing physical endpoint, its primary value lies strictly between the
neighbors.  Necessarily $t>0$, because all clipped values are nonnegative
and a primary-zero target always has a zero-valued neighbor.  The desired
suffix then satisfies precisely
\[
 C_i(D)=g_{si}+t.
 \tag{\ref{r7offset:theorem}$^*$}
\]
Its original cost is strictly bracketed by the two neighboring original
suffix costs.  Let $a$ be the left bracket endpoint.  In the usual
left-to-right Cut convention, request prefix cost
\[
 C_i(\text{suffix starting at }a)-(g_{si}+t).
\]
Every legal returned root lies inside the strict bracket and leaves a
suffix of the desired original cost.  For an ordered union, cached
component costs first locate the component containing this target; only
one ordinary interval Cut is required.  The trimmer's original endpoint
cost is already known from the target.  At most $N-1$ further Evals record
the other observers.  Reusing a physical coordinate needs no new query.

Consequently a clipped primary-zero trim needs neither a query to a signed
utility nor an extremal root.  The source's formal envy-free, unused-parent,
and whole-parent guarantees imply their primary, physical counterparts.
This proves the output claims in part~\ref{r7offset:clipped}.

\subsection{Affine utilities: virtual bottom and nonnegative targets}

For $u$, physical empty has value $-g_{si}$, which can be negative and
parent dependent.  It cannot be declared to have common benchmark zero.
For each observer set
\[
 M_i=-1-\max_s g_{si}.
\]
Place on every parent a virtual empty record with formal value $(M_i,0)$,
strictly below all its physical affine values.  Keep a separate record
for physical empty, with primary value $-g_{si}$.  The virtual record has
no physical material; its bookkeeping coordinate may equal the empty
endpoint, but its value is explicitly virtual.  Initialize the inherited
assignment on distinct virtual records with benchmark $M_i$.  It is
coherently envy-free because each observer sees the same virtual value
on all parents.

\begin{lemma}[Nonnegative affine targets]\label{r7offset:nonnegative}
Every actual trim target and every returned own utility in this execution
has primary value at least zero.  No output or queried root lies in the
interior of a virtual segment.
\end{lemma}
\begin{proof}
At the top-level call, every offered whole parent has nonnegative primary
utility for each participant.  At every contested stage an uncontested
whole parent exists, since $m>k$.  Its utility in the displayed maximum
therefore makes $\beta_i\ge0$.  An easy favorite-whole assignment is
also nonnegative.

Recursive calls require a different justification.  A recursive parent can
have negative whole utility to some participant.  However, its inherited
assignment has already been formed from physical assignments of
nonnegative own utility.  Its inherited benchmark thus has primary value
$b_i\ge0$.  Every recursive contested target satisfies
$\beta_i\ge b_i\ge0$.  In an easy branch the favorite whole parent meets
the benchmark: the participant's inherited suffix belongs to an offered
parent, whose whole utility is at least that suffix's utility, and the
favorite whole is at least as good.  Benchmark updates use current own
utilities, and equal-value-chain reassignments preserve them.  Induction
over calls and stages proves the claimed nonnegativity.

The assertion concerns primary values; a primary-zero target may have any
secondary label.  If a neighboring physical endpoint has the target
primary value, reuse it and insert a label.  Otherwise a genuine trim is
again exactly the original suffix target $g_{si}+t$, now with $t\ge0$.
In particular, $t=0$ can require a positive original cost $g_{si}$.
It is not generally the physical empty set.

No nonnegative target lies strictly inside the virtual segment below
physical empty, whose primary values are below $-g_{si}\le0$.
When $g_{si}=0$ and a primary-zero target's secondary label lies below
the physical-empty label, reuse the same physical empty endpoint with
a different formal record.  Thus every actual trim and output is physical.
\end{proof}

The ordinary Cut conversion and strict-bracket argument are the same as
for clipped positive targets.  For a finite strict realization, shift
observer $i$'s row by $-M_i$.  This changes no utility comparison or target
relation.  The virtual bottom becomes zero, and all physical values are
positive.  The finite-record interpolation described above then realizes
the source's strict monotone interface.  Lemma~\ref{r7offset:nonnegative}
ensures that the physical simulator never queries or returns any artificial
interior.  The source output is therefore physically envy-free for $u$,
dominates every unused whole parent, and includes an original offered
parent whole.  Every own affine value is nonnegative, including on the
whole output.  This proves part~\ref{r7offset:affine}.

\subsection{Query count and exact material accounting}

The source's recursive selected sets have distinct sizes at a stage, and
its final same-size recursive call is omitted.  The chain repair makes no
query.  The number of cuts is bounded by the envelope defined by
\[
 S_0=S_1=0,\qquad
 S_k=S_{k-1}+k(k-1)+\sum_{j=1}^{k-2}S_j
     =2S_{k-1}+2(k-1)
     =2^{k+1}-2k-2.
\]
Each request is either an endpoint reuse or at most one original Cut and
$N-1$ Evals.  Computing offsets, virtual bottoms, and transformed
comparisons uses the cached input values.  The $k=1$ call chooses a greatest
whole parent without a cut.  This proves the query bounds in
Theorem~\ref{r7offset:theorem}.

Each assigned suffix has one distinct original parent.  Its physical
prefix complement, together with the wholly unused parents, is exactly
the residual domain.  This assertion uses original sets and costs; no
additivity of $u$ or $v$ across parents is invoked.  Null components and
null tails remain in their physical assigned or complementary sets.
Uncached input components require their actual separate evaluation costs;
a disconnected union is not treated as one free interval Eval.

\subsection{What zero stripping does and does not preserve}

\begin{lemma}[Clipped-zero stripping]\label{r7offset:stripping}
Consider the nominal clipped output.  If $v_j(D_j)=0$, then
$v_j(D_i)=0$ for every nominal assigned suffix, and
$v_j(W_s)=0$ for every wholly unused parent.  If the caller executes only
the nominal outputs with positive own clipped utility, every executed
participant's affine own utility equals its clipped own utility.  All
its pairwise comparisons against the retained outputs remain valid.
Any nonnegative rank threshold paid by a stripped observer is zero.
\end{lemma}
\begin{proof}
Clipped utilities are nonnegative.  The nominal envy-free and unused-parent
inequalities with left side zero force every indicated right side to
zero.  A rank threshold is at most the nominal own value, so a nonnegative
such threshold is zero.  On a positive clipped own value,
$\max\{u_i(D_i),0\}=u_i(D_i)>0$.  Removing other nominal outputs does not
invalidate any surviving comparison.
\end{proof}

The physical caller must reserve all nominally selected parent roots,
including stripped proposals, and freeze an omitted root at its actual
old allocation while charging its old reserves.  Omitting a donor
operation removes its proposed deficit; it cannot supply the credit or
whole-parent material of that unexecuted operation.  Those physical
obligations are proved in the component-batch argument, not by the utility
lemma alone.

Stripping can destroy the whole-parent guarantee.  With one participant,
two parents of original cost one, and offset two on both, every clipped
utility is zero but a whole output has affine utility minus one.  Such
an output cannot be used as an actual donor.  Atomic extraction and closed
terminal batches consequently use the affine part of
Theorem~\ref{r7offset:theorem}, whose nonnegative-whole assumption permits
every zero-affine output to remain physical and retains the whole-parent
certificate.

\section{Whole-piece augmentation and two-group completion}
\label{r6c:completion}

Let $N$ be a population of $n$ agents with finite nonnegative nonatomic Borel costs,
let $B$ be a nonempty proper subset, and let $X,R$ be disjoint recorded
finite-union domains, with all component values already recorded.
Put $\alpha=2^{n+2}$. The separated-domain hypothesis is
\begin{equation}\label{r6c:separated}
 C_i(X)\ge\alpha C_i(R)\ (i\in B),\qquad
 C_i(R)\ge\alpha C_i(X)\ (i\in N\setminus B).
\end{equation}

We give the full completion procedure to specify its physical
complements and zero-cost behavior. Lemma~\ref{r6core:near-equal}
partitions a recorded domain $U$ into $m$ pieces, each of cost at
least $C_i(U)/(2m)$ for each of $r$ observers, with at most
$r(4rm-1)$ cuts. It uses only this lower bound, not exact equality
for a designated observer. Zero-total observers impose no restriction.

\begin{lemma}[Whole-piece augmentation]\label{r6c:oblige}
Given $k\ge1$ recipients and $m=2^{k+1}$ disjoint recorded pieces,
there is a partial chore-envy-free allocation giving each recipient
a bundle containing one whole input piece. The exact unallocated
complement is retained. At most $2^k-1$ cuts are made.
\end{lemma}
\begin{proof}
Number the recipients $1,\ldots,k$ and set $a_i=2^{i-1}$.
In forward order, $i$ reserves her $a_i$ most costly available
original pieces, breaking ties by labels. The remaining $2^k+1$
pieces are public bundles, initially one whole piece each. Reserved
pieces are not public choices; public identities never change.

Process agents in reverse order. Let $v_i$ be the $(a_i+1)$-st
smallest current public-bundle cost for $i$, and designate a set
$D_i$ of exactly $a_i$ smallest public bundles. Attach to each a
prefix from a distinct $i$-reserved piece, of cost equal to $v_i$
minus the current bundle cost. A zero supplement is empty but its
designation stays in $D_i$. Keep all unused reserved tails in the
complement.

Before $i$ acts, larger-index agents have touched at most
$\sum_{j>i}a_j=2^k-2^i$ public bundles. At least $2^i+1$ original
public pieces remain intact. Each was available when $i$ reserved
her most costly pieces, so it costs no more to $i$ than any of her
reserves. There are more than $a_i+1$ such candidates; hence $v_i$
is bounded by the cost of each reserve and every supplement fits.

After this augmentation all public bundles cost at least $v_i$ to
$i$, and every member of $D_i$ costs exactly $v_i$. Later
augmentations only add cost. Now agents choose in forward order,
each taking a remaining minimum-cost bundle, preferring her own
$D_i$ in a tie. Inductively earlier choices belong to their own
designated sets. Since $|D_i|>\sum_{j<i}|D_j|$, some member of
$D_i$ was neither touched by a smaller-index augmenter nor chosen
earlier. It still costs $v_i$. The choice therefore belongs to
$D_i$ and has cost $v_i$, while every other chosen bundle costs at
least $v_i$ to this observer. This is envy-free.

Every chosen bundle contains its original whole public piece.
Distinct reserved pieces supply disjoint supplements. Unchosen
public bundles, unused reserves and every reserved tail form the
exact complement. There are at most $\sum_i a_i=2^k-1$ supplement
cuts, including all tie and zero-designation cases.
\end{proof}

\begin{lemma}[Two-group completion]\label{r6c:discrepancy}
Suppose \eqref{r6c:separated} holds. Recursive calls on strictly
smaller populations suffice to divide $X\cup R$ completely and
without envy. The preparation uses at most $19\cdot2^n$ cuts and
$19n2^n$ recorded ordinary queries.
\end{lemma}
\begin{proof}
Put $D=N\setminus B$, $b=|B|$ and $c=|D|$. Partition $X$ for
the $b$ observers in $B$ into $m_D=2^{c+1}$ pieces by
Lemma~\ref{r6core:near-equal}, and augment them for the $c$
recipients in $D$ using Lemma~\ref{r6c:oblige}. Symmetrically,
partition $R$ for the observers in $D$ into $m_B=2^{b+1}$ pieces
and augment them for recipients in $B$. Keep both partial outputs
tentative until both are ready. Let $X_D\subseteq X$ and
$R_B\subseteq R$ be their exact complements, including all
unspent reservation tails. Recursively divide $X_D$ among $D$
and $R_B$ among $B$, then add those allocations.

Within either group, the two allocations are envy-free and have
disjoint domains. For $i\in B$ her entire new bundle lies in $R$,
whereas every recipient in $D$ retains a whole original near-equal
$X$ piece of cost at least
\[
 C_i(X)/(2m_D)\ge C_i(R),
\]
since $2m_D\le2^{n+1}\le\alpha$. Thus cross-group envy is absent.
The symmetric argument applies to $i\in D$. This gives envy-freeness
on all of $X\cup R$, independently of earlier allocations. Every
domain is exactly its partial allocation plus its recursive
complement. Both child groups are nonempty and smaller than $n$.

The two splits and augmentations make at most
\[
 8(b^22^c+c^22^b)+2^b+2^c-n-2
\]
cuts. For positive integers $t$, $t^2/2^t\le9/8$, and
$2^b+2^c\le2^n$. Therefore this count is at most $19\cdot2^n$.
Each new endpoint costs at most $n$ queries with current-observer
recording. All input totals and resulting color values are cached
sums, so there is no additional initial-evaluation term. Zero
observers can only reduce the count.
\end{proof}

\section{Improved randomized allocation for general measures}
\label{r6u:appendix}

We prove an expected polynomial query bound using ordinary individual
Robertson--Webb queries. Queries in every rejected attempt are counted.

\begin{theorem}\label{r6u:general}
For $n$ finite nonnegative nonatomic Borel measures, interpreted as cake
values or chore costs, a randomized protocol returns a complete exactly
envy-free allocation into finite unions of intervals almost surely, with
\[
 \E{Q}=O\!\left(n^{12}\log(2n)\right),\qquad
 \E{C}=O\!\left(n^{11}\log(2n)\right).
\]
The bounds hold uniformly over profiles and legal measurable
nonanticipating Cut-response policies, including arbitrary legal replies on
plateaus. The protocol uses exact real arithmetic and independent uniform
real draws. No time or finite-bit implementation bound is asserted.
\end{theorem}

We use the physical compensation framework of Mackenzie and
Suzuki~\cite{MS26}. Three changes yield the smaller sample bound.
A deterministic selection of witness cells leaves a large compensation
domain. Its maximum-determinant selection rule is stable under deletion
of an unselected sample, which controls interpolation at fresh correction
endpoints. Iterated rounding then reduces whole-cell discrepancy from
quadratic to linear in the active population. A finite sequence of fresh
correction attempts converts a constant local success probability into a
uniform trial guarantee without multiplying the endpoint-error bound by
the number of attempts. We give the construction and its full accounting.

For nonnegative measures $\mu_i$, use margins
\[
 g_{ij}(X)=\mu_i(X_i)-\mu_i(X_j)\quad\text{for cake},\qquad
 g_{ij}(X)=\mu_i(X_j)-\mu_i(X_i)\quad\text{for chores}.
\]
Thus exact envy-freeness means $g_{ij}(X)\ge0$. All conditional
statements include the complete preceding query history. Fixed tie rules
and finite linear-algebra, assignment, and matching algorithms make the
choices measurable. The original population is $n$; the active population
$k\le n$ may decrease, but the parameters and endpoint records always
refer to the original population.

\subsection{Ordinary queries and fixed interpolation tables}
\label{r6u:setup}

Every domain is a finite ordered union of intervals with recorded
component values. Record every new endpoint for all $n$ original
observers, including frozen observers; one Cut and these records use
at most $n+1\le2n$ queries. Cached component sums locate a cumulative
target. At a recorded cumulative value reuse its endpoint. Otherwise
the residual target lies strictly between zero and the value of one
component $[a,b]$, and a single ordinary Cut from $a$ realizes it.
Every legal root lies inside $(a,b)$: at $b$ the cumulative value
already strictly exceeds the residual target, and nonnegativity
prevents it from decreasing afterward. No sum-measure or union-cut
oracle is assumed.

Concatenate the components into a virtual interval. Logical cells can
cross already recorded joins; those old coordinates need not refine
every new sampling grid. The physical sets remain unions of the
corresponding subintervals, with gaps excluded. Empty/full targets and
literal complements retain all null fragments.

We use the following elementary circular law. On a recorded union $S$
with $\nu(S)>0$, a uniform start in the $\nu$-value circle followed by
an arc of length $t$ gives a piece $C$ of exact value $t$ with at most
two endpoints. Each endpoint has conditional marginal law
$\nu|_S/\nu(S)$ before their shared start. Each fixed point of $S$
has inclusion probability $t/\nu(S)$. Consequently, for any finite
signed nonatomic measure $\xi$ fixed before the start,
\begin{equation}\label{r6u:circular}
 \E{\xi(C)\mid\mathcal H}=\frac{t}{\nu(S)}\xi(S),\qquad
 \E{|\xi(C)|\mid\mathcal H}\le\frac{t}{\nu(S)}|\xi|(S).
\end{equation}
This remains valid when $\xi$ charges cutter-null intervals. Indeed,
if $F$ is the continuous nondecreasing cumulative function and $Y$ is
a uniform target, any legal root $X$ satisfies $X\le x$ exactly when
$Y\le F(x)$ except at $Y=F(x)$. Membership of a fixed point in a
rotated arc is membership of its cumulative coordinate in that value
arc. Tonelli proves~\eqref{r6u:circular}. Nonunique-root levels are
countable, since each nontrivial level interval contains a distinct
rational point, and fresh random targets and their fixed circular
shifts avoid them almost surely. Distinct target levels have ordered
roots. A circular partition into $a$ specified nonnegative masses
similarly uses at most $a$ endpoints and covers the entire domain.
Zero amounts are empty, and a full amount takes the whole domain,
including null fragments. No independence is asserted within one
circular group.

Normalize the $k$ active measures on a positive buffer $B$ to
$\nu_i(B)=1$, and let $\mu=k^{-1}\sum_i\nu_i|_B$. An exact
$\mu$-sample uses one ordinary Cut: choose an actual active agent
uniformly and then a uniform own-value target. Fresh choices yield
iid samples under every admitted response policy.

For fixed signed measures $\xi_1,\ldots,\xi_k$, a fixed permanent grid,
and $m$ independent $\mu$-samples, let $P$ be their common grid.
Write $\gamma_{ij}=\xi_i-\xi_j$, let $F_{ij}$ be its cumulative
function, and let $\widehat F_{ij}$ be its linear interpolant on $P$.
Define
\[
 D_{ij}=\sum_{E\in P}|\gamma_{ij}(E)|,\qquad
 e_{ij}(x)=|F_{ij}(x)-\widehat F_{ij}(x)|,
 \qquad u_P(x)=\max_{i<j}\frac{e_{ij}(x)}{D_{ij}+e_{ij}(x)},
\]
using $0/0=0$. For the fixed-measure sampling step used later by Snapshot, the integrated
interpolation estimate is
\begin{equation}\label{r6u:integrated}
 \E{\int u_P(x)\,d\mu(x)}\le\frac{2(k-1)}{m+1}.
\end{equation}
Here the underlying measures and permanent boundaries are fixed before
sampling. We prove~\eqref{r6u:integrated}, including its zero cases.
For finite row vectors $x_1,\ldots,x_k$ put
$d_{ij}=\|x_i-x_j\|_1$ and
\[
 s_a=\max_{i<j}\frac{|x_i(a)-x_j(a)|}{d_{ij}}.
\]
Then $\sum_as_a\le k-1$, a special case of the subspace sensitivity
bound~\cite[Lemma 2.6]{MMWY21}. To prove this elementary version,
identify identical rows first, so all remaining edge distances are
positive. If there is one row the assertion is immediate. Otherwise
take a minimum spanning tree on the distinct rows. Every edge $e$
on the path from $i$ to $j$ has $d_e\le d_{ij}$, since replacing a
longer edge by $(i,j)$ would reduce the tree's total length. Hence
\[
 \frac{|x_i(a)-x_j(a)|}{d_{ij}}
 \le \sum_{e\in\operatorname{path}(i,j)}
       \frac{|x_{e^-}(a)-x_{e^+}(a)|}{d_e}
 \le \sum_{e\in T}
       \frac{|x_{e^-}(a)-x_{e^+}(a)|}{d_e}.
\]
Taking the maximum and summing over coordinates gives one unit per
tree edge, at most $k-1$. Identical-row quotients are zero.

Now delete an interior point from a deterministic augmented grid.
For a fixed pair let $a,b$ be the signed differences on its two
incident cells, put $A=|a|+|b|$, and let $V$ be its total variation
over the augmented cells. The interpolation error $e$ at the deleted
point is at most $A$. The remaining cell variation is
$D^{(-z)}=V-A+|a+b|$. If $V>0$, cross multiplication gives
\[
 \frac{e}{D^{(-z)}+e}\le\frac{A}{V},
 \qquad
 e(V-A)\le A(V-A+|a+b|).
\]
The second inequality uses $e\le A$ and $V-A\ge0$. If $V=0$,
both incident differences and the error vanish. The maximum normalized
deletion error is therefore at most the sum of the two incident
coordinate sensitivities. Each cell is incident to at most two
deletion points, so the sum of these maxima over any subset of
interior points is at most $2(k-1)$.

Adjoin an independent $(m+1)$st $\mu$-sample. Atomlessness eliminates
coincidences with other samples and the permanent boundaries almost
surely. By exchangeability, each sampled point is equally likely to
be the distinguished point removed from the augmented grid. Removing
it gives precisely the interpolation error of the original experiment
at an independent sample. Divide the deletion bound by $m+1$ and
average to prove~\eqref{r6u:integrated}. The same argument works
conditional on any independent information that fixes the measures
and permanent grid before sampling.

\begin{lemma}[Aggregate endpoint interpolation]\label{r6u:aggregate}
After $P$, let adaptive groups of endpoints be generated. Before each
group's fresh randomness, its conditional endpoint marginal laws sum
to at most a predictable measure $K_g$. Suppose, pathwise,
$\sum_{g\text{ reached}}K_g\le\Lambda_{\rm sum}\mu$, and let $b>0$.
Except with probability
\begin{equation}\label{r6u:aggregate-failure}
 \frac{2(k-1)\Lambda_{\rm sum}(1+b)}{b(m+1)},
\end{equation}
every pair and every finite union $E$ with boundaries in $P$ or the
generated list satisfy
\[
 |\gamma_{ij}(E)-\widehat\gamma_{ij}(E)|\le bD_{ij}.
\]
The estimate uses the original immutable grid. It includes zero
disagreements, adaptive stopping, rejecting groups, and unions chosen
after all values have been observed.
\end{lemma}

\begin{proof}
Before a group, $u_P$ is a fixed nonnegative measurable function of
its history. Conditional marginal domination bounds its expected
endpoint charge by $\int u_P\,dK_g$. Sum by conditional expectation,
use the pathwise measure bound, and apply~\eqref{r6u:integrated}:
\[
 \E{\sum_{Z\text{ generated}}u_P(Z)}
 \le\frac{2(k-1)\Lambda_{\rm sum}}{m+1}.
\]
For each pair put $E^*_{ij}=\sum_Ze_{ij}(Z)$. The common charge is
at least $E^*_{ij}/(D_{ij}+E^*_{ij})$. If any $E^*_{ij}>bD_{ij}$,
that charge exceeds $b/(1+b)$, including when $D_{ij}=0$. One
Markov inequality proves~\eqref{r6u:aggregate-failure}. Outside the
event all $E^*_{ij}\le bD_{ij}$. Telescope cumulative errors at the
endpoints of each union; original boundaries contribute zero.
Reused endpoints are charged at their original generation.
\end{proof}

\subsection{Stable witnesses and a large compensation domain}
\label{r6u:repair}

The input to Repair consists of disjoint recorded whole endowments
$W_1,\ldots,W_k$ and a disjoint positive buffer $B$, satisfying
\begin{equation}\label{r6u:repair-input}
 g_{ij}(W)\ge-\tau\mu_i(B).
\end{equation}
There is no upper bound on endowment values relative to the buffer.
Normalize $\nu_i=\mu_i/\mu_i(B)$, and draw $m$ independent samples
from $\mu=k^{-1}\sum_i\nu_i|_B$. Before sampling, take each row's
$q$ equally spaced interior quantiles. Their common coarse grid has at
most $kq+1$ logical cells, each of every row's value at most
$\sigma=1/(q+1)$. The sampled refinement has at most $kq+m+1\le2m$
cells. All later interpolation uses this immutable refinement.

For each of $k$ batches, each unordered row pair, and each of $L$
repetitions, choose two representative cells among the coarse parents
not previously removed. Remove the whole parents of both choices before
the next request. Thus there are
\[
 r=k\binom{k}{2}L
\]
requests and at most $2r$ removed parents. The selection rule below is
applied to the original-buffer normalized vectors
$(\nu_i(E),\nu_j(E))$. Its conclusion holds for arbitrary later
normalizations; the selected pair itself need not be invariant under
rescaling the rows.

\begin{lemma}[Maximum-determinant witnesses]\label{r6u:representatives}
For a finite nonempty family of nonnegative vectors $x_E\in\mathbb R^2$,
choose an ordered pair $(F,G)$ maximizing
\[
 \bigl(|\det(x_F,x_G)|,\ \|x_F\|_1+\|x_G\|_1\bigr)
\]
lexicographically, with a fixed geometric tie rule; repeated cells are
allowed. Every linear functional $a$ satisfies
\[
 |a(x_E)|\le |a(x_F)|+|a(x_G)|
            \le2\max\{|a(x_F)|,|a(x_G)|\}.
\]
Deleting unchosen alternatives preserves the chosen pair.
\end{lemma}
\begin{proof}
If the maximum determinant is nonzero, Cramer's rule writes
$x_E=\alpha x_F+\beta x_G$, with $|\alpha|,|\beta|\le1$ by
maximality. Apply $a$ and the triangle inequality. If all determinants
vanish, the nonnegative vectors lie on a common ray. The secondary
criterion, which permits $(F,F)$, selects a vector of maximum total mass;
every other vector is a scalar multiple in $[0,1]$ of it. If that mass
is zero, all vectors vanish. This proves the bound in every case.
Deleting alternatives other than the selected pair preserves both its
maximality and its position in the fixed tie order.
\end{proof}

Let $R$ be the union of the unremoved coarse parents. Since
$4r\le q+1$, every row satisfies
\begin{equation}\label{r6u:retained-mass}
 \nu_i(R)\ge1-2r\sigma\ge\tfrac12.
\end{equation}
In particular, each selection has a nonempty available domain. Let
$\mathcal P_R$ be the finite family of refined grid cells retained in $R$.
Put $w_i=\mu_i/\mu_i(R)$ and
\[
 A_{ij}=\max_{E\in\mathcal P_R}|(w_i-w_j)(E)|.
\]
Every refined cell has $w_i$-value at most
$\eta=2/(s^8\ell)$ under the parameters below; more precisely its
value is at most $2/(q+1)$. In each selected pair choose the cell with
larger absolute $(w_i-w_j)$-value. Every final cell of $R$ was available
at each earlier selection. Lemma~\ref{r6u:representatives} therefore
provides $L$ witnesses per row pair and batch, in distinct removed parents,
with absolute signed value at least $A_{ij}/2$, and hence at least
$A_{ij}/3$.

The following stability argument is needed because $R$ and its
normalizers depend on the sampled grid. A fixed-measure interpolation
estimate cannot simply be conditioned on their realized values.

\begin{lemma}[Deletion stability]\label{r6u:stability}
For each realized coarse-plus-sample grid there is a measurable set
$\mathcal B$ of virtual coordinates such that
\begin{equation}\label{r6u:stability-mean}
 \E{\mu(\mathcal B)}\le\frac{4r+4k^2}{m+1}.
\end{equation}
If $z\notin\mathcal B$ lies in a retained refined cell $E=[a,b]$,
then, simultaneously for every pair $i,j$ and arbitrary positive
normalizers defining $w_i,w_j$,
\[
 |(w_i-w_j)([a,z])|\le2A_{ij},\qquad
 |(w_i-w_j)([z,b])|\le2A_{ij}.
\]
Here $A_{ij}$ is the maximum full-cell disagreement on the retained
original sampled grid. Original grid endpoints have zero interpolation
error and may be omitted from the exceptional set.
\end{lemma}
\begin{proof}
Insert $z$ into the sampled grid and run the same deterministic sequence
of selections. Protect the four endpoint slots of every selected pair.
After all removals, select one more maximum-determinant pair for each
ordered row pair in the retained domain and protect its four endpoints.
Define $\mathcal B$ by the event that the inserted coordinate is one of
these protected endpoints. All selections are finite measurable
comparisons, so this set and its dependence on the samples are measurable.

Suppose $z$ is unprotected and its coarse parent is retained on the
original grid. Deleting $z$ merges two adjacent cells within that parent.
At each selection, neither split cell was selected. Removing those
unchosen alternatives leaves the augmented-grid winner unchanged. If
adding the merged cell changed the winner, that winner would contain
the merged cell, and this coarse parent would be removed. At the first
possible divergence this contradicts its survival in the original run.
Induction therefore gives identical removed parents and identical earlier
witnesses on the two grids. The protected final pair also avoids $z$;
its two cells survive unchanged on the original grid. Apply
Lemma~\ref{r6u:representatives} to either split cell, with the arbitrary
normalizing coefficients. Both final witnesses have disagreement at most
$A_{ij}$ in the original retained grid, proving the two bounds.

For the expectation, adjoin an independent $(m+1)$st sample and regard
each of the $m+1$ samples in turn as the inserted point. Atomlessness
makes them distinct from one another and from fixed coarse endpoints
almost surely. At most $4r+4k^2$ of these coordinates can be protected.
Exchangeability and integration over the distinguished point give
\eqref{r6u:stability-mean}. No independence between the retained domain
and the original samples is used.
\end{proof}

Let $F_i$ be the cumulative $w_i|_R$-value and let $\widehat F_i$ be
its linear interpolant on the original grid. At a nonexceptional point,
its pairwise interpolation error is a convex combination of the signed
prefix and negative signed suffix discrepancies. Consequently
\begin{equation}\label{r6u:good-endpoint}
 |(F_i-F_j)(z)-(\widehat F_i-\widehat F_j)(z)|\le2A_{ij}.
\end{equation}
The statement is uniform over the normalization constants, so it applies
to the actual sample-dependent $R$. Cells outside $R$ have zero rows and
zero error. Logical cells crossing recorded physical joins are interpreted
using volume along the recorded domain, as in Section~\ref{r6u:setup}.

\subsection{Assignment slack and whole-cell payments}
\label{r6u:payments-section}

For each pair and batch, list its witnesses as $E_1,\ldots,E_L$ and
write $\gamma_a=(w_i-w_j)(E_a)$. Independently choose a uniform
$T\in\{0,\ldots,L\}$. Start with the negative witnesses, add positive
ones of index at most $T$, and remove negative ones of index at most $T$.
The resulting signed value is
\[
 \sum_{\gamma_a<0}\gamma_a+\sum_{a\le T}|\gamma_a|.
\]
Its successive increments are at least $A_{ij}/3$ when $A_{ij}>0$.
Union the pair pieces within each batch, then apply an independent uniform
permutation of batches to old labels to obtain seeds $Z_j$. All seeds are
disjoint and lie outside $R$. Each seed has every $w_i$-value at most
$\binom{k}{2}L\eta$; all seeds together have every $\nu_i$-value at
most $r\sigma<1/8$. Their labeled distribution is exchangeable even
when different batches have different distributions.

Set $V_{ij}=w_i(W_j\cup Z_j)$. Choose a maximum-sum assignment
$\pi$ for cake, a minimum-sum assignment for chores. Cake has edge
$j\to i$ of weight
$V_{i,\pi(j)}-V_{i,\pi(i)}+2\lambda A_{ij}$; chores has edge
$i\to j$ of weight
$V_{i,\pi(i)}-V_{i,\pi(j)}+2\lambda A_{ij}$.
Reject a positive cycle. We now prove that its conditional probability,
for every preparation, is at most
\begin{equation}\label{r6u:isolation}
 \frac{k^2(k-1)[12\lambda\ln(2k)+1]}{L+1}.
\end{equation}
For an unordered pair $\{i,h\}$ and column $j$, let $\Phi_{ij}$ be
the optimum after deleting row $i$ and column $j$, using the same
maximization or minimization convention. Define
\[
 d_{ih,j}=|V_{ij}+\Phi_{ij}-V_{hj}-\Phi_{hj}|.
\]
Condition on the batch permutation and every threshold except that
of pair $\{i,h\}$ in the batch assigned to column $j$. The
deleted-column scores do not depend on this threshold. The signed
comparison changes monotonically, with step at least $A_{ih}/3$.
For $A_{ih}>0$, any interval of length $4\lambda A_{ih}/u$
therefore contains at most $12\lambda/u+1$ possible scores.
Set
\[
 X_{ih,j}=\min\{1,2\lambda A_{ih}/d_{ih,j}\},\qquad
 Z_{ih,j}=(X_{ih,j}-1/(2k))_+,
\]
where a zero gap gives $X=1$; set both variables to zero when
$A_{ih}=0$. For $0<u\le1$, the conditional tail bound is
\[
 \Prb{X_{ih,j}\ge u}\le(12\lambda/u+1)/(L+1).
\]
Integrating this bound from $1/(2k)$ to one and averaging over the
conditioning gives
\begin{equation}\label{r6u:truncated-expectation}
 \E{Z_{ih,j}}\le
 \frac{12\lambda\ln(2k)+1-1/(2k)}{L+1}.
\end{equation}

If a positive cycle exists, choose a simple one. Let $\Delta\ge0$
be the loss in assignment value for cake, or increase in assignment
cost for chores, when each row takes the column specified by the
cycle. Optimality makes this quantity nonnegative. The ordinary edge
weights sum to $-\Delta$, so positivity implies
$\Delta<2\lambda\sum_{\rm cycle}A_{ih}$.
For an edge transferring column $j$ from its current recipient $h$
to $i$, the chosen assignment is optimal among those assigning $j$
to $h$, while the cyclic competitor is feasible among those assigning
$j$ to $i$. In the maximizing case this gives
\[
 \operatorname{OPT}-\Delta
 \le V_{ij}+\Phi_{ij}
 \le\operatorname{OPT}=V_{hj}+\Phi_{hj}.
\]
In the minimizing case all inequalities reverse after replacing
$\operatorname{OPT}-\Delta$ by $\operatorname{OPT}+\Delta$.
In either case $d_{ih,j}\le\Delta$.
If $\Delta>0$, then
\[
 \sum_{\rm cycle}X_{ih,j}
 \ge\sum_{\rm cycle}\min\{1,2\lambda A_{ih}/\Delta\}\ge1.
\]
If $\Delta=0$, some cycle edge has $A_{ih}>0$ and gap zero, so
the same conclusion follows. The cycle has at most $k$ edges;
its $Z$-sum is at least $1/2$. Its pair-column triples are distinct:
even a two-cycle uses different columns. The sum over all
$k\binom{k}{2}$ triples must therefore be at least $1/2$.
Markov and~\eqref{r6u:truncated-expectation} prove
\eqref{r6u:isolation}. No winning-assignment event was conditioned on.

If there is no positive cycle, add a dummy source with zero-weight
edges to all vertices and let $t_i$ be the maximum weight of a
simple path ending at $i$, including the zero path. Appending an
edge to a maximizing path and deleting any resulting nonpositive
cycles cannot decrease its weight. Thus these nonnegative potentials
give the ideal payment inequalities
\begin{equation}\label{r6u:ideal-payments}
 \begin{cases}
 V_{i,\pi(i)}+t_i-V_{i,\pi(j)}-t_j\ge2\lambda A_{ij},&\text{cake},\\
 V_{i,\pi(j)}+t_j-V_{i,\pi(i)}-t_i\ge2\lambda A_{ij},&\text{chores}.
 \end{cases}
\end{equation}
Initial $R$-normalized envy is at most $2\tau=1/(2s^2)$, by
\eqref{r6u:retained-mass}. Each seed has value at most $1/(4s^2)$.
For cake put $M_i=\max_jV_{ij}$. The identity assignment has total
shortfall at most $3k/(4s^2)$ from these row maxima, so a maximizing
assignment satisfies
\[
 \sum_i(M_i-V_{i,\pi(i)})\le3k/(4s^2).
\]
The ordinary part of an edge entering $i$ is bounded by its assigned
shortfall. For chores, put $m_i=\min_jV_{ij}$. A minimizing assignment
similarly satisfies
$\sum_i(V_{i,\pi(i)}-m_i)\le3k/(4s^2)$, and an ordinary edge
leaving $i$ is bounded by that row's regret. A simple path charges each
row once. Since $A_{ij}\le\eta$ and $2\lambda\eta\le1/(4s^2)$,
\begin{equation}\label{r6u:target-size}
 0\le t_i\le3k/(4s^2)+(k-1)2\lambda\eta<k/s^2,
 \qquad \sum_it_i<k^2/s^2<1/16.
\end{equation}
This argument treats credits for cake and taxes for chores separately.

\begin{lemma}[Whole-cell payments]\label{r6u:whole-payments}
There are disjoint unions $Y_j$ of retained grid cells such that
\[
 0\le\epsilon_j=t_j-w_j(Y_j)\le\min\{t_j,\eta\},
 \qquad |(w_i-w_j)(Y_j)|\le4kA_{ij}.
\]
They are computable from the recorded finite table without further queries.
The protocol may enumerate the finitely many cell-ownership tables and
select the first satisfying these inequalities. The proof below establishes
that such a table exists; this finite choice also makes the selection
measurable.
\end{lemma}
\begin{proof}
Write $a_{iE}=w_i(E)$ and start with the fractional allocation
$x_{Ej}=t_j$ and unused fraction $x_{E0}=1-\sum_jt_j$ in every cell.
Keep all cell-sum equations and the $k$ owner equations
$\sum_Ea_{jE}x_{Ej}=t_j$. Initially also keep every discrepancy equation
\[
 \sum_E(a_{iE}-a_{jE})x_{Ej}=0.
\]
Fix any coordinate that reaches zero or one. Float the remaining
coordinates in a kernel direction until a new coordinate reaches a
boundary, or until the remaining columns are independent.

Let $v$ be the number of fractional coordinates, $c$ the number of split
cells, and $d$ the number of discrepancy equations still kept. At a state
with independent columns,
\[
 v\le c+k+d,\qquad 2c\le v.
\]
Every fractional coordinate appears in at most $k$ discrepancy equations,
since its recipient is fixed. If $v>4k$, the displayed rank bound gives
$d\ge v/4$. Summing supports therefore shows that some discrepancy
equation involves at most $4k$ currently fractional coordinates. Drop
that equation and continue floating. Its total eventual change, including
all later rounding, is at most $4kA_{ij}$: each remaining coefficient
has magnitude at most $A_{ij}$, each coordinate changes by at most one,
and all other real-color coordinates remain fixed forever. When at most
$4k$ fractional coordinates remain, the same bound applies to each
undropped equation. This process is finite: every step fixes a coordinate
or drops an equation. All owner equations remain exact.

Round the remaining fractions using sorted slots~\cite{ST93}. For each
recipient, sort fractional cells by decreasing own value, place their
fractional amounts consecutively into unit slots, and require each full
slot to receive one contributing cell. The fractional slot incidences
satisfy Hall's condition, since each cell has total capacity at most one.
An integral matching therefore fills all full slots with distinct cells.
Consecutive-slot comparison bounds each recipient's load within one cell
of its fractional target. If it exceeds that target, delete matched cells
until it first falls below it; the resulting deficit is at most $\eta$.
Keep all previously integral allocations.

Crucially, the matching uses only positive fractional incidences, so no
zero real-color coordinate acquires a cell, and no fixed one-coordinate
changes. The robust bound for every dropped equation therefore continues
to hold after this rounding and trimming. Nonnegative loads also bound
each deficit by $t_j$. This proves all three assertions.
\end{proof}

\subsection{Fresh correction attempts and their physical boundaries}
\label{r6u:corrections}

Process recipients in a fixed order. Before correcting $j$, remove all
pending whole-cell payments and completed accepted payments from $R$,
leaving the actual domain $S$. A positive deficit is filled by a fresh
circular piece $C_j\subseteq S$ with exact own value $\epsilon_j$;
a zero deficit uses the empty piece. Put $P_j=Y_j\cup C_j$, record its
endpoints, and check
\[
 w_j(P_j)=t_j,\qquad |w_i(P_j)-t_j|\le\lambda A_{ij}\quad(i\in[k]).
\]
On failure restore the pre-attempt physical domain and make another fresh
attempt for this recipient, retaining and charging every query already
made. Stop at the first passing attempt, or reject the whole Repair trial
after $\ell$ unsuccessful attempts. A rejected attempt creates no physical
hole. The analytic exceptional set $\mathcal B$ is used only in the proof;
no membership oracle for it is needed.

Pending payments satisfy $w_i(Y_j)\le t_j+4kA_{ij}$; accepted ones
satisfy $w_i(P_j)\le t_j+\lambda A_{ij}$. Our parameters give
\begin{equation}\label{r6u:supply}
 \sum_jt_j<1/16,\qquad k\lambda\eta<1/16,\qquad16k\le\lambda.
\end{equation}
Thus $w_i(S)>7/8$ at every reached attempt, including one that fails.
The target $\epsilon_j\le t_j<1/16$ fits inside $S$. This is a physical
supply invariant; it does not rely on successful interpolation.

\begin{lemma}[Maximum-cell envelope]\label{r6u:envelope}
For recipient $j$ and retained cell $E\in\mathcal P_R$, let
\[
 H_j(E)=\max_{i:A_{ij}>0}\frac{|(w_i-w_j)(E)|}{A_{ij}},
 \qquad h_j|_E=H_j(E)/|E|,
\]
with empty maximum zero and with volume interpreted on the recorded
logical cell. For $N\le2m$ retained cells,
\begin{equation}\label{r6u:envelope-bound}
 0\le H_j(E)\le1,\qquad \int_Rh_j\le N,\qquad
 |\widehat\gamma_{ij}(C)|\le A_{ij}\int_Ch_j,
\end{equation}
where $\widehat\gamma_{ij}$ is the interpolated signed measure.
At every fresh attempt,
\begin{equation}\label{r6u:envelope-expectation}
 \E{\int_{C_j}h_j\mid\mathcal H}\le2\epsilon_jN\le2\eta N.
\end{equation}
\end{lemma}
\begin{proof}
Every cell ratio is at most one; integration bounds total envelope mass
by the cell count. The absolute interpolated pairwise density is bounded
by $A_{ij}h_j$, including $A_{ij}=0$, when that density vanishes.
The circular inclusion law~\eqref{r6u:circular} and $w_j(S)>7/8$
give the conditional expectation, even on cutter-null subintervals.
\end{proof}

Only grid boundaries, the new circular endpoints, and endpoints of
previous accepted corrections occur in the actual candidate's virtual
boundary. Whole-cell payments and removed coarse parents use grid
boundaries. There are at most $2k$ non-grid endpoints altogether. This
count includes inherited holes from every earlier accepted correction,
but excludes physically restored rejecting attempts. If all these points
avoid $\mathcal B$, telescoping~\eqref{r6u:good-endpoint} gives
\begin{equation}\label{r6u:interpolation-failure}
 |\gamma_{ij}(C_j)-\widehat\gamma_{ij}(C_j)|\le4kA_{ij}.
\end{equation}
Counting only the latest two endpoints would not prove this estimate.
If also $\int_{C_j}h_j<\lambda/4$, then
\begin{equation}\label{r6u:error-budget}
 |w_i(P_j)-t_j|
 \le4kA_{ij}+4kA_{ij}+(\lambda/4)A_{ij}\le\lambda A_{ij}.
\end{equation}
Thus the actual check passes. The use of a maximum over observers in the
envelope avoids an additional union bound over $i$.

We bound all retries without a factor $\ell$ in the endpoint exception.
Set $\delta_0=1/(32768k^2)$ and suppose
$\mu(\mathcal B)\le\delta_0$. Each fresh endpoint has conditional law
at most $(8/7)w_j|_R\le(16/7)\nu_j|_B$. The two endpoints therefore
hit $\mathcal B$ with conditional probability at most
\[
 a=(32k/7)\delta_0=1/(7168k).
\]
By Lemma~\ref{r6u:envelope} and Markov, the probability that an attempt
has no exceptional endpoint but fails its physical check is at most
\begin{equation}\label{r6u:estimated-failure}
 b=8\eta N/\lambda\le8\eta(2m)/\lambda=1/4.
\end{equation}
These assertions hold after any previous attempts with no exceptional
endpoint; their full query transcripts may affect later replies.

Before the first exceptional endpoint, continuation requires a good but
rejected attempt, whose conditional probability is at most $b$. The
probability of ever seeing a first exceptional endpoint in a recipient's
stopped sequence is consequently at most $a/(1-b)\le4a/3$. Formally,
the bound $B_*=4a/3$ is invariant under the recurrence
$a+bB_*\le B_*$. Exhausting $\ell$ attempts without an exceptional
endpoint has probability at most $b^\ell\le4^{-\ell}$ by iterated
conditioning. The first bad endpoint among all recipients, or exhaustion
at some recipient before that event, therefore has probability at most
\begin{equation}\label{r6u:retry-failure}
 k(4a/3+4^{-\ell})\le1/5376+k/(4s^2).
\end{equation}
All previous accepted endpoints are good on this event's complement,
which justifies the induction used in~\eqref{r6u:interpolation-failure}.
We do not condition on a winning assignment or a successful repair.

On acceptance, the ideal assignment inequalities and the checked payment
errors imply, for $X_i=W_{\pi(i)}\cup Z_{\pi(i)}\cup P_i$,
\begin{equation}\label{r6u:repair-margin}
 g_{ij}(X)/\mu_i(R)\ge\lambda A_{ij}
                 \ge|w_i(P_j)-t_j|.
\end{equation}
The seeds consume less than one eighth of the buffer in every active row.
The disjoint accepted payments consume at most
$\sum_jt_j+k\lambda\eta<1/8$ of $R$ in every normalized row,
and $R\subseteq B$. Hence all additions consume less than one quarter
of $B$. Unused parts of removed parents, as well as unused compensation
material, remain in the literal final residue.

\subsection{Parameters and the complete Repair bound}
\label{r6u:parameters}

For the original population $n$ fix
\begin{equation}\label{r6u:parameters-display}
\begin{gathered}
 s=2^{40}\max\{4,n\},\qquad \ell=\lceil\log_2(2s)\rceil,\\
 q=s^8\ell,\qquad m=s^9\ell/128,\qquad L=s^4\ell,\\
 \sigma=1/(q+1),\qquad \eta=2/(s^8\ell),\qquad
 \lambda=s,\qquad \tau=1/(4s^2),\\
 \epsilon_0=s^{-2},\qquad K=8/\tau,\qquad
 \theta=\tau/(768m)=1/(24s^{11}\ell).
\end{gathered}
\end{equation}
Here $q,m,L,\ell$ are integers; divisibility of $s$ by $128$ makes the
formula for $m$ exact. These parameters are kept when the active population
shrinks. We have $s\ge2^{42}$, $1\le\ell\le s$,
$\ln(2k)\le\ell$, and $\kappa=k/s\le2^{-40}$. In particular,
\begin{align*}
 kq+1&\le m,&4r&\le q+1,& r\sigma&\le k^3/s^4<1/8,\\
 L\eta&=2/s^4,&\binom{k}{2}L\eta&\le1/(4s^2),&
 2\lambda\eta&\le1/(4s^2),\\
 k^2/s^2&<1/16,&k\lambda\eta&<1/16,&16k&\le\lambda,\\
 2k\ell&\le m,&16k(k+2)&\le m,&32k^4\ell&\le m.
\end{align*}
These verify the deterministic supplies and query allowances above.

Markov applied to~\eqref{r6u:stability-mean} bounds the probability that
$\mu(\mathcal B)>\delta_0$ by
\[
 p_{\rm grid}\le\frac{4r+4k^2}{(m+1)\delta_0}
 \le\frac{2k^3L}{(m+1)\delta_0}\le2^{23}\kappa^5.
\]
The assignment bound~\eqref{r6u:isolation}, with $\lambda=s$ and
$L=s^4\ell$, is at most $13\kappa^3$. Together with
\eqref{r6u:retry-failure}, this gives the unconditional estimate
\begin{equation}\label{r6u:failure-budgets}
 p_{\rm reject}\le
 2^{23}\kappa^5+13\kappa^3+1/5376+k/(4s^2)<1/64.
\end{equation}
The final inequality already follows from $\kappa\le1/256$ and
$k/(4s^2)\le1/1024$. There is no random reserve-mass or reserve-count
check. The analytic bad-grid event, assignment failure, and stopped
correction events cover rejection for the actual finite trial.

The coarse quantiles and mixture samples use at most $2m$ Cuts.
Every correction attempt uses at most two; all $k\ell$ attempts
therefore use at most $m$ more. Recording each endpoint for all original
observers gives at most $6nm$ total queries and $3m$ Cuts per trial,
even when every correction batch exhausts its attempts.

\begin{proposition}[Repair interface]\label{r6u:repair-contract}
Fix the parameters~\eqref{r6u:parameters-display}, an active set of
$3\le k\le n$ agents, and the entire history before a fresh trial.
Let $W_1,\ldots,W_k,B$ be disjoint recorded finite unions, with
$\mu_i(B)>0$ for every active $i$ and
$g_{ij}(W)\ge-\tau\mu_i(B)$ for all active $i,j$.
One Repair trial either rejects or returns a permutation $\pi$,
disjoint additions $Y_i\subseteq B$, and the literal residue
$B'=B\setminus\bigcup_iY_i$, such that
\[
 X_i=W_{\pi(i)}\cup Y_i\quad\hbox{is EF among active agents},\qquad
 \mu_i\!\left(\bigcup_jY_j\right)\le\tfrac14\mu_i(B)
 \quad(i\text{ active}).
\]
In particular $3\mu_i(B)/4\le\mu_i(B')\le\mu_i(B)$.
Every old endowment stays whole, all pieces have recorded endpoints,
and the additions include both seeds and reserve payments.
The conditional acceptance probability exceeds $63/64$; every trial
uses at most $3m$ Cuts and $6nm$ queries.

A rejection commits no allocation: the input sets are retained,
although the query history grows. An accepted output is a proposed
partition which the caller may retain without committing it. These
conventions also apply to Snapshot and Shrink below.
\end{proposition}
\begin{proof}
Take $Y_i=Z_{\pi(i)}\cup P_i$ in the physical construction.
Equation~\eqref{r6u:repair-margin} gives active EF; the seed and reserve
spending estimate gives the displayed bound. The remainder is its
literal complement, so no null fragments are lost. The rejection
estimate~\eqref{r6u:failure-budgets} and the cost calculation above hold conditional on the
entire history at the trial's start, including rejected work.
They use no completion invariant. Rejection discards the candidate
allocation while retaining the already charged query transcript.
\end{proof}

\subsection{Snapshot and prescribed shrinking}
\label{r6u:primitives}

The geometric constructions below produce the two contracts used by
completion. We first prove the contrast-preservation fact underlying
Snapshot.

\begin{lemma}[Contrast preservation]\label{r6u:contrast}
Fix a valid repair input and its preceding history. Suppose that after
every preparation, but before the fresh seed thresholds and
their independent column permutation, one can select rows $i,h$ and
old labels $u,v$ with
\[
 (w_i-w_h)(W_u)-(w_i-w_h)(W_v)\ge\Lambda
\]
for a fixed $\Lambda>0$. With probability at least $31/64$ over the
whole trial, repair returns an EF allocation $X$, residue $T$, and
distinct $a,b$ with $g_{ab}(X)\ge(\Lambda/8)\mu_a(T)$.
\end{lemma}

\begin{proof}
Fix a preparation and its selected rows and labels.
Exchanging seed columns $u,v$ preserves their conditional distribution
and reverses their contribution to the contrast. It is nonnegative
with probability at least $1/2$, including ties. Let $S$ denote this preservation event and $G$ a repair return.
Preparation is deterministic on each sampled table, and its mass guards
always hold. Averaging the conditional symmetry and using the unconditional
Repair bound gives
\[
 \Prb{S\cap G}\ge\tfrac12-\Prb{G^c}>31/64.
\]
This inequality is over the whole trial; it does not assert likely
success conditional on each selected table.

On $G$, set $d_{ij}=g_{ij}(X)/\mu_i(R)$,
$\Delta=\max_{i,j}d_{ij}$, and $e_{ij}=w_i(P_j)-t_j$.
By~\eqref{r6u:repair-margin},
$|e_{ij}|\le d_{ij}$. The ideally paid matrix
$H_{ij}=V_{i,\pi(j)}+t_j$ has the row identities
\[
 H_{ij}=w_i(X_i)-(d_{ij}+e_{ij})\quad\text{for cake},\qquad
 H_{ij}=w_i(X_i)+(d_{ij}-e_{ij})\quad\text{for chores}.
\]
In either case a row has range at most $2\Delta$. Column payments
cancel in a two-row, two-column contrast, and column permutations
preserve it. Thus on $S\cap G$, $\Lambda\le4\Delta$.
Choose a maximizing pair $a,b$. The retained-mass bound~\eqref{r6u:retained-mass} and $T\subseteq B$
give
\[
 g_{ab}(X)=\mu_a(R)\Delta
 \ge(1/2)\mu_a(B)\,\Lambda/4
 \ge(\Lambda/8)\mu_a(T).
\]
Its strictly positive margin makes the pair distinct.
\end{proof}

\begin{proposition}[Snapshot]\label{r6u:snapshot}
Fix the parameters~\eqref{r6u:parameters-display} and a recorded domain
$U$ with $\mu_i(U)>0$ for each of $3\le k\le n$ active agents.
Conditional on the entire history before a fresh trial, Snapshot
either rejects, returns a complete EF allocation of $U$ among the
active agents, or returns a partition $(X_1,\ldots,X_k,T)$ of $U$
with $X$ EF among them, $\mu_i(T)>0$ for every active $i$, and
$g_{ab}(X)\ge\theta\mu_a(T)$ for distinct active $a,b$.
All pieces are finite unions with recorded endpoints. Rejection
commits no allocation, as in Proposition~\ref{r6u:repair-contract}.
The conditional acceptance probability exceeds $1/4$, and the
deterministic cost cap is $6m$ Cuts and $12nm$ total queries,
including all candidates and the possible inner Repair trial.
\end{proposition}

For Snapshot, normalize all active values on $U$ to one, fix a cutter
$c$ before sampling, and build the ordinary coarse-plus-sample table.
Let $D_j=\sum_E|\nu_j(E)-\nu_c(E)|$ and $D=\max_jD_j$ over
$M\le2m$ cells. Choose $i^*$ attaining the maximum with a fixed tie
rule. The comparison family is fixed before sampling even though its
maximizing row is selected afterward.

\paragraph{Zero and small disagreement.}
If $D=0$, a fresh circular cutter-equal partition is returned only
after actual complete EF checks. Cell agreement is not itself used
as a certificate of agreement between cells.

Suppose $0<D<\tau/16$ and put $t=(1-KD)/k$. Since $KD<1/2$,
$1/(2k)<t\le1/k$. The fixed signed estimate
$\widehat\sigma=\widehat\nu_{i^*}-\widehat\nu_c$ has total mass zero
and total variation $D$. Let $S^+$ be its positive cells and put
$u=\nu_c(S^+)$. If $u\ge t$, take a fresh circular piece of
cutter-value $t$ in $S^+$ as $W_1$. If $u<t$, take all of $S^+$
and fill the shortfall by a fresh circular piece of cutter-value
$t-u$ in its complement. With a second independent circular start,
partition $U\setminus W_1$ into $W_2,\ldots,W_k$ of cutter-value
$t$ and a buffer $B$ of cutter-value $KD$. The amounts sum to
$1-t$, so this is a complete physical partition. The two groups
need at most $k+2$ new endpoints. Record their values but retain
the original interpolants.

Conditional on any fixed table and preceding candidate history,
$\widehat\sigma(S^+)=D/2$ and
$\widehat\sigma(U\setminus S^+)=-D/2$. The circular inclusion law
therefore gives
\[
 \E{\widehat\sigma(W_1)}
 =\begin{cases}
    tD/(2u),&u\ge t,\\
    D(1-t)/(2(1-u)),&u<t,
   \end{cases}
 \quad\ge tD/2.
\]
The second case follows by adding the retained $D/2$ to
$(t-u)/(1-u)$ times $-D/2$. The fresh second start gives
\[
 \E{\widehat\sigma(W_2)\mid\text{first-step history}}
 =-\frac{t}{1-t}\widehat\sigma(W_1).
\]
Thus $\chi=\widehat\sigma(W_1)-\widehat\sigma(W_2)$ satisfies
$\E{\chi}\ge D/(4k)$, while $\chi\le D$ because its two pieces
are disjoint and the signed estimate has variation $D$.
Consequently
\[
 D/(4k)\le\E{\chi}
 \le D/(8k)+D\,\Prb{\chi\ge D/(8k)},
\]
so the estimated witness event has conditional probability at least
$1/(8k)$.

Generate all $16k$ alternatives on the same table, recording every
endpoint without changing the interpolants, and then select the first
passing the actual checks below. Conditional on each table and the
complete preceding candidate history, the estimated witness chance
is at least $p=1/(8k)$. Iterated conditioning bounds failure of all
candidates by $(1-p)^{16k}\le1/4$: Bernoulli's inequality gives
$(1-p)^{8k}\le1/(1+8kp)=1/2$, then square. No independence of
candidate outcomes under a history-dependent root policy is assumed.

The first circular domain has cutter value at least $t>1/(2k)$
when $u\ge t$, and at least $1-t\ge2/3$ when $u<t$; the second
has value $1-t$. Since $\nu_c\le k\mu$, every endpoint has
conditional marginal law at most $2k^2\mu$ before its group's shared
start, also for $k=3$. The complete list has at most
$16k(k+2)\le32k^2$ endpoints. Applying
Lemma~\ref{r6u:aggregate} with the resulting pathwise bound
$\Lambda_{\rm sum}\le64k^4$ and accuracy $\eps_0$ gives exception
probability
\begin{equation}\label{r6u:snapshot-error}
 \frac{128k^5(1+\eps_0)}{\eps_0(m+1)}
 \le\frac{32768k^5}{s^7\ell}\le\frac{32768}{s^2}<1/64.
\end{equation}
This event counts rejecting alternatives and also covers the smaller
zero branch. If $D=0$, its conclusion gives exact zero cutter
differences on every resulting piece, so the complete partition is
actually equal for every observer. In the positive case an
estimated-good alternative passes the actual checks
\begin{equation}\label{r6u:snapshot-checks}
\begin{gathered}
 \nu_c(W_j)=t,\quad\nu_c(B)=KD,\quad
 \nu_{i^*}(W_1)-\nu_{i^*}(W_2)\ge D/(32k),\\
 (K-2)D\le\nu_i(B)\le(K+2)D,\qquad
 \max_j\nu_i(W_j)-\min_j\nu_i(W_j)\le4D\le\tau\nu_i(B).
\end{gathered}
\end{equation}
Indeed, for row $i$ the estimated cutter difference has zero total
mass and variation $D_i\le D$, so it assigns any set absolute value
at most $D/2$. Its actual difference is at most
$(1/2+\eps_0)D$. The buffer thus lies within $2D$ of $KD$, and
two cutter-equal endowments differ by at most $4D$. The witness
loses at most $2\eps_0D$ and retains at least $D/(16k)>D/(32k)$
because $2\eps_0\le1/(16k)$. Finally
$\tau(K-2)D\ge4D$ since $K-2\ge4/\tau$.

The checked buffer is positive for every row and the range condition
implies~\eqref{r6u:repair-input} for both preference directions.
After every fresh repair preparation, the cutter still
values $W_1,W_2$ equally, and therefore the contrast is
\[
 (w_{i^*}-w_c)(W_1)-(w_{i^*}-w_c)(W_2)
 =\frac{\nu_{i^*}(W_1)-\nu_{i^*}(W_2)}{\nu_{i^*}(R)}
 \ge\frac1{48k(K+2)}.
\]
We used $\nu_{i^*}(R)\le\nu_{i^*}(B)\le(K+2)D$. Lemma~\ref{r6u:contrast} now applies to that
fresh repair after any checked candidate history.

\paragraph{Large disagreement.}
If $D\ge\tau/16$, the positive and negative recorded cutter
differences of row $i^*$ each sum to $D/2$. Choose cells $F,G$
with
\[
 \nu_{i^*}(F)-\nu_c(F)\ge D/(2M),\qquad
 \nu_c(G)-\nu_{i^*}(G)\ge D/(2M).
\]
They are distinct. Set $W_1=F$, $W_2=G$, and all other endowments
empty, with buffer $B=U\setminus(F\cup G)$. Its every value is at
least $1-2\sigma\ge1/2$, and the endowment range is at most
$\sigma\le\tau/2$. Thus the input is valid deterministically, without
new cuts.

After fresh repair preparation write $r_i=\nu_i(R)$
and $\gamma=\nu_{i^*}/r_{i^*}-\nu_c/r_c$. If $r_{i^*}\le r_c$,
\[
 \gamma(F)=\frac{\nu_{i^*}(F)-\nu_c(F)}{r_{i^*}}
       +\nu_c(F)\left(\frac1{r_{i^*}}-\frac1{r_c}\right)
 \ge\frac{D}{2Mr_{i^*}}\ge\frac{D}{3M}.
\]
If $r_c\le r_{i^*}$, the same argument gives
$-\gamma(G)\ge D/(3M)$. Both use
$r_i\le\nu_i(B)\le1$. Compare the chosen cell, with the
appropriate orientation, to an empty third endowment. Such an
endowment exists for every $k\ge3$, so Lemma~\ref{r6u:contrast}
applies here as well.

\begin{proof}[Completion of Proposition~\ref{r6u:snapshot}]
The small and large branches yield respective margins at least
\[
 \frac{\mu_a(T)}{384k(K+2)}
 \quad\text{and}\quad
 \frac{D}{24M}\mu_a(T).
\]
Both are at least $\theta\mu_a(T)$, since
\[
 \tau k(K+2)\le10k\le2m,\qquad
 D\ge\tau/16,\qquad M\le2m.
\]
The first inequality handles the small branch, and the last two
handle the large branch. The remaining geometric conditions follow
from~\eqref{r6u:parameters-display}:
\[
 K\ge6,\qquad \sigma\le\tau/2,\qquad 1-2\sigma\ge1/2.
\]

A complete return or a checked positive input is obtained with
probability at least $1-1/4-1/64=47/64$. The zero branch uses its
actual check and interpolation; the small branch uses amplification
and the same interpolation event; the large branch supplies a valid
input deterministically. After selecting a checked positive input,
use exactly one fresh repair. Its conditional contrast-success
probability is at least $31/64$ after the entire observed candidate
history. Therefore Snapshot returns with probability at least
$(47/64)(31/64)>1/4$. Return only after checking actual EF and
completeness or the stated margin. A rejection commits none of the
alternative partitions. Each noncomplete return leaves at least
three quarters of its positive checked buffer, so its residue is
positive for every active observer.

The table costs at most $2m$ cuts, all alternatives at most
$16k(k+2)\le m$, and its one inner repair at most $3m$. Recording
every endpoint gives the complete cap of $6m$ Cuts and $12nm$
queries, including all rejecting work.
\end{proof}

\begin{lemma}[One-sided individual-query refinement]\label{r6u:refinement}
Let $k$ measures be positive on a recorded domain $U$, normalized
to $\nu_i(U)=1$. For $0\le f\le1/k$ and $0<\delta<1$, one can
construct disjoint pieces $W_1,\ldots,W_k$ such that
$f-\delta\le\nu_i(W_j)\le f$ for all $i,j$, retaining their
complement as a dummy. The number of Cuts is at most $k^3J$, and
the total number of queries at most $(n+1)k^3J$, where
\[
 J=\max\{0,\lceil\log_2(k^2/\delta)\rceil\}.
\]
\end{lemma}

\begin{proof}
On each cell of the recorded ordered domain, initialize fraction $f$
for each real color and fraction $1-kf$ for the dummy. Preserve the
cell-sum equations and the $k^2$ real-color moments
$\sum_E\nu_i(E)x_{Ej}=f$. Floating with fixed boundary coordinates
works exactly as in the reserve plan: a nonzero homogeneous direction
allows a finite move to the next boundary, and each move fixes a
new coordinate. At a terminal floating state, $a$ fractional cells
give at least $2a$ fractional columns and at most $a+k^2$
independent equations. Hence at most $k^2$ cells remain fractional.
Integral assignments stay fixed permanently.

Refine each fractional cell at every observer's own median whenever
that observer has positive mass there; skip zero-mass rows. Each
median is a strict interior value target on a recorded ordered union
and uses one ordinary Cut. Its union representation is handled by
cached component sums. Every child lies on one side of every row's
median and has at most half that row's parent value, regardless of
plateaus and the chosen legal roots. A zero-mass row has the same
property without a cut. Record each new endpoint for all original
observers and give every child its parent's fractional vector, which
preserves all moments. Float again.

Repeat for $J$ refinement rounds. Any cell still fractional at the
end has passed through all $J$ rounds, because integral cells never
reenter. Each original cell has every normalized value at most one,
so every surviving fractional cell has value at most $2^{-J}$.
There are at most $k^2$ of them; their union has each row's value
at most $k^2 2^{-J}\le\delta$. Give all remaining fractional cells
to the dummy. Each real color loses a nonnegative amount at most
$\delta$ from its exact fractional moment $f$. This proves both
one-sided inequalities and retains every physical fragment.

Each round refines at most $k^2$ cells by at most $k$ individual
Cuts. Recordings cost at most $n$ evaluations per cut, proving the
stated bounds. All floating operations are finite exact linear algebra.
The depth can be computed by integer doubling and comparisons, without
a logarithm oracle.
\end{proof}

\begin{proposition}[Prescribed Shrink interface]\label{r6u:shrink-contract}
Fix the parameters~\eqref{r6u:parameters-display}, $3\le k\le n$
active agents positive on a recorded domain $U$, and a factor
\[
 1\le G\le(2/\theta)(8/\tau)^{k-1}
\]
chosen before the fresh trial. One Shrink trial either rejects or
returns a partition $(H_1,\ldots,H_k,U')$ of $U$ into recorded
finite unions, where $H$ is EF among the active agents and
\[
 \frac3{8G}\mu_i(U)\le\mu_i(U')\le\frac1G\mu_i(U)
 \qquad(i\text{ active}).
\]
Rejection commits no allocation. Conditional on the entire preceding
history, acceptance has probability greater than $63/64$.
Every trial, including its carving and inner Repair, costs at most
$4m$ Cuts and $8nm$ queries.
\end{proposition}
\begin{proof}
Normalize $\nu_i=\mu_i/\mu_i(U)$. For the prescribed $G$, apply
Lemma~\ref{r6u:refinement}
with $f=(1-1/(2G))/k$ and $\delta=\tau/(4G)$. The physical buffer
$B=U\setminus\bigcup_jW_j$ then satisfies
\[
 \frac1{2G}\le\nu_i(B)
 \le\frac1{2G}+\frac{k\tau}{4G}\le\frac1G,
\]
using $k\tau\le2$. Each endowment range is at most
$\delta\le(\tau/2)\nu_i(B)$, so repair applies with positive
buffer normalizations in both preference directions. A shrinking
trial consists of this deterministic carving and one fresh trial of
Proposition~\ref{r6u:repair-contract}; its failure probability is
less than $1/64$. On return all
endowments are allocated and at most one quarter of $B$ was consumed,
giving
\begin{equation}\label{r6u:shrink}
 \frac3{8G}\mu_i(U)\le\mu_i(U')\le\frac1G\mu_i(U).
\end{equation}
Rejected trials also pay for carving. It suffices to use depth
\[
 J_0=\max\{0,\lceil\log_2(4k^2G/\tau)\rceil\}.
\]
Only factors up to
\[
 G_{\max}=(2/\theta)(8/\tau)^{k-1}
\]
are used. Since $2/\theta\le s^{32}$, $8/\tau\le s^{12}$, and
$4k^2/\tau\le s^{14}$, for the active populations $k\ge3$ used by the caller,
\begin{equation}\label{r6u:shrink-cost}
 J_0\le(12k+34)\ell\le32k\ell,\qquad k^3(32k\ell)\le m.
\end{equation}
Use the fixed integer depth $32k\ell$, which is at least $J_0$.
Thus every such shrinking trial costs at most $4m$ Cuts and $8nm$
queries. This is a logarithmic-precision carve, not an uncounted
sequence of repairs.
\end{proof}

\subsection{Global completion, zero residues, and the total bound}
\label{r6u:completion}

A committed state consists of recorded finite unions
$(W_1,\ldots,W_n,U)$ partitioning $[0,1]$, with $W$ globally EF,
a nonempty active set $A$, and frozen set $F=[n]\setminus A$.
Every old active bundle remains a movable whole label; frozen bundles
stay fixed. The guards are
\begin{equation}\label{r6u:guards}
 g_{fi}(W)\ge\mu_f(U)\quad\text{for cake},\qquad
 g_{if}(W)\ge\mu_i(U)\quad\text{for chores}\quad(f\in F,i\in A).
\end{equation}
Initially $U=[0,1]$, all bundles are empty, and everyone is active.
Over all original observers define
\[
 \mathcal D_i(W,U)=\{j\ne i:g_{ij}(W)\ge\mu_i(U)\},\qquad
 \Phi(W,U)=\sum_i|\mathcal D_i(W,U)|\le n(n-1).
\]

\begin{proposition}[Composition and completion]\label{r6u:completion-contract}
For $n\ge3$ with parameters~\eqref{r6u:parameters-display}, start
from the initial state above. For either preference direction, the
Repair, Snapshot, and Shrink
interfaces of Propositions~\ref{r6u:repair-contract},
\ref{r6u:snapshot}, and~\ref{r6u:shrink-contract} yield a protocol
that preserves the committed-state invariant and returns a complete
EF allocation almost surely. It uses at most $n(n-1)$ nontrivial
iterations. Conditional on the history starting any such iteration,
its expected number of full trials is less than eight, and
\[
 \E{Q}\le100n^3m,\qquad
 \E{C}\le49n^2m.
\]
Here a full trial is one Snapshot attempt, one Shrink attempt, or
one standalone Repair attempt; the first two include their inner
Repair, and all rejected work is counted.
\end{proposition}
\begin{proof}
First, an EF partial allocation $L$ of $U$ among the active agents
can be added under its original owners. Active margins then add.
For cake, active own values do not decrease, while a frozen
observer's guarded margin toward recipient $i$ loses at most
$\mu_f(L_i)\le\mu_f(U)-\mu_f(U\setminus\bigcup_jL_j)$.
For chores, frozen observers see only nonnegative additions to active
bundles, while an active observer's guarded margin toward a frozen
agent loses at most its own added cost, with the same residue bound.
Thus global EF and the guards persist with the literal new residue.
This argument also applies to a complete additive return.

The active nondomination graph has edge $i\to j$ for distinct active
agents exactly when $j\notin\mathcal D_i$. If it has a proper sink
strongly connected component $S$, freeze $S$ for cake and its active
complement for chores. The absent outgoing nondomination edges are
precisely the newly required guards. No material moves, the active
population decreases, and $\Phi$ is unchanged. This extends the
domination framework of~\cite{Psomas26} using total old bundles.

Handle active zero residues at committed EF states before normalization.
If $\mu_i(U)=0$ for a chore agent, giving all of $U$ to that agent
is itself an EF allocation of $U$, so the additive argument completes.
A zero-value cake agent can freeze with its old bundle when at least
two agents remain:
its new guard is $g_{ij}(W)\ge0=\mu_i(U)$.
Apply the one-agent base before freezing the last recipient. If all
remaining active measures vanish, give the residue to one of them.
The additive argument protects frozen observers even when they value
that residue.
One active agent takes everything left. Two divide the ordered residue
into cutter-equal halves; the other chooses a larger half for cake or
a smaller half for chores. Add these pieces under their original
owners. This is complete EF and needs at most one new Cut and its
records.

Apply these reductions until completion or until a nontrivial iteration
starts at a saved EF anchor $(W,U)$ with
$k\ge3$, positive active residue values, and a strongly connected
active graph. Obtain an accepted Snapshot $X$ on $U$ by fresh retries
of Proposition~\ref{r6u:snapshot}, retaining the same domain after
each rejection.
A complete return is added to $W$ and finishes. Otherwise let $T$
be its positive residue and $d_{ij}=g_{ij}(X)$, with certificate
$d_{aj}\ge\theta\mu_a(T)$.
The saved anchor is not changed by this partial return or by the
following Shrink returns; they are retained as temporary disjoint
partitions until the final Repair has passed.

At most two successful prescribed shrinking requests produce a
snapshot permutation $\alpha$, final residue values $\rho_i>0$,
and an edge $(b,h)$ of the saved graph such that
\begin{equation}\label{r6u:transfer}
 d_{i,\alpha(i)}\le\tau\rho_i,\qquad
 d_{b,\alpha(h)}\ge2\rho_b.
\end{equation}
To see this, in certificate column $j$ consider
$x_i=d_{ij}/\mu_i(T)\ge0$ and the $k$ factors
$G_e=(2/\theta)(8/\tau)^e$, $0\le e<k$.
A row forbids only a factor with $3\tau/8<G_ex_i<2$. This interval
has multiplicative width smaller than $8/\tau$, so each row forbids
at most one factor, and the diagonal row forbids none. Choose a safe
factor and use Proposition~\ref{r6u:shrink-contract} on $T$, retrying
the same factor and domain after a rejection. Write $\rho_i$ for the
accepted residue values at this stage. The two alternatives obey
\[
 \begin{aligned}
 Gx_i\le3\tau/8&\ \Longrightarrow\quad
 d_{ij}\le\frac{3\tau}{8G}\mu_i(T)\le\tau\rho_i,\\
 Gx_i\ge2&\ \Longrightarrow\quad
 d_{ij}\ge\frac2G\mu_i(T)\ge2\rho_i.
 \end{aligned}
\]
The certificate row is large because $Gx_a\ge2$, and the diagonal
row is small. On a saved-graph path from $a$ to $j$, choose an edge
$i\to h$ crossing
from a large to a small row. If $h=j$, identity suffices. Otherwise
take a simple path $j=u_0\to\cdots\to u_t=h$.
If each path edge has snapshot loss
$d_{u_r,u_{r+1}}\le\tau\rho_{u_r}$, set
$\alpha(u_r)=u_{r+1}$ for $r<t$, $\alpha(h)=j$, and fix
the other labels. The closing own loss is small because $h$ is a
small row in column $j$; hence all own losses are small, and
$(b,h)=(i,h)$ is the desired large comparison. If
some path edge $p\to q$ has loss greater than $\tau\rho_p$,
instead apply Shrink once more to the current residue with factor
$2/\tau$, which is at most $G_{\max}$, and use identity.
Its new residue values $\rho'_i$ satisfy
$0<\rho'_i\le\tau\rho_i/2$, so
$d_{pq}>\tau\rho_p\ge2\rho'_p$; take $(b,h)=(p,q)$ and
rename $\rho'$ as $\rho$. No shrink follows a nonidentity
transfer, so previously small losses are not inadvertently enlarged.

Let $H_i$ be the sum of the EF shrinking allocations under their old
owners and let $T_0$ be their final residue. Form
\[
 D_i=W_i\cup H_i\cup X_{\alpha(i)},\qquad\rho_i=\mu_i(T_0).
\]
The pieces $H$ are EF among active agents, because their disjoint
Shrink margins add. For either preference direction the exact identity is
\[
 g_{ir}(D)=g_{ir}(W)+g_{ir}(H)
                 +d_{i,\alpha(r)}-d_{i,\alpha(i)}.
\]
Consequently $g_{ir}(D)\ge-\tau\rho_i$ and
$g_{bh}(D)\ge(2-\tau)\rho_b$. These disjoint whole endowments and
the positive buffer $T_0$ meet every precondition of
Proposition~\ref{r6u:repair-contract}; no global EF assertion about
$D$ is needed. Retry that same Repair input until it returns
$Z_i=D_{\beta(i)}\cup Y_i$. Extend $f=\beta^{-1}$ by identity on
frozen labels and put $Z_i=W_i$ there. Here $Y_i$ includes all final
Repair additions. The exact physical identity is
\begin{equation}\label{r6u:whole-labels}
 Z_{f(j)}=W_j\cup B_j,\quad
 B_j=H_j\cup X_{\alpha(j)}\cup Y_{f(j)}\ (j\in A),\quad
 B_j=\varnothing\ (j\in F).
\end{equation}
These additions are disjoint and partition $U\setminus U'$.
Thus, for every original observer, including frozen ones,
\begin{equation}\label{r6u:physical-balance}
 \sum_j\mu_i(B_j)+\mu_i(U')=\mu_i(U).
\end{equation}

Active-active EF follows from Repair. Before claiming global EF,
compare each active $i$ only with the active recipient $f(i)$.
For a frozen $i$, $f(i)=i$ and $B_i=\varnothing$. We obtain for
every original observer
\begin{equation}\label{r6u:own-transport}
 \mu_i(Z_i)\ge\mu_i(W_i)+\mu_i(B_i)\quad\text{for cake},\qquad
 \mu_i(Z_i)\le\mu_i(W_i)+\mu_i(B_i)\quad\text{for chores}.
\end{equation}
For chores this allows own cost to increase by at most the old
label's additions.
Frozen-frozen comparisons are unchanged. For cake, the own-value
bound preserves active comparisons to frozen bundles.
A frozen observer comparing
to $f(j)$ has margin at least
$\mu_f(U)-\mu_f(B_j)\ge\mu_f(U')$, preserving its guard.
For chores, frozen observers see only nonnegative additions to active
old bundles, so their comparisons persist. By the own-cost bound,
an active observer's margin toward
any frozen agent is at least
$\mu_i(U)-\mu_i(B_i)\ge\mu_i(U')$.
Both residue inequalities use~\eqref{r6u:physical-balance} and
nonnegativity. This proves global EF and both guards before committing
$(Z,U')$ and counting targets.

For any original observer and old proper target $j$, these same own
inequalities and~\eqref{r6u:whole-labels} give
\[
 g_{i,f(j)}(Z)\ge
 \begin{cases}
 \mu_i(U)-\mu_i(B_j)\ge\mu_i(U'),&\text{cake},\\
 \mu_i(U)-\mu_i(B_i)\ge\mu_i(U'),&\text{chores}.
 \end{cases}
\]
If the new row residue is positive, the image cannot be its observer,
so injectivity preserves all proper-target counts. If it is zero,
global EF saturates that count at $n-1$ instead; an old target may
then map to its observer and must not be counted as a proper image.
For the witness, compare $Z_b$ with the active recipient $f(b)$
of $D_b$. For cake this gives $\mu_b(Z_b)\ge\mu_b(D_b)$;
for chores it gives
$\mu_b(Z_b)\le\mu_b(D_b)+\mu_b(Y_{f(b)})$.
Since $Z_{f(h)}=D_h\cup Y_{f(h)}$, nonnegativity and the final
Repair bound $\mu_b(\bigcup_iY_i)\le\rho_b/4$ give in both cases
\[
 g_{b,f(h)}(Z)\ge(7/4-\tau)\rho_b>\rho_b\ge\mu_b(U').
\]
Repair leaves $\mu_b(U')\ge3\rho_b/4>0$. The target is proper
and new, because $h$ was outside the anchor's target set and $f$ is
injective. Thus the global potential increases by at least one.
This comparison is with the saved anchor, not with a potentially
larger temporary potential after the separate Snapshot/Shrink pieces
were added under their old owners.

All normalizations are positive when used. Noncomplete Snapshot
returns a positive checked buffer after quarter spending, Shrink
has~\eqref{r6u:shrink}, and final repair leaves three quarters of
its input buffer. A complete return is committed additively and
saturates targets. The protocol performs zero-residue reductions only
at committed EF anchors. The lower bounds above exclude an active
zero residue inside the pending iteration; in particular, no zero
exit is applied to the uncorrected transferred endowments.

Each nontrivial iteration begins with a proper nondomination edge and
thus an unsaturated potential. Each nonterminal iteration increases
the same integer potential, while a terminal iteration completes from
an unsaturated state. Hence there are at most $n(n-1)$ iterations
over all population changes. Each request fixes its input before
retrying and uses the uniform conditional acceptance bound of its
interface after the entire preceding history: greater than $1/4$ for
Snapshot and $63/64$ for Shrink or Repair. For a bound $p$, iterated
conditioning gives $\Prb{N>r\mid\mathcal H}\le(1-p)^r$ for the
number $N$ of attempts, and hence $\E{N\mid\mathcal H}\le1/p$.
Thus every request returns almost surely, without independence
between requests or attempts. One iteration requests one Snapshot,
at most two shrinks,
and one final repair, so its expected trial count conditional on its
starting history is at most
\[
 4+3(64/63)=148/21<8.
\]
Every full trial has deterministic cap $6m$ Cuts and $12nm$ queries,
including all rejected work and any inner repair. Therefore
\begin{equation}\label{r6u:global-cost}
 \E{Q}\le100n^3m,\qquad
 \E{C}\le49n^2m.
\end{equation}
More explicitly, the full ordinary-query implementation satisfies
\begin{align*}
 \E{Q}&\le96n^2(n-1)m+(n+1)\bigl(n(n-1)+1\bigr)+2n,\\
 \E{C}&\le48n(n-1)m+n(n-1)+1.
\end{align*}
The additional terms cover initialization and normalization at every
reached committed boundary, including its possible final two-agent split.
For $m\ge1$ these quantities are at most the bounds in
\eqref{r6u:global-cost}. Summing conditional iteration bounds over at most
$n(n-1)$ reached iterations justifies the expectation even when their
inputs depend on all earlier accepted and rejected trials.

All internal computation in a finite trial terminates: floating fixes a new coordinate or drops an equation at each step,
refinement depth is explicitly bounded,
and the remaining operations are finite linear algebra and graph or
matching algorithms. Every accepted partition has finitely many recorded endpoints. Old bundle
values are sums of cached components; rejected queries remain in the
charged transcript even when their tentative pieces are restored. Almost
sure termination therefore produces a complete allocation into finite
unions, including all null fragments.
\end{proof}

For $n\ge1$, $s\le2^{42}n$ and $\ell\le45\log_2(2n)$. Thus
\[
 m\le c_0n^9\log_2(2n),\qquad c_0=(2^{42})^9\,45/128.
\]
The full bounds are consequently
$\E{Q}\le100c_0n^{12}\log_2(2n)$ and
$\E{C}\le49c_0n^{11}\log_2(2n)$.
Proposition~\ref{r6u:completion-contract} proves
Theorem~\ref{r6u:general}; the direct bases cover $n=1,2$.

\section{A lower bound linear in the success probability}
\label{r6lb:appendix}

Fix $n\ge2$ and $p\in(0,1]$. Consider a jointly measurable ordinary
Eval/Cut protocol whose probability of returning a complete exactly
envy-free allocation is at least $p$ on every normalized finite-step
profile with densities in $[1/2,3/2]$. Returned shares may have arbitrarily
many components, provided each return uses finitely many intervals.
The protocol's random tape is drawn independently of the input from an
arbitrary probability space. In this appendix, let $Q$ count only answered
calls on the uniform profile. The resulting lower bound also applies to
the paper's total query count, which is at least $Q$.

\begin{theorem}\label{r8lb:quantitative}
For cake and for chores,
\[
 \E{Q}\ge
 \max\left\{0,\frac{n(np-1)}2,\frac{n^2p}4-1\right\}.
\]
In particular, for every fixed $p>0$ the bound is $\Omega_p(n^2)$,
with a coefficient linear in $p$.
\end{theorem}

The expectation is unconditional. No termination assumption is made
on unsuccessful tapes. For $p=1$, the first term gives $n(n-1)/2$.
The classical deterministic cake lower bound of
Procaccia~\cite{Procaccia09} has the same quadratic order; the theorem
extends the guarantee to randomized failure, chores, and unrestricted
finite fragmentation.

\subsection{Replay and the uniform baseline}
We stop at the first illegal request. This preserves every correct
return and cannot increase the cost. Each answered Eval contributes
its queried interval to a determining list. An answered Cut starting
at $a$ and returning $z$ contributes $[a,z]$. Strict positivity makes
the root unique, so preserving that integral preserves the answer
and legality. A reported unavailable target contributes its bracket;
such a flag is charged as an answer. If an illegal unavailable-target
request is instead uncounted, retain its bracket as one additional
determining interval. Valuation-independent syntax errors need no
such interval.

For a cutoff $T$, write $S_T$ for a correct return by the cutoff,
$s_T=\Prb{S_T}$, and $J_T$ for the indicator of an uncounted illegal
terminal request before the cutoff. If $e_i$ is row $i$'s number of
determining intervals, then
\[
 \sum_i e_i\le Q+J_T,\qquad \E{J_T}\le1-s_T.
\]
Equal values on this list preserve the entire capped execution on
the same tape, including its terminal request and returned output.
This follows by induction on the answered history. If $\E{Q}=\infty$,
there is nothing to prove. Otherwise the probability $e_T$ of still
having a pending request at the cutoff tends to zero. Terminal
query-free divergence is already a retained failure.

\subsection{A narrow bump inside a wider moat}
Identify the endpoints only to define arcs on the unit circle;
queries and allocations remain ordinary subsets of $[0,1]$.
Fix
\[
  \frac1n<w<1,\qquad 0<h<w.
\]
Independently of the tape, choose a uniform center $X$. Let $B_h(X)$
and $W_w(X)$ be the concentric circular arcs of lengths $h$ and $w$,
and put
\[
 d_X=\frac12\mathbf1_{B_h(X)}-\frac{h}{2w}\mathbf1_{W_w(X)}.
\]
The two terms have equal integral. Both $1+d_X$ and $1-d_X$ are
normalized finite-step densities in $[1/2,3/2]$. The plus sign is used
for cake and the minus sign for chores.

In the first experiment choose one independent uniform row $I$ and
replace its density by this perturbed density. In the second choose
a uniform row and its cyclic successor and perturb both identically. Every fixed
profile in either experiment belongs to the promised input class.

An interval whose endpoints avoid $W_w(X)$ contains either all or
none of this arc, up to endpoints. Its deviation integral is therefore
zero. A fixed endpoint belongs to $W_w(X)$ with probability $w$.
Let $D$ be the event that a determining interval in a perturbed row
has an endpoint in the moat. For every fixed baseline tape,
\[
 \Prb{D\mid\omega}\le\frac{2w}{n}(Q(\omega)+J_T(\omega))
 \quad\text{in the one-row experiment},
\]
and the coefficient is $4w/n$ in the two-row experiment. This is a
union bound over the complete baseline transcript; it does not
condition an adaptive query on earlier nondetection. Outside $D$,
the all-outcome replay argument applies exactly.

\subsection{Decoding a successful uniform allocation}
Every successful uniform allocation has all bundle lengths equal to
$1/n$. Suppose the narrow bump avoids the finitely many output
boundaries, and let $a$ be its owner. For a perturbed row $i\ne a$,
write $M_j=|A_j\cap W_w(X)|$. In the cake orientation,
\[
 v_i(A_a)-v_i(A_i)
 =\frac h2-\frac{h}{2w}M_a+\frac{h}{2w}M_i
 \ge\frac h2\left(1-\frac1{nw}\right)>0.
\]
The reversed density gives the same strictly positive gap
$v_i(A_i)-v_i(A_a)$ for chores. Thus every perturbed row must own
$X$ if the old allocation remains correct. A single independent
uniform row owns $X$ with probability $1/n$; two distinct rows cannot
both own it.

The strict inequality $w>1/n$ avoids any exceptional case in which
a bundle coincides with the moat. We take $w\downarrow1/n$ only
after completing the argument.

\subsection{Retaining failures and choosing one width}
For every measurable set $A$,
\[
 \left|\int_A d_X\right|\le h/2.
\]
Indeed, the positive and negative terms separately have total mass
$h/2$. Every envy comparison therefore changes by at most $h$,
including in the two-row experiment: each comparison uses only
one observer's valuation.

On a complete but incorrect uniform return, let $\Delta>0$ be
the difference between the longest and shortest bundle lengths.
A shortest owner's comparison with a longest bundle witnesses this
gap for cake; a longest owner's comparison with a shortest bundle
does so for chores. If $\Delta>h$, the replayed output remains
incorrect. Malformed returns and nonreturns remain failures by replay.

The endpoint list on each returned tape is finite, although its
expected length need not be finite. If it has length $N$, the
probability that the narrow bump meets it is at most
$\min\{1,hN\}$. Dominated convergence gives a total endpoint loss
tending to zero as $h\downarrow0$. The probability of an incorrect
uniform return with $0<\Delta\le h$ also tends to zero. Consequently,
for each fixed $w$ and any $\eta>0$, one deterministic $h\in(0,w)$
loses probability at most $\eta$ in both tests. The width is fixed
before the center, perturbed rows, and tape are drawn.

\subsection{Expected query accounting}
The fixed-input success guarantee integrates to success at least
$p$ under either hard distribution. At any fixed cutoff, the preceding
argument, with geometric loss at most $\eta$, gives
\[
 p\le\frac{s_T}{n}+\frac{2w}{n}\bigl(\E{Q}+1-s_T\bigr)+\eta+e_T
 \quad\text{and}\quad
 p\le\frac{4w}{n}\bigl(\E{Q}+1-s_T\bigr)+\eta+e_T.
\]
The losses can be removed without assuming that $s_T$ is fixed across
cutoffs. Indeed, for $0\le s\le1$, $n\ge2$, and $w\ge1/n$,
\[
 \frac{s}{n}+\frac{2w}{n}(q+1-s)
 \le \frac1n+\frac{2w}{n}q+\frac{2(w-1/n)}n.
\]
After multiplication by $n^2$, the difference is
$(n-2)(1-s)+2(nw-1)s\ge0$. First choose a cutoff with $e_T$ arbitrarily
small, then choose one sufficiently narrow bump at that cutoff.
Letting both losses vanish yields
\[
 p\le\frac1n+\frac{2w}{n}\E{Q}+\frac{2(w-1/n)}n,
 \qquad
 p\le\frac{4w}{n}\bigl(\E{Q}+1\bigr).
\]
Finally let $w\downarrow1/n$. The two inequalities become
$\E{Q}\ge n(np-1)/2$ and $\E{Q}\ge n^2p/4-1$.
Taking their maximum with zero proves the theorem.

All events used here are measurable. Circular arc membership is a
Borel distance test; actual density values and cumulative integrals
are jointly measurable in their center and endpoints. Strictly
positive cumulative functions give measurable unique Cut replies.
Finite transcript and finite-output spaces are countable disjoint
unions of their fixed-length Borel parameter spaces. Endpoint-loss,
output-gap, and comparison events are therefore measurable.
Tonelli justifies integration first over the tape or first over the
fixed input. No regular conditional probability on the tape space
is required.

\section{Bounded-error lower bounds for mixed resources}
\label{r6m:appendix}

We prove Theorem~\ref{thm:mixed-mc} for an oracle that always returns the
leftmost legal root. Reflection gives the always-rightmost case. The hard
input has a final interval that every acceptable allocation must split.
The construction moves the beginning of that interval past the coordinates
used by many executions, while preserving all their query answers.

\begin{lemma}[Relative approximation on positive-net profiles]
\label{r6m:relative-equivalence}
For $0<\alpha\le1$, the signed relative comparison
\[
 y-x\le(1-\alpha)\max\{|x|,|y|\}
\]
is equivalent to $x\ge\alpha y$ if $y\ge0$, and to $x\ge y/\alpha$
if $y<0$. It always holds when $x\ge y$. For a complete allocation with
$v_i(I)>0$ for all $i$, condition~\eqref{eq:signed-relative-ef} is equivalent
to $v_i(A_i)\ge\alpha v_i(A_j)$ for all $i,j$.
\end{lemma}
\begin{proof}
If $x\ge y$, the left side is nonpositive and the right side is
nonnegative. The corresponding threshold is at most $y$, so it is also
satisfied. Suppose $x<y$. If $0\le x<y$, the maximum is $y$, and the
comparison is exactly $x\ge\alpha y$. If $x<y\le0$, the maximum is $-x$,
and the comparison is exactly $x\ge y/\alpha$. If $x<0<y$, neither
condition can hold: $y-x$ exceeds $\max\{|x|,|y|\}$, whereas
$\alpha y>0>x$. The zero endpoints follow from the same inequalities.

For the allocation statement, positive total value implies that some
bundle has positive value to each observer. Relative EF then forces
that observer's own value to be positive. Comparisons with nonnegative
bundles are the usual cake inequalities, and comparisons with negative
bundles hold automatically. Conversely, the cake inequalities force a
positive own value and imply every relative comparison. Summing them
also gives $v_i(A_i)\ge\alpha v_i(I)/(n-1+\alpha)$.
\end{proof}

\subsection{Model and finite-horizon statement}
\label{r6m:model}

Fix $n\ge3$ and $0\le\eps<\alpha\le1$. Each valuation $v_i$ is a finite
atomless signed Borel measure on $[0,1]$, normalized by $v_i([0,1])=1$.
Write $F_i(x)=v_i([0,x])$. An allocation is correct if
\begin{equation}
 \alpha v_i(A_j)-v_i(A_i)\le\eps\qquad(i,j\in[n]).
 \label{r6m:eq:fairness}
\end{equation}
By Lemma~\ref{r6m:relative-equivalence}, $\eps=0$ gives the signed
multiplicative guarantee; $\alpha=1$ gives additive $\eps$-EF.
There is no common bound on total variation.

An Eval query returns one interval value. A signed query
$\operatorname{Cut}(i,a,b,t)$ is legal when
\begin{equation}
 0\le a\le b\le1,\qquad
 \min\{0,F_i(b)-F_i(a)\}\le t\le
 \max\{0,F_i(b)-F_i(a)\}.
 \label{r6m:eq:legal}
\end{equation}
It returns the least $x\in[a,b]$ satisfying $F_i(x)-F_i(a)=t$.
Continuity ensures that this root exists. The rule is fixed independently
of the protocol and its random tape.

We use the measurability assumptions of Section~\ref{sec:model}: on each
fixed profile, the existence, coordinates and replies of every numbered
call, and the event and finite encoding of a return, are measurable.
A \emph{pointwise-legal partial strategy} makes only legal calls on every
profile and tape. It may abort, return an incorrect allocation, or fail
to return. Malformed outputs are discarded. Every retained output is a
complete allocation into finite unions of intervals, with no restriction
on its number of endpoints. Endpoints themselves have zero value; a fixed
half-open convention makes their ownership immaterial.

\begin{theorem}[Finite-horizon hardness for leftmost roots]
\label{r6m:finite-horizon}
Let $\mathcal A$ be a measurable pointwise-legal partial strategy making
at most $h$ oracle calls on every profile and tape. For every
$\delta\in(0,1)$, there is one profile of finite piecewise-constant signed
densities, with all net totals one, such that
\[
 \Prb{\mathcal A\text{ returns a correct allocation}}\le\delta
\]
under the always-leftmost oracle. Output endpoints may be chosen freely
and need not have been queried.
\end{theorem}

The profile may depend on the strategy, its tape law, $h,\alpha,\eps$,
and $\delta$. It is fixed before the tape used to execute the protocol
is drawn. We establish the theorem by finitely many changes to the profile.

\subsection{A final interval that must be split}
\label{r6m:frontier}

Set $K=1+1/\alpha$. Maintain a number $\ell<1$, called the
\emph{frontier}, and a finite partition of $[0,\ell]$ into intervals
with constant density vectors $d=(d_1,\ldots,d_n)$. On each such interval,
either every $d_i\le0$, or one entry is $p>0$ and every other entry is
at most $-Kp$. On the remaining interval $(\ell,1]$, every density is
constant and strictly positive. All net totals are one, and
\begin{equation}
 F_i(x)<1\quad(x<1),\qquad F_i(1)=1.
 \label{r6m:eq:barrier}
\end{equation}
Uniform densities with $\ell=0$ satisfy these conditions.

\begin{lemma}[Unsplit final interval]
\label{r6m:tail-obstruction}
Every allocation that assigns $(\ell,1]$ to one agent violates
\eqref{r6m:eq:fairness}. This holds even if the intervals in $[0,\ell]$
are divided into arbitrary measurable pieces.
\end{lemma}
\begin{proof}
Let $c$ own $(\ell,1]$ and consider the average
\[
 E_c=\frac1{n-1}\sum_{i\ne c}
       \bigl(\alpha v_i(A_c)-v_i(A_i)\bigr).
\]
Initially assigning the entire resource to $c$ gives $E_c=\alpha$.
For a constant-density interval $J\subseteq[0,\ell]$, let $x_{Jk}$
be the fraction of its length assigned to $k$. Moving that fraction
from $c$ to $k\ne c$ decreases $E_c$ by
\begin{equation}
 \frac{x_{Jk}|J|}{n-1}
 \left(\alpha\sum_{i\ne c}d_i+d_k\right).
 \label{r6m:eq:coefficient}
\end{equation}
The parenthesized coefficient is nonpositive. This is immediate when
all entries are nonpositive or the positive entry belongs to $c$.
If the positive entry $p$ belongs to $k$, there is another index
different from $c,k$, since $n\ge3$, and
\[
 \alpha\sum_{i\ne c}d_i+d_k
 \le(\alpha+1)p-\alpha Kp=0.
\]
If the positive entry belongs to a third agent, the coefficient is
at most $\alpha p-(\alpha+1)Kp\le0$.
Summing \eqref{r6m:eq:coefficient} over intervals and recipients gives
$E_c\ge\alpha$. Some nonowner therefore has comparison gap at least
$\alpha>\eps$.
\end{proof}

\begin{lemma}[Advancing the frontier]
\label{r6m:advance}
For any $\ell<r<1$ and agent $j$, the frontier can be moved to $r$
while preserving all the stated conditions, every cumulative value
on $[0,\ell]\cup\{1\}$, and the entire valuation of agent $j$.
There is also such an update that assigns every agent density $-1$
on $(\ell,r]$, without preserving a designated agent.
\end{lemma}
\begin{proof}
Let $d_j>0$ be agent $j$'s current density on $(\ell,1]$.
Keep this row unchanged. For every $i\ne j$, use density $-Kd_j$
on $(\ell,r]$, and density
\[
 \frac{1-F_i(\ell)+Kd_j(r-\ell)}{1-r}>0
\]
on $(r,1]$. The new resolved interval has exactly the required density
pattern. Each changed cumulative first decreases from a value below one,
then increases linearly to one. Thus \eqref{r6m:eq:barrier} persists.
For the second update, use $-1$ on $(\ell,r]$ for every row and
restore its net total by a positive constant density on $(r,1]$.
Both operations introduce one resolved interval and keep all densities
finite.
\end{proof}

\subsection{Preserving every first-root reply}
\label{r6m:replay}

Call a query \emph{protected} if its bracket endpoints and its returned
root, when it is a Cut query, lie in $[0,\ell]\cup\{1\}$.

\begin{lemma}[First-root replay]
\label{r6m:selector}
Every protected query retains its legality and answer under subsequent
frontier updates. For a fixed continuous cumulative function, the
least-root response is a Borel function on the legal query domain.
\end{lemma}
\begin{proof}
All cumulative values at query endpoints remain fixed, so Eval answers,
Cut bracket totals, and legality are preserved. Let $z$ be an old
least-root reply. If $z\le\ell$, the cumulative function on $[a,z]$
is unchanged, so $z$ remains the first root. If $z=1$ and $a<1$,
the target absolute cumulative level is one; by
\eqref{r6m:eq:barrier}, this level is never reached earlier.
The degenerate query $a=b=1,t=0$ is unchanged directly.
This covers negative targets, full-bracket targets, zero targets,
and zero-total brackets.

For measurability, fix $F$. Its legal query domain is Borel.
For rational $q$, existence of a root in $[a,\min\{b,q\}]$ is
equivalent, when this interval is nonempty, to
\[
 \min_{x\in[a,\min\{b,q\}]}|F(x)-F(a)-t|=0.
\]
The minimum is continuous in its interval endpoints and target:
substitute $x=a+s(u-a)$ and minimize over the fixed compact interval
$s\in[0,1]$. The event that the least root is smaller than $z$
is the union of these Borel events over rational $q<z$.
\end{proof}

A new query to agent $j$ can always be protected. Compute its reply
on the current profile and choose $r<1$ beyond all its coordinates
smaller than one. The first update in Lemma~\ref{r6m:advance}
keeps row $j$ unchanged everywhere, so it preserves this new reply,
as well as every earlier protected query.

Likewise, a finite output can be protected by moving the frontier
past all its endpoints smaller than one using the all-negative update.
The remaining interval then has one owner, so
Lemma~\ref{r6m:tail-obstruction} makes the output incorrect.
Every later update preserves that output's values: each interval
endpoint lies in the fixed prefix or at one, where cumulative values
are unchanged. This preservation of failures is essential when a
protocol is allowed to fail with positive probability.

\subsection{One fixed hard input for randomized protocols}
\label{r6m:fixed-input}

\begin{proof}[Proof of Theorem~\ref{r6m:finite-horizon}]
Fix the tape probability space. On each current profile, describe the
$k$th call by its row and finite list of coordinates, including a Cut
reply. If the $k$th call does not exist, use the formal descriptor
$\operatorname{Eval}(1,0,0)$. This is only notation for the analysis;
the protocol does not detect nontermination or execute extra calls.
The descriptors are measurable by the assumed measurability of call
existence and transcripts. Identical real transcripts and tapes
determine identical internal computation, including a return or a
nonreturn after the last call.

Maintain a measurable set $H$ of tapes whose executions already fail
on the current profile. Their actual queries are protected, and the
endpoints of any returned allocation are protected as well. All future
frontier updates preserve these executions and their output values,
so no tape in $H$ becomes successful. Initially $H=\varnothing$.

Let $S=\Omega\setminus H$ have mass $u>0$. We capture a fixed fraction
of this mass in one batch of at most $h+1$ updates. At stage
$k\in\{1,\ldots,h\}$, the first $k-1$ descriptors of the surviving
tapes are already protected. Evaluate their $k$th descriptors on the
current profile. Some row $j$ occurs on a measurable subset of at least
$1/n$ of the surviving mass. On each tape in this subset, the maximum
of $\ell$ and the descriptor coordinates smaller than one is strictly
less than one. Continuity from below of probability therefore gives a
fixed $r\in(\ell,1)$ that bounds these coordinates on at least half
the subset. Retain those tapes and perform the update preserving
row $j$.

The selected query's answer is unchanged because its entire row is
unchanged. Previous answers replay by Lemma~\ref{r6m:selector}.
If the descriptor is formal, the preceding real transcript and tape
already determine the absence of another call. Thus all first $k$
descriptors are now protected. Each stage retains at least a
$1/(2n)$ fraction of the current mass. After $h$ stages, the surviving
mass is at least $u/(2n)^h$, and every actual query on those executions
is protected.

For each surviving tape with a finite return, take the maximum of
$\ell$ and its output endpoints smaller than one; for a nonreturn,
use $\ell$. These are measurable quantities strictly below one,
even though the number of endpoints is unbounded. Choose a further
fixed $r<1$ that bounds them on at least half the surviving mass,
and make the all-negative frontier update. All actual transcripts
replay, so every finite output is unchanged and leaves $(r,1]$
unsplit. It is incorrect by Lemma~\ref{r6m:tail-obstruction}.
Nonreturns remain nonreturns. Add the retained tapes to $H$.
The newly captured mass is at least
\begin{equation}
 \gamma u,\qquad \gamma=\frac1{2(2n)^h}.
 \label{r6m:eq:capture}
\end{equation}
The same argument applies when $h=0$, using only the output update.

After $B$ batches, the uncaptured mass is at most $(1-\gamma)^B$.
Choose a finite $B$ for which this is at most $\delta$.
There have been at most $(h+1)B$ updates, so the final profile has
finitely many constant-density intervals and finite total variation.
At least $1-\delta$ of its tapes fail.

Every row choice and coordinate quantile was selected using the fixed
tape law on one fixed intermediate profile. Only finitely many such
profiles and measurable sets were used. The resulting input is fixed
before the execution tape is drawn; no input is chosen in response
to that realized tape. No measurability over a space of profiles or
infinite limit of profiles is needed.
\end{proof}

\subsection{Expected queries and rightmost roots}
\label{r6m:expected}

\begin{proof}[Proof of Theorem~\ref{thm:mixed-mc}]
Suppose the original protocol has probability at least $p>0$ of a
legal correct finite return on every input with the always-leftmost
oracle. Let $\mathcal S$ be its success event and $Q$ its number of
attempted calls. Fix an integer $q\ge0$.

Simulate the protocol, aborting before its $(q+1)$st attempted call.
Reject malformed syntax, invalid rows, and invalid bracket endpoints.
Before each simulated Cut, Eval its bracket and check
\eqref{r6m:eq:legal}; abort if the target is illegal, and otherwise
issue the Cut. Hide the extra Eval reply from the simulated protocol.
At a finite return, discard malformed allocations and normalize endpoint
ownership, without testing any value inequality or limiting the number
of endpoints. Permit all query-free computation after the $q$th call;
a subsequent nonreturn is retained.

This is a pointwise-legal partial strategy with at most $2q$ actual
calls. The fixed least-root rule returns exactly the original answers
on legal simulated histories. Its correct-return event is therefore
$\mathcal S\cap\{Q\le q\}$. For any $0<\delta<p$,
Theorem~\ref{r6m:finite-horizon} gives a fixed profile with
\[
 \Prb{\mathcal S,\ Q\le q}\le\delta.
\]
On that same profile, $\Prb{\mathcal S}\ge p$, and hence
\begin{equation}
 \E{Q}\ge(q+1)\Prb{\mathcal S,\ Q>q}
          \ge(q+1)(p-\delta).
 \label{r6m:eq:expected}
\end{equation}
Since $q$ is arbitrary, no uniform finite bound on expected queries
exists.

For an always-rightmost protocol, reflect the input by $x\mapsto1-x$.
A Cut on $[a,b]$ with target $t$ corresponds to a Cut on
$[1-b,1-a]$ with target $B-t$, where $B$ is the bracket value.
The legal target interval is preserved and spatial order reverses.
One Eval obtains $B$ and one least-root Cut gives the greatest root
on the reflected input. Eval queries and output intervals reflect
directly. A uniformly bounded always-rightmost protocol would
therefore yield an always-leftmost protocol with at most twice the
query cost, the same fairness guarantee, and the same success
probability, contradicting the preceding conclusion.
\end{proof}

Taking $\eps=0$ proves the multiplicative claim; taking $\alpha=1$
proves the additive claim. At $\eps\ge\alpha$, assigning the entire
resource to one agent satisfies \eqref{r6m:eq:fairness} without queries,
so the tolerance range is sharp on this normalization. The result
concerns a uniform bound over inputs. It neither rules out a finite
protocol on each individual input nor asserts that a single input
must have infinite expected cost. The construction uses unbounded
variation across inputs and gives no impossibility for a class with
a fixed common variation bound or for variation-normalized error.

\section{Formal verification artifact}\label{sec:formal}
The accompanying artifact~\cite{FHArtifact26} formalizes the four main
results in Lean. The statements use adaptive Eval/Cut protocols and finite
interval representations of allocations. The upper-bound developments
construct the physical allocations and account for every oracle query;
the lower-bound developments include the transcript and probability
arguments establishing a fixed hard input.

\begin{center}\small
\renewcommand{\arraystretch}{1.12}
\begin{tabular}{@{}>{\raggedright\arraybackslash}p{.23\linewidth}>{\raggedright\arraybackslash}p{.71\linewidth}@{}}
\toprule
Main result & Verified statement\\\midrule
Deterministic chores & A complete exact EF allocation from the concrete bounded protocol, with the explicit query envelope and the leading exponent $n\log_2 n+O(n)$.\\
\midrule[.25pt]
Randomized upper bounds & A measurable query protocol for cake and chores, almost-sure termination, exact EF on return, and the stated polynomial expected query and Cut bounds, including rejected attempts.\\
\midrule[.25pt]
Randomized lower bound & The explicit expected-query lower bound, linear in the required success probability, including arbitrary finite outputs whose endpoints need not have been queried.\\
\midrule[.25pt]
Mixed impossibility & The multiplicative and additive impossibility statements for signed valuations, including prescribed leftmost-root and rightmost-root replies.\\
\bottomrule
\end{tabular}
\end{center}

The formal algorithms make conservative implementation choices: a fixed
sufficient clearance length and explicit endpoint-recording costs for
chores, and a separate charge for each normalization step in the randomized
protocol. These choices preserve the main bounds. The artifact does not
claim that every sharper auxiliary ledger in the paper is formalized
verbatim. Its endpoint manifest identifies the public Lean statements
corresponding to the four theorems; those statements record the precise
hypotheses.

The source package pins Lean 4.35.0, mathlib, and the fixed-point library
used for the chore construction. The verifier rejects proof placeholders
and project axioms and checks every included theorem's transitive axiom
dependencies. Only \texttt{propext}, \texttt{Classical.choice}, and
\texttt{Quot.sound} are allowed. The package includes build instructions,
source hashes, and independent correspondence reviews.

\section*{AI methodology}
\phantomsection\label{sec:ai-methodology}
OpenAI Codex, using the Astra model, materially supported literature discovery
and comparison, conjecture and proof development, proof checks, exact
experiments, Lean formalization, and drafting throughout this article.
Fable 5.2 also contributed proposed upper-bound improvements that informed
the revised proofs. The artifact preserves sources and review reports.
Appendix~\ref{sec:formal} describes the formal verification coverage.

\end{document}